\documentclass{llncs}            

\usepackage[breaklinks=true]{hyperref} 
\hypersetup{
	colorlinks=true,
	citecolor=teal,
	linkcolor=blue,  
}

\usepackage{latexsym,amsfonts}
\usepackage{amsmath}
\usepackage{amssymb}
\usepackage{array}
\usepackage[english]{babel}

\usepackage{graphicx,color}
\usepackage{latexsym,amsfonts}
\usepackage{amsmath}
\usepackage{amssymb}
\usepackage{bbm}

\usepackage{algorithmic}
\usepackage{algorithm}
\usepackage{tikz}
\usepackage{url}
\usepackage{fullpage}

\usepackage[T1]{fontenc}

\newcounter{lpnumber} 
\newenvironment{linearprogram}[1]
{ \stepcounter{lpnumber}
	\begin{gather} #1 \tag{LP\arabic{lpnumber}} \\[-5ex] \notag
	\end{gather}
	\hspace{1.5cm} subject to \\[-3ex]
	\align }
{ \endalign }
\newcommand{\minimize}[1]{\text{minimize} \ #1}
\newcommand{\maximize}[1]{\text{maximize} \ #1}

\newcommand{\opt}{\mathsf{opt}}

\newcommand{\Nbr}{\mathsf{Nbr}}
\newcommand{\wt}{\mathsf{wt}}
\newcommand{\conv}{\mathsf{conv}}
\newcommand{\vote}{\mathsf{vote}}

\newcommand{\cost}{\mathsf{cost}}

\newtheorem{new-claim}{Claim}
\newtheorem{obs}{Observation}
\newtheorem{Definition}{Definition}

\begin{document}

	\title{Quasi-popular Matchings, Optimality, and Extended Formulations}
	\author{Yuri Faenza\inst{1} and Telikepalli Kavitha\inst{2}\thanks{Supported by the DAE, Government of India, under project no. 12-R\&D-TFR-5.01-0500. Part of this work was done while visiting MPI for Informatics, Saarland Informatics Campus, Germany.}}
	\institute{IEOR, Columbia University, New York, USA. \email{yf2414@columbia.edu} \and Tata Institute of Fundamental Research, Mumbai, India. \email{kavitha@tifr.res.in}}
	
	\maketitle

	\begin{abstract}
		Let  $G = (A \cup B,E)$ be an instance of the stable marriage problem where every vertex ranks its neighbors 
		in a strict order of preference. A matching $M$ in $G$ is {\em popular} if $M$ does not lose a head-to-head election against any matching. 
		Popular matchings are a well-studied generalization of stable matchings, introduced with the goal of enlarging the set of admissible solutions, while maintaining a certain level of fairness. Every stable matching is a min-size popular matching.  Unfortunately, when there are edge costs,
		it is NP-hard to find a popular matching of minimum cost  --
		even worse, the min-cost popular matching problem is hard to approximate up to any factor.  \\ 
		Let $\opt$ be the cost of a min-cost popular matching. Our goal is to efficiently compute a matching of cost at most $\opt$
		by paying the price of mildly relaxing  popularity. 
		Our main positive results are two bi-criteria algorithms that find in polynomial time a {\em near-popular} or ``quasi-popular''  matching of cost at most $\opt$. Moreover, one of the algorithms finds a quasi-popular matching of cost at most that of a min-cost popular fractional matching, which could be much smaller than $\opt$. 
		
		Key to the other algorithm are 
		new results for certain polytopes. 
		In particular, we give a polynomial-size extended formulation for an integral polytope sandwiched between the popular and quasi-popular matching polytopes. We complement these results by showing that it is NP-hard to find a quasi-popular matching of minimum cost, 
		and that both the popular and quasi-popular matching polytopes have near-exponential extension  complexity.

		\smallskip

		This version of the paper goes beyond the conference version~\cite{FK20} in the following two points: (i)~the algorithm for finding a quasi-popular matching of cost at most that of a min-cost popular fractional matching is new;
		(ii)~the proofs from Section~\ref{sec:ec-P} and Section~\ref{sec:lb-facets} are now self-contained (the conference version used constructions from \cite{CFKP18} to show these lower bounds).
		
	\end{abstract}

	\vspace{-.5cm}
	
	\section{Introduction}\label{intro} Consider a bipartite graph $G = (A \cup B, E)$ on $n$ vertices and $m$ edges where every vertex ranks its neighbors in a strict order of preference. Such an instance,
	commonly referred to as a {\em marriage} instance, is a classical model in two-sided matching markets. 
	A matching $M$ is stable if there is no {\em blocking pair} with respect to $M$: a pair $(a,b)$ blocks $M$ if both $a$ and $b$ prefer each other to their respective assignments in $M$. The notion of stability was introduced by Gale and Shapley~\cite{GS62} in 1962 who showed that stable matchings always exist in $G$ and one such matching can be efficiently computed.  
	
	A broad class of objectives can be captured by defining a function $\cost: E \rightarrow \mathbb{R}$ and asking for a stable matching whose 
	sum of edge costs is minimized. Thus the min-cost stable matching problem includes several stable matching problems such as finding one
	with max-utility or with min-regret, or one with given forced and forbidden edges. 
	More generally, a cost function allows a decision-maker to ``access'' the whole family of stable matchings (possibly of exponential size), while the Gale-Shapley algorithm will always return the same stable matching (i.e., the one that is optimal for one side of the bipartition). 
	There are several polynomial time algorithms to compute a min-cost stable matching
	and special variants of this problem~\cite{Fed92,Fed94,ILG87,Rot92,TS98,VV89}. 
	
	Stable matchings and their extensions are used in many optimization problems in computer science, economics, and operations research, such as matching students to schools/colleges~\cite{AS03,BCCKP18} and medical interns to hospitals~\cite{CRMS,NRMP}.  
	Stability is a strict condition. 
	For instance, all stable matchings have the same size~\cite{GS85} which may be only half the size of a max-size matching in $G$. While
	matching students to schools/colleges or medical interns to hospitals, one seeks larger matchings. 
	On the other hand, one does not want to ignore vertex preferences and impose a max-size matching. Thus, what we seek is a weaker notion of stability that captures ``overall stability'' (rather than forbid blocking edges)
	and achieves more social good, i.e., a better value (than the best stable matching) with respect to the given cost function.
	
	\smallskip
	
	\noindent{\bf Popularity.}
	A natural relaxation of stability is {\em popularity}. Motivated by problems in cognitive science~\cite{Gar75}, it has been  extensively studied in the computer science community: 
	see~\cite{Cseh} for a survey.  Consider an election between two matchings $M$ and $N$: here each vertex casts a vote for the matching where it gets assigned the more preferred partner (being left unmatched is its
	least preferred state)
	and it abstains from voting if its assignment is the same in $M$ and $N$. 
	Let $\phi(M,N)$ 
	(resp., $\phi(N,M)$) be the number of votes for $M$ (resp., $N$) in this election. We say $N$ is \emph{more popular} than $M$ if 
	$\phi(N,M) > \phi(M,N)$. 
	
	\begin{definition}
		\label{def:popular}
		A matching $M$ is popular if there is no matching in $G$ that is more popular than $M$, i.e.,  $\phi(M,N) \ge \phi(N,M)$ for all matchings $N$ in $G$.
	\end{definition}
	
	A popular matching never loses a head-to-head election against any matching. That is, it is a {\em weak 
		Condorcet winner}~\cite{Con85,condorcet} in the voting instance where matchings are the candidates and vertices are voters. 
	So, no election  can force a migration from a popular matching to some other matching. 
	
	Consider $G = (A \cup B, E)$ where $A = \{a_1,a_2\}$, 
	$B = \{b_1,b_2\}$, and $E = \{(a_1,b_1),(a_1,b_2),(a_2,b_1)\}$. Suppose $a_1$ prefers $b_1$ to $b_2$ and similarly, $b_1$ prefers $a_1$ to 
	$a_2$. This instance admits only one stable matching $S = \{(a_1,b_1)\}$;  the max-size matching $M = \{(a_1,b_2),(a_2,b_1)\}$ is not stable but it is popular.
	In an election between $S$ and $M$, the vertices $a_1,b_1$ vote for $S$ and the vertices $a_2,b_2$ vote for $M$.
	
	The notion of popularity was introduced by G\"ardenfors~\cite{Gar75} in 1975 
	who showed that every stable matching is popular; in fact, every stable matching is a min-size popular matching~\cite{HK11}. Thus, we can obtain larger matchings (as in the example above) by relaxing stability to popularity. There are efficient algorithms to compute a max-size popular 
	matching~\cite{HK11,Kav14} in $G$ and the size of a max-size popular matching is at least $2|M_{\max}|/3$, where $M_{\max}$ 
	is a max-size matching in $G$. 
	
	Though computing a min-size/max-size popular matching is easy, it is NP-hard to decide if $G$ admits a popular matching that is neither a min-size nor a max-size popular matching~\cite{FKPZ18}. It was also shown in~\cite{FKPZ18} that it is NP-hard to compute a min-cost popular matching in $G = (A \cup B,E)$; moreover, this problem is NP-hard to approximate to any multiplicative factor even in the restricted case when every edge has cost 0 or 1.

	\medskip
	
	\noindent{\bf  Relaxing Popularity.} 
	Though popularity is a natural notion of global stability and a min-cost popular matching is 
	more optimal (wrt its cost) than a min-cost stable matching, the fact that it is NP-hard to approximate to any factor a min-cost popular matching represents a computational barrier. For the sake of regaining the computational tractability lost when relaxing stability to popularity, 
	it makes sense to also relax popularity and replace it with {\em near-popularity}.
	This poses the following question: how to define a matching that is ``close'' to being popular? The literature has already provided a suitable concept: the \emph{unpopularity factor} of a matching, introduced in~\cite{McC06} and studied, e.g., in~\cite{BHHKW15,HK13,Kav14,KMN09,Rua}. Given a matching $M$, its unpopularity factor is: 
	\[ u(M) = \max_{\substack{N\in{\cal M}_G\\N\ne M}}\frac{\phi(N,M)}{\phi(M,N)},\]
	\noindent where ${\cal M}_G$ is the set of matchings in $G$. Thus, in an election between $M$ and any other matching,
	the ratio $|\{$vertices {\em against} $M\}|/|\{$vertices {\em for} $M\}|$ is bounded from above by $u(M)$.
	Note that $u(M) \in \mathbb{Q}_{\ge 0} \cup \{\infty\}$. The function {\em unpopularity factor} on ${\cal M}_G$, when it is not $\infty$, 
	captures therefore the gamut of different matchings $M$ in $G$ that are \emph{Pareto-optimal}. A matching $M$ is Pareto-optimal if there is no matching (other than $M$) where every vertex is matched to a partner at least as good as in $M$. 
	A matching $M$ is popular iff $u(M) \le 1$, while a matching $M$ is Pareto-optimal iff $u(M) < \infty$.

	Interestingly, all known algorithms~\cite{BHHKW15,HK13,Kav14} for computing unpopular matchings with bounded unpopularity factor bound 
	$u(M)$, where $M$ is the output matching, by a value 2 or more\footnote{The symmetric difference $M \oplus N$ of 2 matchings $M$ and $N$ may have {\em short} alternating paths wrt $M$ which force $u(M)~\ge~2$. Forbidding such paths for an unpopular matching $M$ is necessary to ensure $u(M) < 2$.} (see Section~\ref{sec:background}). Thus, matchings $M$ with $u(M)\leq 2$ can be regarded as being close to popular. 
	
	Observe that no matching wins more than $1/2$-fraction of the votes cast in its head-to-head election against a popular matching. Similarly, no matching wins more than $2/3$-fraction of the votes cast in its head-to-head election against a  matching
	with unpopularity factor at most 2.
	
	\begin{Definition}
		\label{def:quasi-pop}
		A matching $M$ in $G = (A \cup B, E)$ is {\em quasi-popular} if $u(M) \le 2$.
	\end{Definition}
	
	Summarizing, for the sake of efficiency in computation, when comparing two matchings, we are ready to relax {\em never losing} 
	(the definition of popular matchings) to {\em losing within a factor bounded by~2} (the definition of quasi-popularity). Note that
	if we scale the votes in favor of $M$ by 2, then a quasi-popular matching $M$ never loses an election. We show that relaxing popularity to quasi-popularity allows us to design efficient algorithms for finding good matchings in $G$. 
	
	\subsection{Our Results and Techniques} Given the discussion above, our objective would be to show an efficient algorithm for the min-cost quasi-popular matching problem. However, as we show here, this problem is as hard as the starting problem -- i.e., computing a min-cost popular matching. 
	
	\begin{theorem}\label{thm:hardness}
		Given a marriage instance $G = (A \cup B, E)$ with a function $\cost: E \rightarrow \{0,1\}$, it is NP-hard to compute a min-cost
		quasi-popular matching in $G$. Moreover, it is NP-hard to   approximate its cost within any multiplicative factor.
	\end{theorem}
	
	Thus finding or approximating a min-cost popular or quasi-popular matching is hard. 
	Our next attempt is to overcome this hardness by considering the following intriguing question:
	can we find a quasi-popular matching of cost at most that of a min-cost popular matching? 
	Surprisingly (in the midst of all these hard problems), this problem is tractable and this is our main result.

	\begin{theorem}
		\label{thm:algo}
		Given a marriage instance $G = (A \cup B, E)$ with a function $\cost: E \rightarrow \mathbb{R}$, there is a polynomial time algorithm 
		to compute a quasi-popular matching $M$ such that $\cost(M) \le \opt$, where $\opt$ is the cost of a min-cost popular matching in $G$.
	\end{theorem}

	We give two proofs of Theorem~\ref{thm:algo}, based on two different algorithms. 
	The first approach is as follows. Let ${\cal M}_1$ (respectively, ${\cal M}_2$) be the set of popular (resp., quasi-popular) matchings in $G$. 
	We will show a set ${\cal M}^*$ such that (1)~${\cal M}_1 \subseteq  {\cal M}^* \subseteq {\cal M}_2$ and 
	(2)~$\conv({\cal M}^*)$ admits a formulation  
	of size $O(m+n)$ in $\mathbb{R}^{m+n}$, where
	$\conv({\cal M}^*)$ is the convex hull of edge incidence vectors of matchings in ${\cal M}^*$.
	Linear programming over this formulation  
	yields Theorem~\ref{thm:algo}. 
	
	This result is proved in Section~\ref{sec:algo}, where
	we show that $\conv({\cal M}^*)$ is a linear projection of one of the faces
	${\cal F}_{G^*}$ of a new extension\footnote{Using the terminology from polyhedral combinatorics, we say that a polytope $Q$ that linearly projects to a polytope $P$ is an \emph{extension} of $P$, and that a linear description of $Q$ is an \emph{extended formulation} for $P$.} of the {\em dominant matching} polytope of a supergraph $G^*$ of $G$.
	
	\begin{Definition}
		\label{def:dominant}
		A popular matching $M$ in $G$ is {\em dominant} if $M$ is more popular than every strictly larger matching, i.e.,
		$\phi(M,M') > \phi(M',M)$ for every strictly larger matching $M'$.
	\end{Definition}
	
	\smallskip
	
	Every dominant matching is a max-size popular matching (by Definition~\ref{def:dominant}). 
	Dominant matchings always exist in a marriage instance $G$~\cite{HK11}. 
	Dominant matchings coincide with the linear image of stable matchings in a related bipartite graph $G'$ on $n$ vertices and $2m$ edges~\cite{CK16,FKPZ18}. Hence 
	an extension of the dominant matching polytope ${\cal D}_G$ in $\mathbb{R}^{2m}$ (as the linear image of the stable matching polytope of $G'$) was known. We obtain the above polytope ${\cal F}_{G^*}$ as a face of a new
	extension of ${\cal D}_{G^*}$ of size $O(m+n)$ in $\mathbb{R}^{m+n}$ (see Theorem~\ref{thm:extra-dominant-extn} and Theorem~\ref{thm:dominant-extn}). The main tool for proving the integrality of this new 
	extension is a compact extended formulation of the popular matching polytope in a marriage instance where every stable matching is perfect~\cite{HK17}.
	
	The second algorithm shows actually a result stronger than Theorem~\ref{thm:algo}: we show that we can obtain in polynomial time a quasi-popular matching of cost at most that of a min-cost \emph{popular fractional matching}. The set of all popular fractional matchings includes all popular matchings. We defer a discussion on popular fractional matchings and a description of their polytope to Section~\ref{prelims}.

	\medskip
	
	\noindent{\bf  Polyhedral Results.} 
	It is a natural question whether the positive/negative results on popular and quasi-popular matchings are mirrored by the sizes of the descriptions of the associated polytopes. Let ${\cal P} = \conv({\cal M}_1)$ (respectively, ${\cal Q} = \conv({\cal M}_2)$) be the popular (resp., quasi-popular) matching polytope 
	of $G$, for  ${\cal M}_1$ and ${\cal M}_2$ as defined above (just below Theorem~\ref{thm:algo}). 
	We prove that both ${\cal P}$ and ${\cal Q}$ have near-exponential 
	{\em extension complexity}  (see Section~\ref{sec:lb-xc} for definitions). These results follow from our hardness result and a new reduction that we show here, along with known lower bounds in polyhedral combinatorics. In particular, we show that a face of ${\cal P}$ (resp., ${\cal Q}$) is an extension of certain independent set polytopes for which a lower bound on the extension complexity is
	known~\cite{Goos-et-al}.
	
	\begin{theorem}
		\label{thm:xc}
		The extension complexity of the polytope ${\cal P}$ (also, of the polytope ${\cal Q}$) is $2^{\Omega\left(\frac{m}{\log m}\right)}$.
	\end{theorem}
	
\begin{figure}[h]
		\begin{minipage}{.45\textwidth}
			\begin{center}\includegraphics[scale=.18]{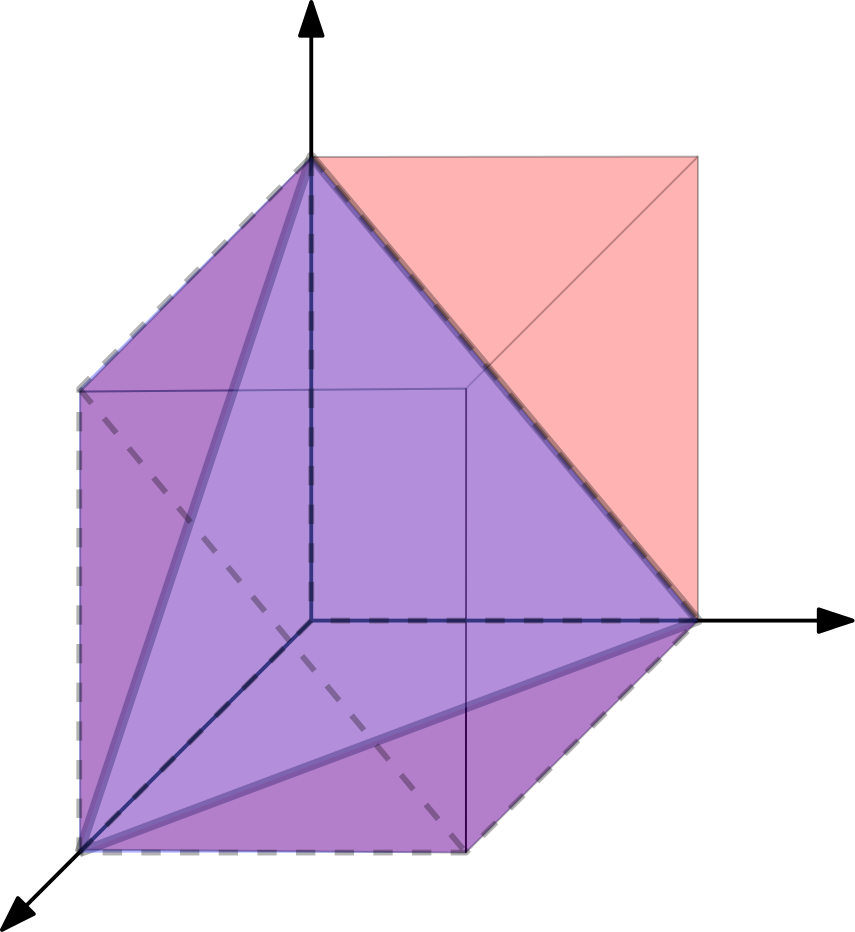}\caption{The dashed wedge is sandwiched between the tetrahedron and the cube. Similarly, $\conv({\cal M}^*)$, which has a compact extended formulation, is sandwiched between the popular matching polytope ${\cal P}$ and the quasi-popular matching polytope ${\cal Q}$ -- neither of which has a compact extended formulation.}
				\label{fig:sandwich}\end{center}
		\end{minipage}\begin{minipage}{.1\textwidth} \hspace{1cm}
	\end{minipage}\begin{minipage}{.45\textwidth}
	\begin{center}\includegraphics[scale=.155]{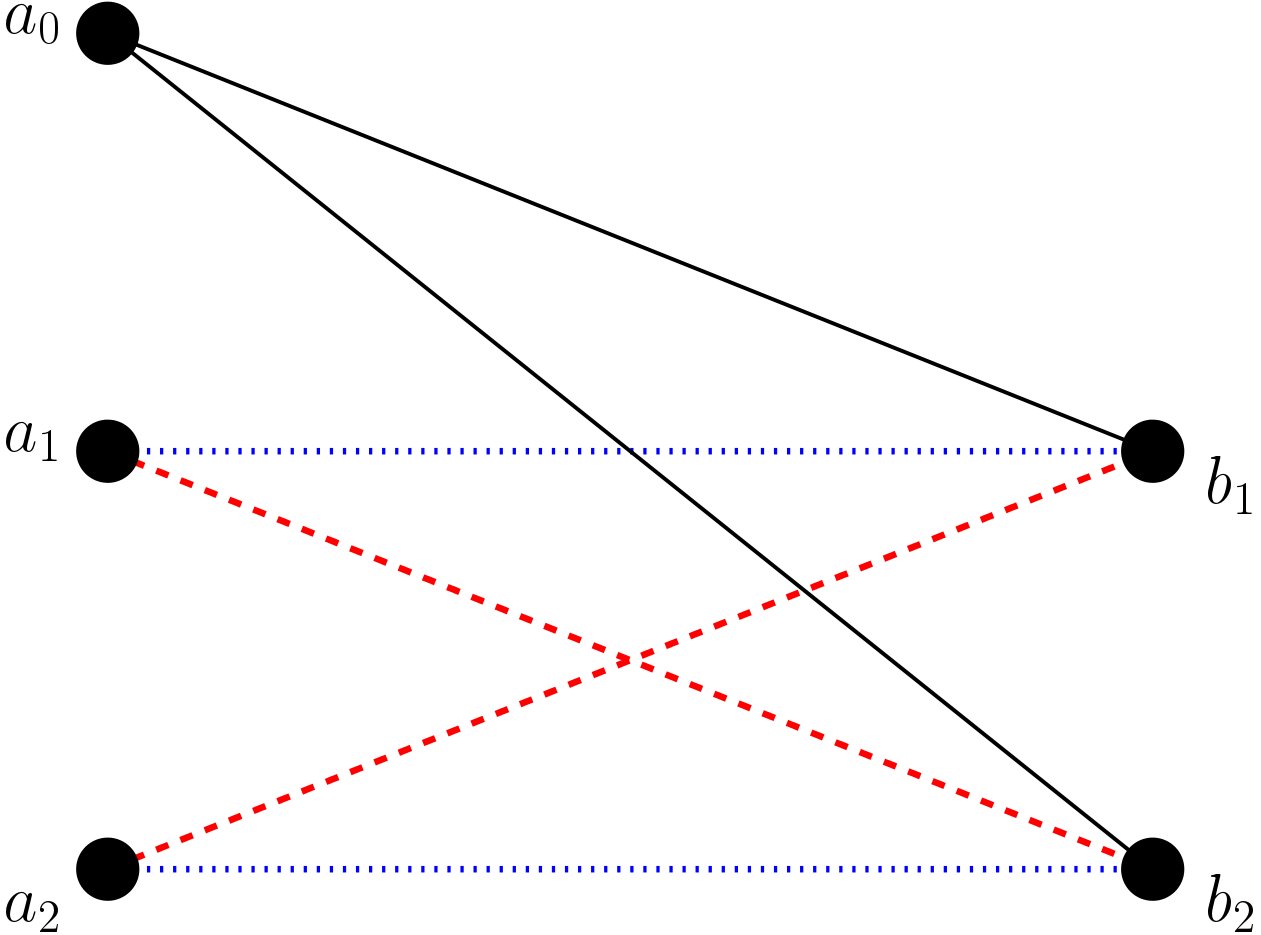}
		\caption{An instance where the popular fractional matching polytope is not integral.\vspace{.8cm}}
		\label{fig:non-int}
	\end{center}
\end{minipage}
\end{figure}

	The proof of Theorem~\ref{thm:algo} shows that $\conv({\cal M}^*)$ is an ``easy-to-describe''
	integral polytope sandwiched between two hard ones, see Fig.~\ref{fig:sandwich} for a representation. As $\conv({\cal M}^*)$ has a formulation of size $O(n+m)$, it is, in a way, a polytope with ``smallest extension complexity'' sandwiched between ${\cal P}$ and ${\cal Q}$. Theorem~\ref{thm:xc} actually shows that the face $F'$ of ${\cal P}$ containing all and only the popular matchings of maximum size has near-exponential extension complexity. This is in sharp contrast with the dominant matching polytope ${\cal D}_G$ (contained in $F'$), that admits a compact extended formulation. However, unlike stable matchings whose polytope in $\mathbb{R}^m$ has a linear number of facets~\cite{VV89,Rot92}, we show that ${\cal D}_G$ does not admit a polynomial-size description in $\mathbb{R}^m$. In fact, a complete linear description of ${\cal D}_G$ in  the original space was not known so far. We give one here, and show that ${\cal D}_G$ has an exponential number of facets (in the size of the graph). Recall that ${\cal M}_G$ is the set of matchings in $G = (A \cup B, E)$ and $|E| = m$.

	\begin{theorem}
		\label{thm:dom}
		The dominant matching polytope ${\cal D}_G$ admits a formulation of size $O(|{\cal M}_G|)$ 
		in $\mathbb{R}^m$. Moreover, there exists a constant $c > 1$ such that ${\cal D}_G$ has $\Omega(c^m)$ facets.
	\end{theorem}
	
	Our results on the dominant matching polytope are proved in Section~\ref{sec-new:polytopes}.
	
\medskip

\noindent{\bf  Optimal Quasi-popular Matchings.}
In order to prove Theorem~\ref{thm:hardness}, we first prove structural results on quasi-popular matchings. Stable matchings (by definition) and popular matchings (see \cite{HK11}) have simple {\em forbidden structures} in terms of blocking edges---to come up with such forbidden structures for quasi-popular matchings seems much more complex. Hence, we do not pursue this combinatorial approach. 
Instead, we extend the LP-method used for popular fractional matchings~\cite{HK17,Kav16,KMN09} to design an appropriate dual certificate or {\em witness} (a vector in $\mathbb{R}^n$: see Section~\ref{prelims} for details) for quasi-popularity, and deduce Theorem~\ref{thm:hardness} (proved in Section~\ref{sec:hardness}) from this.

\medskip

\noindent{\bf The Popular Fractional Matching Polytope.}
The popular fractional matching polytope  ${\cal L}_G$ is a relaxation of the popular matching polytope that contains, roughly speaking,
all points of the matching polytope that satisfy a linearization of the popularity constraint.
The polytope ${\cal L}_G$ need not be integral. An instance
from \cite{HK17} is included in Fig.~\ref{fig:non-int}, where $a_1$ is the top choice of $b_1, b_2$ while $a_2$ is their second  choice and $a_0$ is their last choice;
similarly, $b_1$ is the top choice of $a_0,a_1,a_2$ and $b_2$ is their second choice.

The blue dotted matching $\{(a_1,b_1),(a_2,b_2)\}$ is the only popular matching here. 
The red dashed matching $\{(a_1,b_2),(a_2,b_1)\}$ is not popular since the matching $\{(a_0,b_2),(a_1,b_1)\}$ is more popular. However there is a convex combination of the red dashed and blue dotted matchings that is popular: this is the
half-integral matching $\vec{x}$ where $x_e = 1/2$ for all $e$ in $\{(a_1,b_1),(a_1,b_2),(a_2,b_1),(a_2,b_2)\}$. 

The polytope ${\cal L}_G$ is half-integral~\cite{HK17} and
a compact extended formulation of ${\cal L}_G$ was given in \cite{KMN09}.
Thus, we can find a min-cost popular fractional matching $\vec{q}$ in $G$ in polynomial time. 
We show that $\vec{q}$ can be expressed as the convex combination of the edge incidence vectors of two quasi-popular matchings in $G$, thus obtaining in polynomial time a quasi-popular matching of cost at most $\opt^*$, where $\opt^*$ is the cost of $\vec{q}$.

We remark that all our results -- both positive and negative -- hold for bipartite graphs. Our proof techniques borrow and extend tools from
various approaches to matchings and popular matchings from the literature.

\medskip

\noindent{\bf Conference version.} This version of the paper goes beyond the conference version~\cite{FK20} in the following two points:
(i)~the algorithm for finding a quasi-popular matching of cost at most that of a min-cost popular fractional matching is new;
(ii)~the proofs from Section~\ref{sec:ec-P} and Section~\ref{sec:lb-facets} are now self-contained (the conference version used constructions from \cite{CFKP18} to show these lower bounds).

\subsection{Background and Related Results}
\label{sec:background}
Algorithmic questions for popular matchings were first studied in the one-sided preferences model, where it is only vertices in $A$ that have preferences over their neighbors and cast votes (vertices in $B$ are objects). Popular matchings need not exist here and an efficient algorithm was given in 
\cite{AIKM07} to determine if an instance admits a popular matching or not. McCutchen~\cite{McC06} introduced the measure of unpopularity 
factor and showed that computing a matching with {\em least unpopularity factor} in the one-sided preferences model is NP-hard. Interestingly,
in the one-sided preferences model, $u(M) \ge 2$ for any unpopular matching $M$~\cite{McC06}.

When vertices on both sides have strict preferences, i.e., in a marriage instance,  stable matchings always exist and hence popular 
matchings always exist. As mentioned earlier, efficient algorithms are known to compute max-size popular matchings~\cite{HK11,Kav14}.
These matchings compute dominant matchings (see Definition~\ref{def:dominant}). A linear time algorithm was given in \cite{CK16} to decide if $G$ has a popular matching with a given edge $e$: it was shown that
it was enough to check if there was either a stable matching or a dominant matching with the edge $e$.  A compact extended formulation of the popular fractional matching polytope in the one-sided preferences model was given in \cite{KMN09}, where it was  
shown that popular {\em mixed} matchings always exist and can be computed in polynomial time.
This formulation was extended to the two-sided preferences model in \cite{Kav16} and analyzed in \cite{HK17} where half-integrality of the popular fractional matching polytope was shown.

\smallskip

\noindent \emph{Bounded Unpopularity Factor.}
A size-unpopularity factor trade-off in a marriage instance $G = (A\cup B, E)$ was shown in \cite{Kav14}.
Matchings with low unpopularity factor in dynamic matching markets 
were studied in \cite{BHHKW15}. 
Popular matchings were also studied in the roommates model where the graph $G$ need not be bipartite -- popular matchings need not exist 
here and it was shown in \cite{HK13} that $G$ always admits a matching with unpopularity factor $O(\log n)$  and there are
indeed instances where every matching has unpopularity factor $\Omega(\log n)$. 

\smallskip

\noindent \emph{Hardness Results.}
It was recently shown~\cite{FKPZ18,GMSZ18} that it is NP-hard to decide if a roommates instance admits a popular matching or not. Several
hardness results for popular matchings in a marriage instance $G$ were shown in \cite{FKPZ18}: these include (i) the hardness
of deciding if $G$ admits a popular matching that is neither a stable nor a dominant matching and (ii)~the hardness of deciding 
if $G$ admits a popular matching that contains/forbids two given edges $e$ and $e'$. 
This showed that it is NP-hard to compute a max-utility popular matching when edge utilities are non-negative; this problem admits a 2-approximation and it is NP-hard to approximate it to a better factor~\cite{FKPZ18}. 

When all popular matchings in $G$ have the same size, some NP-hardness results for stable/popular matchings were shown in \cite{CFKP18}: these include deciding if $G$ admits a popular matching that is {\em not} dominant and if $G$ has a stable matching that is also dominant.

\smallskip

\noindent \emph{Bi-criteria Approximation.}
Using the notation of bi-criteria approximation algorithms (see e.g.~\cite{k-means}), our algorithm (from Theorem~\ref{thm:algo}) is a $(1,2)$ approximation for the min-cost popular matching problem, where the first entry denotes the ratio to the cost of the min-cost popular matching, and the second the unpopularity factor. Bi-criteria approximation algorithms have been developed for e.g.~bounded-degree spanning tree~\cite{goemans,singh} and $k$-means~\cite{k-means} problems. To the best of our knowledge, it is the first time that such an algorithm is proposed for a matching problem under preferences. 

\smallskip

\noindent\emph{Polytopes and Complexity.}
Linear programming is a classical tool for solving combinatorial optimization problems in general, and matching problems in particular, see e.g.~\cite{CCZ,Sch03}. Much research has been devoted to give complete linear  descriptions (of polynomial size) for polytopes associated to those problems, or showing lower bounds on the size of any such description. 
Although one expects those bounds to be related to the complexity of the associated optimization problems, this is not always the case: 
e.g., the matching polytope does not have a compact extended formulation~\cite{rothvoss}. It is also not true that NP-hardness proofs imply strong lower bounds on the extension complexity: 
there exists an $O(\sqrt{n})$-approximated extended formulation of polynomial size for the independent 
set polytope of a graph on $n$ vertices~\cite{Bazzi-et-al} despite the hardness results~\cite{Hastad} on the problem.

\section{Preliminaries}\label{prelims} Our input is a bipartite graph $G = (A \cup B, E)$ on $n$ vertices and $m$ edges, where each vertex ranks its neighbors in a strict preference order. The earliest characterization of popular matchings was given in \cite{HK11}: this was a combinatorial characterization.

The first characterization of popular matchings in terms of a dual certificate or a succinct witness of popularity was given in \cite{Kav16}. 
The popular fractional matching polytope in a marriage instance was analyzed in \cite{HK17} where its half-integrality was shown. We now give a brief overview of these polyhedral results. 

\subsection{A Polyhedral Overview of Popular Matchings}
\label{sec:poly-overview}
Given any matching $M$ in $G$, for any edge $(a,b) \notin M$, define $\vote_a(b,M)$ as follows: (here $M(a)$ is $a$'s partner in
the matching $M$ and $M(a) = \mathsf{null}$ if $a$ is unmatched)
\begin{equation*} 
	\vote_a(b,M) = \begin{cases} +   & \text{if\ $a$\ prefers\ $b$\ to\ $M(a)$};\\
		- &  \text{if\ $a$\ prefers\ $M(a)$\ to\ $b$.}			
	\end{cases}
\end{equation*}

We similarly define $\vote_b(a,M)$. Label every edge $(a,b) \notin M$ by $(\vote_a(b,M),\vote_b(a,M))$.  Thus, every edge outside
$M$ has a label in $\{(\pm, \pm)\}$, and it is labeled $(+,+)$ if and only if it blocks $M$.

Let $\tilde{G}$ be the graph $G$ augmented with self-loops. That is, we assume each vertex is its own last choice neighbor. 
So we can henceforth regard any matching $M$ in $G$ as a perfect matching $\tilde{M}$ in $\tilde{G}$ by adding self-loops for all 
vertices left unmatched in $M$.
The following edge weight function $\wt_M$ in $\tilde{G}$ will be useful to us. For any edge $(a,b)$ in $G$, define:
\begin{equation*}
	\wt_M(a,b) = \begin{cases}  2 & \text{if\ $(a,b)$\ is\ labeled\ $(+,+)$;}\\
		-2 & \text{if\ $(a,b)$\ is\ labeled\ $(-,-)$;}\\
		0 & \text{if $(a,b) \in M$ or labeled otherwise.}
	\end{cases}                            
\end{equation*}

For any $u \in A \cup B$, let $\wt_M(u,u) = 0$ if $u$ is left unmatched in $M$, else $\wt_M(u,u) = -1$.
Let $N$ be any matching in $G$. Then $\wt_M(\tilde{N})$ is the difference in the number of votes for $N$ and for $M$ in their head-to-head election,
i.e., $\wt_M(\tilde{N}) = \phi(N,M)-\phi(M,N)$.

Hence $M$ is popular in $G$ if and only if every perfect matching in the graph $\tilde{G}$ (with edge weights given by $\wt_M$) has weight 
at most 0. Equivalently, $M$ is popular
if  and only if $\tilde{M}$ is an optimal solution to the max-weight perfect matching problem in $\tilde{G}$ (since $\wt_M(\tilde{M}) = 0$). 
In other words, every popular matching is a {\em self-optimal} matching.

\smallskip

\noindent{\em Remark.} Suppose we replace ``perfect matching'' with ``matching'' in the above statement (so $\tilde{M}$ and $\tilde{G}$ can be replaced by $M$ and $G$, respectively).
Observe that this characterizes {\em stable} matchings, i.e., a matching $M$ is a max-weight matching in $G$ with edge weights given by $\wt_M$ 
if and only if $M$ is stable.\footnote{This view of stable matchings was shared with us by Tam\'as Fleiner.} This is because $\wt_M(M) = 0$, so
we need to have $\wt_M(e) \le 0$ for every edge $e$ if $M$ is a max-weight matching under $\wt_M$. Equivalently, blocking edges are forbidden for $M$.

\smallskip

Going back to the max-weight perfect matching problem in $\tilde{G}$ with edge weights given by $\wt_M$,
consider the LP in variables $\alpha_u$ for $u \in A \cup B$ that is dual to the max-weight perfect matching LP.
The characterization of popular matchings given below follows from LP-duality and total unimodularity of the system. 
\begin{theorem}[\cite{Kav16,KMN09}]
	\label{thm:witness}
	A matching $M$ in $G = (A \cup B, E)$ is popular if and only if there exists a vector $\vec{\alpha} \in \{0, \pm 1\}^n$ such that
	$\sum_{u \in A \cup B} \alpha_u = 0$,
	\[ \alpha_a + \alpha_b \ \ \ge \ \ \wt_M(a,b)\ \ \ \forall\, (a,b)\in E\ \ \ \  \text{and} \ \ \ \ \alpha_u \ \ge\ \wt_M(u,u)\ \ \ \forall\, u \in A \cup B.\] 
\end{theorem}

For any popular matching $M$, a vector $\vec{\alpha} \in \{0, \pm 1\}^n$ as given in Theorem~\ref{thm:witness} will be called $M$'s {\em witness}. A popular matching may have several witnesses. Any stable matching $M$ has $\vec{0}$ as a witness, since $\wt_M(e) \leq 0$ for all edges $e$ in $\tilde{G}$. 

\medskip

\noindent{\bf Popular Fractional Matchings.}
Rather than fixing a matching $M$, we could generalize the above characterization to any point $\vec{x}$ in the matching polytope. This would extend
the notion of popularity to fractional matchings. 
Roughly speaking, a fractional matching $\vec{x}$ is popular 
if and only if $\vec{x}$ is a {\em self-optimal} fractional matching, i.e., $\vec{x}$ is a max-weight perfect fractional matching in $\tilde{G}$ 
under the edge weight function $\wt_x$, which is the extension of $\wt_M$, described below.

Given a vertex $u$ and a pair $v,v'$ of $u$'s neighbors, let $\vote_u(v,v')$ be $u$'s vote for $v$ versus $v'$. So $\vote_u(v,v')$ is $1$ if $u$ prefers  $v$ to $v'$, 
it is $-1$ if $u$ prefers  $v'$ to $v$, otherwise it is $0$ (i.e., $v = v'$). 

For a point $\vec{x}$ in the matching polytope of $G$, first make $\vec{x}$ perfect in $\tilde{G}$ by setting
$x_{(u,u)} = 1 - \sum_{e \in \delta(u)}x_e$ for each $u \in A \cup B$. The following definition will be useful:
$$\vote_u(v,\vec{x}) =  \sum_{v'\in A \cup B}\,x_{(u,v')}\cdot\vote_u(v,v') = \sum_{v':\, v'\,\prec_u\, v}x_{(u,v')} - \sum_{v':\, v' \succ_u\, v}x_{(u,v')},
$$
where $\{v': v'\, \prec_u v\}$ consists of all neighbors of $u$ that are
ranked worse than $v$ in $u$'s preference list and the set $\{v': v' \succ_u v\}$ consists of all those that are ranked better; 
the vertex $u$ is a member of the former set. For any $(a,b) \in E$, its weight $\wt_x(a,b)$ is equal to $\vote_a(b,\vec{x}) + \vote_b(a,\vec{x})$.
Thus
\begin{equation}
	\label{eq:wt-x}
	\wt_x(a,b) \ \ = \ \ \left(\sum_{b':\, b'\,\prec_a\, b}x_{(a,b')} \ - \ \sum_{b':\, b' \succ_a\, b}x_{(a,b')}\right) \ + \  \left(\sum_{a':\, a'\,\prec_b\, a}x_{(a',b)} \ - \ \sum_{a':\, a' \succ_b\, a}x_{(a',b)}\right).
\end{equation}

For any $u \in A \cup B$, the weight of the self-loop $(u,u)$ is 
$\wt_x(u,u) \ = \ \vote_u(u,\vec{x}) \ = \ - \sum_{e \in \delta(u)}x_e$, where $\delta(u)$ is the set of
edges incident to $u$ in $G$.
A fractional matching $\vec{x}$ is popular if and only if every perfect fractional matching in $\tilde{G}$ 
with edge weights given by $\wt_x$ has weight at most 0. Note that, when $\vec{x}$ is the incidence vector of a matching $M$, we have $\wt_x = \wt_M$.
Section~\ref{sec:fractional-popular} discusses popular fractional matchings in more detail.

The popular fractional matching polytope ${\cal L}_G$ is the smallest convex set that contains
all popular fractional matchings in $G$.
We can formulate ${\cal L}_G$ as follows. Let $\tilde{E}$ be the edge set of $\tilde{G}$ and
let $\tilde{\delta}(u) = \delta(u) \cup \{(u,u)\}$ for all vertices $u$.
\begin{eqnarray*}
	\wt_x(N) & \ = \ & \sum_{e \in N}\wt_x(e)  \ \le \  0 \ \ \ \ \ \forall\,\text{perfect\ matchings\ $N$\ in\ $\tilde{G}$,}\\
	\sum_{e \in \tilde{\delta}(u)}x_e & \ = \ & 1 \ \ \ \ \ \forall\, u \in A \cup B \ \ \ \ \text{and} \ \ \ \ x_e \ \ge \ 0 \ \ \forall\, e \in \tilde{E}.
\end{eqnarray*}  

The above formulation of ${\cal L}_G$ in the original space involves exponentially many constraints.
The following compact extended formulation ${\cal E}_G$ of ${\cal L}_G$ follows from \cite{KMN09} (and was explicitly
spelt out in \cite{Kav16}).

\begin{minipage}{.5\textwidth}
	\begin{eqnarray}
		\sum_{u \in A \cup B}\alpha_u  & \ = \ &  0 \label{sec5-constr3-bis}\\
		\alpha_a + \alpha_b & \ \ge \ & \wt_x(a,b) \ \ \ \ \ \ \ \, \forall (a,b) \in E\label{sec5-constr1}\\
		\alpha_u  & \ \ge \ & \wt_x(u,u) \ \ \ \ \ \ \ \, \forall u \in A \cup B \label{sec5-constr3}
	\end{eqnarray}
\end{minipage}\begin{minipage}{.5\textwidth}
\begin{eqnarray}
	\sum_{e\in\tilde{\delta}(u)}x_e & \ = \ &  1 \ \ \ \ \ \ \, \forall u\in A\cup B \label{sec5-constr5}\\  
	x_e \ & \ \ge \ & \ 0 \, \ \ \ \ \ \ \forall e \in \tilde{E}. \label{sec5-constr5-bis} \\
	\nonumber
\end{eqnarray}
\end{minipage}

\medskip

\noindent{\em Remark.} As observed in \cite{HK17}, setting $\alpha_u = 0$ for all $u$ makes the above formulation a description of the stable matching polytope that is equivalent to Rothblum's formulation~\cite{Rot92}.

\smallskip

The LP that gives rise to the above formulation minimizes $\sum_u\alpha_u$ subject to constraints~\eqref{sec5-constr1}--\eqref{sec5-constr5-bis}. 
This LP is {\em self-dual} (it is the same as its dual); this property was used in \cite{HK17} to show the half-integrality of the polytope ${\cal L}_G$.
Interestingly, in the special case when $G$ admits a perfect stable matching, i.e., when all vertices are matched in some (equivalently, every~\cite{GS85}) stable matching, 
it was shown in~\cite{HK17} that ${\cal L}_G$ is integral. This result will be used in Section~\ref{sec:algo}.

The above formulation allows us to extend the notion of {\em witness} to popular fractional matchings.

\begin{definition}
	\label{def:frac-witness}
	If $(\vec{x},\vec{\alpha})$ satisfies constraints~\eqref{sec5-constr3-bis}-\eqref{sec5-constr5-bis} then $\vec{\alpha}$ is called a witness of $\vec{x}$'s popularity.
\end{definition}  

Hence, for any $\vec{x}$ in the matching polytope of $G$ (this is given by constraints~\eqref{sec5-constr5}-\eqref{sec5-constr5-bis}), $\vec{\alpha}$ is a witness of
$\vec{x}$'s popularity if constraints~\eqref{sec5-constr3-bis}-\eqref{sec5-constr3} are satisfied. For example, consider the half-integral matching $\vec{x}$ in the
instance from \cite{HK17} illustrated in Fig.~\ref{fig:non-int}. Let $\alpha_{a_0} = \alpha_{a_1} = \alpha_{b_2} = 0$, $\alpha_{b_1} = 1$, and $\alpha_{a_2} = -1$. 
It is easy to check that $(\vec{x}, \vec{\alpha})$ satisfies~\eqref{sec5-constr3-bis}-\eqref{sec5-constr3}. Thus $\vec{x}$ is popular and $\vec{\alpha}$ is a witness of $\vec{x}$'s popularity.

\smallskip

\noindent{\em Remark.} 
When $\vec{x}$ is integral, the three constraints \eqref{sec5-constr3-bis}-\eqref{sec5-constr3} are 
the same as the three constraints in Theorem~\ref{thm:witness} that characterized popular matchings in terms of witnesses.

\subsection{Dominant Matchings}
\label{sec:dominant-prelims}

\begin{figure}[h]
	\centerline{\resizebox{0.23\textwidth}{!}{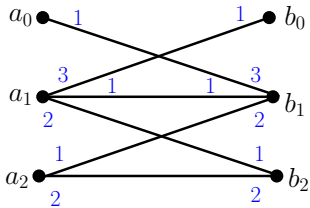}}
	\caption{The numbers on edges denote vertex preferences. So $a_1,b_1$ are each other's top choice neighbors and their second choices are $b_2,a_2$, respectively, and their last choices are $b_0,a_0$ respectively, and so on.}
	\label{fig:dominant-example}
\end{figure}

Recall the definition of a dominant matching: this is a popular matching that is more popular than all strictly larger matchings in $G$.
Consider the instance in Fig.~\ref{fig:dominant-example} on six vertices $a_0,a_1,a_2,b_0,b_1,b_2$. The preferences of vertices are indicated in the figure.
This instance has 2 max-size popular matchings: $D = \{(a_1,b_2),(a_2,b_1)\}$
and $S = \{(a_1,b_1),(a_2,b_2)\}$. The matching $S$ is not dominant since it does not defeat the larger matching 
$N = \{(a_0,b_1),(a_1,b_0),(a_2,b_2)\}$ while the matching $D$ defeats $N$ and is dominant.

For any matching $M$, let $G_M$ be the graph obtained from $G$ by removing edges labeled $(-,-)$. 
Observe that in the example discussed above, there is no $D$-augmenting path in $G_D$
while there is an $S$-augmenting path $a_0 - b_1 - a_1 - b_0$ in $G_S$.
In fact, the absence of such augmenting paths is what characterizes dominant matchings in the set of popular matchings.

\begin{theorem}[\cite{CK16}]
	\label{thm:dominant-comb}
	A popular matching $M$ is dominant iff $G_M$ has no $M$-augmenting path.
\end{theorem}

Any maximum-size popular (hence any dominant) matching matches the same subset of vertices and in fact, a
vertex left unmatched in a dominant matching is left unmatched in every popular matching in $G$~\cite{Hirakawa-MatchUp15}. 
Call a vertex $v$ {\em popular} if it is matched in any max-size popular matching, else call $v$ {\em unpopular}. Similarly, call a vertex $v$ {\em stable} if it is matched in some (equivalently, every~\cite{GS85}) stable matching in $G$, else call $v$ 
{\em unstable}. It is known that every popular matching in $G$ matches all stable vertices~\cite{HK11}. 

The LP-based characterization of dominant matchings given below follows from their combinatorial characterization in \cite{CK16}. 
\begin{theorem}
	\label{thm:dominant-witness}
	A popular matching $M$ in $G = (A \cup B, E)$ is dominant if and only if $M$ admits a witness $\vec{\alpha}$ such that 
	$\alpha_v \in \{\pm 1\}$ for every popular vertex $v$. 
\end{theorem}

We will call a witness $\vec{\alpha}$ such that $\alpha_v \in \{\pm 1\}$ for all popular vertices $v$ a {\em dominant} witness.
For the dominant matching $D$ in the above example, the vector $\vec{\alpha}$ defined as $\alpha_{a_1} = \alpha_{b_1} = 1$, $\alpha_{a_2} = \alpha_{b_2}~=~-1$, and $\alpha_{a_0} = \alpha_{b_0} = 0$ is a dominant witness (observe that $a_0,b_0$ are unpopular vertices).
The characterization of dominant matchings in terms of such witnesses will be useful to us in Section~\ref{sec:algo}. This characterization will lead to a new compact extended
formulation of the dominant matching polytope in Section~\ref{sec-new:polytopes} as a face of the extension ${\cal E}_G$ of the popular fractional matching polytope.

\section{Our Algorithm}\label{sec:algo} We prove Theorem~\ref{thm:algo} in this section.
Our input instance is $G = (A \cup B, E)$ and without loss of generality, assume $|A| \ge |B|$. The first step in our algorithm
is an {\em augmentation} of the graph $G$ into $G^* = (A \cup B, E^*)$, with $E^*\supseteq E$.
The new edges in $E^*$ are obtained by introducing edges between certain pairs of {\em unstable} vertices
and each new edge has cost 0. 
Given a popular matching $M$ in $G$, we will use these new edges to obtain an extension $M^*$ in $G^*$ such that
$M^*$ is a {\em dominant} matching in $G^*$, see Lemma~\ref{lemma2}. 

Moreover, $M^*$ satisfies some stronger condition on witnesses than a dominant matching (as given in Theorem~\ref{thm:dominant-witness}). 
Hence, we will call $M^*$ \emph{extra-dominant}, see Definition~\ref{constr-dom}. 
The second step of the algorithm proves that every extra-dominant matching in $G^*$ projects to a quasi-popular matching in $G$, see Lemma~\ref{lemma3}.

The first two steps imply that in order to prove Theorem~\ref{thm:algo}, it suffices to efficiently find a min-cost extra-dominant matching (in $G^*$). We show how to do this by providing a compact extended formulation of the extra-dominant matching polytope. 
In fact, we first describe an  
extension of the dominant matching polytope of $G^*$, and then show that one of its faces is an  
extension of the extra-dominant matching polytope of $G^*$. 

Lemma~\ref{lemma2} and Lemma~\ref{lemma3} are our technical lemmas here and their proofs are given in Section~\ref{sec:lemmas}. 
The proofs of correctness of our extended formulations are given in Section~\ref{sec:polytope-proof}.

\subsection{The Popular Subgraph} The augmentation of $G$ into $G^*$ is based on a certain subgraph of $G$ called its 
\emph{popular subgraph}.
Call an edge $e$ in  $G = (A \cup B, E)$ {\em popular} if there is some popular matching in $G$ that contains $e$.
Let $E_0\subseteq E $ be the set of popular edges in $G$. The set $E_0$ can be computed in linear time, see~\cite{CK16}. 
Call $G_0 = (A \cup B, E_0)$ the {\em popular subgraph} of $G$. 

The subgraph $G_0$ need not be connected: let $C_1,\ldots,C_h$ be the set of connected components in $G_0$. 
Call a component $C_i$ {\em non-trivial} if $|C_i| \ge 2$. The following observation will be useful.

\smallskip

\begin{obs}
	\label{obs1}
	For any non-trivial connected component $C_i$ in the popular subgraph $G_0$, the number of unstable vertices in $A \cap C_i$ equals the number of unstable vertices in 
	$B \cap C_i$. 
\end{obs}

\smallskip

Indeed, every max-size popular matching $M$ restricted to $C_i$ is a perfect matching, since all max-size popular matchings match the same set of vertices, while vertices left unmatched in any max-size popular matching are left unmatched in all popular matchings (see the discussion in Section~\ref{sec:dominant-prelims}). 
Thus $|A \cap C_i| = |B \cap C_i|$. The number of stable vertices in $A \cap C_i$ equals the number of stable vertices in $B \cap C_i$, since every stable matching matches stable vertices
in $C_i$ among themselves. Hence Observation~\ref{obs1} follows.

\begin{lemma}
	\label{lemma1}
	Let $C_i$ be any connected component in the popular subgraph $G_0$, $M$ a popular matching in $G$, and $\vec{\alpha}\in\{0,\pm 1\}^n$ a witness for $M$. Then one of the following two statements has to hold:
	\begin{itemize}
		\item $\alpha_u = 0$ for all vertices $u$ of $C_i$ and no unstable vertex of $C_i$ is matched in $M$.
		\item $\alpha_u \in\{\pm 1\}$ for all vertices $u$ of $C_i$ and every unstable vertex of $C_i$ is matched in $M$.
	\end{itemize}
\end{lemma}

\proof{} Let $M$ be a popular matching in $G$.
Consider the max-weight perfect matching LP (this is \eqref{LP1} given below) in $\tilde{G} = (A \cup B, \tilde{E})$ with edge weight function $\wt_M$ described in Section~\ref{prelims}. For any $u \in A \cup B$, let $\tilde{\delta}(u) = \delta(u) \cup \{(u,u)\}$. 
The linear program \eqref{LP2} is the dual LP.

\begin{linearprogram}
	{
		\label{LP1}
		\maximize \sum_{e \in \tilde{E}} \wt_M(e)\cdot x_e  
	}
	\qquad\sum_{e \in {\tilde{\delta}}(u)}x_e \ & = \ \ 1  \ \ \, \forall\, u \in A \cup B \notag\\
	x_e  \ & \ge \ \ 0   \ \ \ \forall\, e \in \tilde{E}. \notag
\end{linearprogram}

\begin{linearprogram}
	{
		\label{LP2}
		\minimize \sum_{u \in A \cup B}\alpha_u
	}
	\alpha_{a} + \alpha_{b} \ & \ge \ \ \wt_{M}(a,b) \ \ \ \, \forall\, (a,b)\in E \notag\\
	\alpha_u \ & \ge \ \ \wt_M(u,u) \ \ \ \forall\, u \in A \cup B. \notag
\end{linearprogram}

Any popular matching in $G$ is an optimal solution to \eqref{LP1} and any witness of $M$ is an optimal solution to \eqref{LP2}. 
So if $(a,b)$ is a popular edge,
then complementary slackness implies that $\alpha_a + \alpha_b = \wt_M(a,b)$, where
$\vec{\alpha} \in \{0, \pm 1\}^n$ is a witness of $M$. Since $\wt_M(a,b) \in \{0, \pm 2\}$, $\alpha_a$ and $\alpha_b$ have the same
parity. So either $\alpha_a = \alpha_b = 0$ or $\alpha_a,\alpha_b \in \{\pm 1\}$.

Let $u$ be an unstable vertex in $G$. Let $S$ be a stable matching in $G$.
The perfect matching $\tilde{S}$ contains therefore the self-loop $(u,u)$. Since $\tilde{S}$ is an
optimal solution to \eqref{LP1} (recall that every stable matching is popular), we have $\alpha_u = \wt_M(u,u)$ by complementary slackness.
Thus $\alpha_u =0$ if and only if $u$ is left unmatched in~$M$ (otherwise $\alpha_u = -1$).

For every connected component $C_i$ in $G_0$, it follows from our first observation above that either (i)~$\alpha_u = 0$ for all vertices $u \in C_i$ 
or (ii)~$\alpha_u = \pm 1$ for all vertices $u \in C_i$. For every unstable vertex $u$,
we have $\alpha_u = \wt_M(u,u)$ (by our second observation above). Hence in case~(i), all unstable vertices in 
$C_i$ are left unmatched in $M$ and in case~(ii), all unstable vertices in $C_i$ are matched in $M$. \qed \endproof

Let $C_1,\ldots,C_k$ be the non-trivial connected components in $G_0$.
Consider $C_i$ where $i \in \{1,\ldots,k\}$.
We know from Observation~\ref{obs1} that the number of unstable vertices in $A \cap C_i$ is the same as the number of unstable vertices in 
$B \cap C_i$.
Let $a_1,\ldots,a_t$ be the unstable vertices in $A \cap C_i$ and let $b_1,\ldots,b_t$ be the unstable vertices in $B \cap C_i$.
Compute an arbitrary pairing of $a_1,\ldots,a_t$ with $b_1,\ldots,b_t$ --- this
partitions unstable vertices of $C_i$ into disjoint pairs, say $(a_j,b_j): 1 \le j \le t$.  
Let $S_i$ be the set of these $t$ pairs.

We also consider the trivial or singleton components in $G_0$: each such component consists of a single {\em unpopular} vertex, i.e., 
a vertex that is left unmatched in every popular matching. Since $|A \cap C_i| = |B \cap C_i|$ for every non-trivial component $C_i$ and 
$|A| \ge |B|$, there are at least as many unpopular vertices in $A$ as in $B$. We compute an arbitrary pairing $S_0$ between unpopular vertices in $A$ and in $B$. If $|A| > |B|$ then some unpopular vertices in $A$ are left out of $S_0$. 

The instance $G^* = (A \cup B, E^*)$ is defined as follows:
$E^* = \cup_{i=0}^k S_i \cup E$. So $G^*$ is the graph $G$ augmented with new edges $(a,b)$ in $\cup_{i=0}^k S_i$.
Preference lists of vertices in $G^*$ are the same as in $G$, except for unstable vertices in $G$, some of whom have acquired 
a new neighbor in $G^*$. For any unstable vertex $u$ with a new neighbor $v$ in $G^*$, the vertex $v$ is at the tail 
of $u$'s preference list, i.e., $v$ is $u$'s least preferred neighbor in $G^*$.

\medskip

\noindent{\bf The Matching $M^*$.} Let $M$ be any popular matching in $G$.
Lemma~\ref{lemma1} tells us that for any edge $(a,b) \in \cup_{i=1}^k S_i$, either both $a$ and $b$ are matched in $M$ or neither is matched
in $M$. This is also true for $(a,b) \in S_0$, since in this case both $a$ and $b$ are left unmatched in $M$.  Define $M^*$ as follows:
\[ M^*  = M \cup \{(a,b) \in \cup_{i=0}^k S_i\ \text{such\ that}  \ a \ \text{and}\ b\ \text{are\ unmatched\ in}\ M\}.\]

It is easy to see that $M^*$ is a $B$-perfect matching in $G^*$. Let $U^*$ be the set of unstable vertices in $G^*$.
So $U^*\subseteq A$, in fact, $U^* \subseteq U_A$, where $U_A$ is the set of 
unstable vertices in $G$ that are in $A$.

A stable matching in $G^*$ is $B$-perfect and a matching of maximum size in $G^*$. Hence, all popular matchings in $G^*$ match the same set of vertices: this is the set $(A \cup B) \setminus U^*$. Thus $U^*$ is the set of unpopular vertices in $G^*$. We now define a subclass of dominant matchings in $G^*$. 

\smallskip

\begin{Definition}
	\label{constr-dom}
	A popular matching $T$ in $G^* = (A \cup B, E^*)$ is extra-dominant if $T$ admits a witness 
	$\vec{\alpha}$ such that $\alpha_v \in \{\pm 1\}$ for every popular vertex $v$ and $\alpha_a = -1$ for all 
	$a \in U_A\setminus U^*$.
\end{Definition}

\smallskip

We call a witness $\vec{\alpha}$ as described in Definition~\ref{constr-dom} extra-dominant. Note that every extra-dominant matching is also dominant, by the LP-based characterization of dominant matchings (given by Theorem~\ref{thm:dominant-witness}). The following lemmas illustrate the relationship between (quasi-)popular matchings in $G$ and extra-dominant matchings in $G^*$.

\begin{lemma}
	\label{lemma2}
	
	For any popular matching $M$ in $G$, the matching $M^*$ is extra-dominant in $G^*$. 
\end{lemma}

\begin{lemma}
	\label{lemma3}
	For any extra-dominant matching $T$ in $G^*$, $T\cap E$ is a quasi-popular matching in $G$. 
\end{lemma}

\subsection{Extended Formulations of the Dominant and Extra-Dominant Matching Polytopes}\label{sec:polytope} We will now show extended formulations of the dominant and extra-dominant matching polytopes of $G^*$.
For any fractional matching $\vec{x}$ in $G^*$, define the weight function $c_x$ on any edge $(a,b)$ of $G^*$ as follows:
\begin{equation*}
	c_x(a,b) = \begin{cases}  1 & \text{if\ $a \in U^*$;}\\
		\wt_x(a,b) & \text{otherwise.}
	\end{cases}                            
\end{equation*}
See \eqref{eq:wt-x} for the definition of $\wt_x(a,b)$. Observe that $c_x$ is an affine function of $\vec{x}$.

Call an edge $e$ in  $G^*$ {\em dominant} if there is some dominant matching in $G^*$ that contains $e$.
Let $E^*_D$ denote the set of dominant edges in $G^*$. The set $E^*_D$ can be determined in linear time, see~\cite{CK16}.
Consider the polytope ${\cal F}_{G^*}$ in variables $\vec{x}$ and $\vec{\alpha}$ defined by constraints 
\eqref{new-constr3-bis}--\eqref{new-constr2} given below. Here $\delta^*(u)$ is the set of edges incident to vertex $u$ in $G^*$.

\begin{minipage}{.45\textwidth}\begin{eqnarray}
		\sum_{u \in A \cup B}\alpha_u \ & = & \ 0 \label{new-constr3-bis}\\
		\alpha_a + \alpha_b \ & \ge & \ c_x(a,b) \ \ \ \   \forall (a,b) \in E^*\label{new-constr1}\\
		\alpha_u \ & \ge & \ -1 \ \ \ \ \ \ \ \ \ \ \forall u \in A\cup B\label{new-constr3} \\ 
		\alpha_a \ & =  & \ 0 \ \ \ \ \ \ \ \ \ \ \ \ \ \forall a \in U^*\label{new-constr4}\end{eqnarray}
\end{minipage}
\begin{minipage}{.5\textwidth}\begin{eqnarray}
		\sum_{e \in \delta^*(u)}x_e \  & =    & 1\ \ \  \ \ \ \ \ \ \ \forall u\in (A\cup B)\setminus U^* \label{constr5}\\
		x_e \ & =  & \ 0 \, \ \ \ \ \  \ \ \ \forall e \in E^* \setminus E_D^*\label{new-constr4-bis}\\
		x_e \ & \ge & \ 0 \ \ \ \ \ \ \  \ \,  \forall e \in E^*_D \label{constr5-first}\\ 
		\alpha_a \ & =   & -1 \ \ \ \  \ \ \  \ \forall a \in U_A \setminus U^*.\label{new-constr2}
	\end{eqnarray}
\end{minipage}

\begin{theorem}
	\label{thm:extra-dominant-extn}
	The polytope ${\cal F}_{G^*}$ is an extension of the extra-dominant matching polytope of $G^*$. 
\end{theorem}

Thus linear programming on ${\cal F}_{G^*}$  gives a min-cost extra-dominant matching in $G^*$.
Let ${\cal C}_{G^*}$ be the polytope defined by constraints (\ref{new-constr3-bis})-(\ref{constr5-first}).
Theorem~\ref{thm:dominant-extn} will help us prove Theorem~\ref{thm:extra-dominant-extn}. 

\begin{theorem}
	\label{thm:dominant-extn}
	${\cal C}_{G^*}$ is an extension of the dominant matching polytope of $G^*$. Moreover, the extreme points 
	of ${\cal C}_{G^*}$ are all and only the vectors $(I_M,\vec{\alpha})$, where $I_M$ is the incidence vector of a dominant matching $M$ in $G^*$ and $\vec{\alpha}$ is a dominant witness of $M$. 
\end{theorem}

The proof of Theorem~\ref{thm:dominant-extn} uses the extension of the popular fractional matching polytope described in Section~\ref{sec:poly-overview}.
Recall the integrality of this extension ${\cal E}_H$ in a marriage instance $H$ that admits a {\em perfect} 
stable matching~\cite[Theorem~2]{HK17}. 
We apply this with $H=G^* \setminus U^*$ and the polytope ${\cal C}_{G^*}$ is realized as a face of an extension of ${\cal E}_H$.
The proofs of Theorem~\ref{thm:extra-dominant-extn} and Theorem~\ref{thm:dominant-extn} are given in Section~\ref{sec:polytope-proof}.

\paragraph{\bf The Algorithm.}
We are given a function $\cost$ on the edges of $G$. 
We set $\cost(e) = 0$ for every new edge $e$ in $G^*$, i.e., $\cost(e) = 0$ for all $e \in \cup_{i=0}^k S_i$. 
All the {\em old} edges in $G^*$, i.e., those in $E$, inherit their edge costs from $G$. 
For any matching $T$ in $G^*$, 
note that $\cost(T\cap E) = \cost(T)$.

Our algorithm that proves Theorem~\ref{thm:algo} stated in Section~\ref{intro} is summed up below.
\begin{enumerate}
	\item Identify all unstable vertices in $G = (A \cup B, E)$ by running Gale-Shapley algorithm.
	\item Determine the popular subgraph of $G$ by computing all popular edges in $G$ using \cite{CK16}.
	\item Augment the graph $G$ into the graph $G^*$ by adding edges in $\cup_{i = 0}^k S_i$ as described earlier.
	\item Compute a min-cost extra-dominant matching $T$ in $G^*$ by solving a linear program over ${\cal F}_{G^*}$.
	\item Return $T \cap E$.
\end{enumerate}

It follows from Lemma~\ref{lemma2}
and Theorem~\ref{thm:extra-dominant-extn} that $\cost(T) \le \opt$ where $\opt$ is the cost of a min-cost popular matching in $G$.
We know from  Lemma~\ref{lemma3} that  $T\cap E$ is a quasi-popular matching in $G$. 
Since $\cost(T \cap E) = \cost(T) \le \opt$, the correctness of our algorithm follows.

\subsection{Proofs of Our Technical Lemmas}\label{sec:lemmas}

\proof{ (of Lemma~\ref{lemma2}).}
Since $M$ is popular in $G$, it has a witness $\vec{\alpha} \in \{0, \pm 1\}^n$ such that $\sum_u\alpha_u = 0$ and $(I_M,\vec{\alpha})$ 
satisfies edge covering constraints as given in Theorem~\ref{thm:witness}, where $I_M$ is the incidence vector of $M$.
For any non-trivial connected component $C_i$ in the subgraph $G_0$, it follows from Lemma~\ref{lemma1} that  
either (i)~$\alpha_u = 0$ for all $u \in C_i$ or (ii)~$\alpha_u \in \{\pm 1\}$ for all $u \in C_i$.

We will now define a witness $\vec{\beta}$ that proves the popularity of $M^*$ in $G^*$. 
For any non-trivial connected component $C_i$ in $G_0$ do: 
\begin{itemize}
	\item if $\alpha_u \in \{\pm 1\}$ for all $u \in C_i$ then set $\beta_u = \alpha_u$ for all $u \in C_i$.
	\item if $\alpha_u = 0$ for all $u \in C_i$ then set $\beta_a = -1$ for all $a \in A \cap C_i$ and $\beta_b = 1$ for all $b \in B\cap C_i$.
\end{itemize}

We will now set $\beta$-values for vertices that are unpopular in $G$. These vertices are outside $\cup_{i=1}^kC_i$. 
We set $\beta_b = 1$ for all unpopular vertices $b\in B$ and $\beta_a = -1$ for all those unpopular vertices $a\in A$ that have an edge incident to them in $S_0$. For each unpopular vertex $a\in A$ that does not appear in $S_0$ (so $a$ is unmatched in $M^*$), we set $\beta_a = 0$. Let us check that $\vec{\beta}$ satisfies Theorem~\ref{thm:witness}.

Observe first that $\sum_{(a,b) \in S_0}(\beta_a + \beta_b) = 0$. 
For each edge $(a,b) \in M$, we have $\alpha_a + \alpha_b = \wt_M(a,b) = 0$ and it is easy to see from our assignment of $\beta$-values that
$\beta_a + \beta_b = \alpha_a + \alpha_b = 0$. Thus $\sum_{u \in A \cup B}\beta_u = 0$.

\medskip

\noindent{\em Edge-Covering Constraints.}
We will now show that $\beta_a + \beta_b \ge \wt_{M^*}(a,b)$ for all edges $(a,b)$ in $G^*$ along with 
$\beta_u \ge \wt_{M^*}(u,u)$ for all vertices $u$, where the function $\wt_{M^*}(e)$ was defined in Section~\ref{prelims}.
The constraints $\beta_u \ge \wt_{M^*}(u,u)$ for all vertices $u$ are easy to see: either (i)~$\beta_u = -1$ which implies that 
$u$ is matched in $M^*$ and so $\wt_{M^*}(u,u) = -1$ or (ii)~$\beta_u \ge 0 \ge \wt_{M^*}(u,u)$.

We will now show edge covering constraints hold for all edges in $G^*$. It is easy to see that 
$\beta_a + \beta_b \ge \wt_{M^*}(a,b)$ for all the {\em new} edges $(a,b)$ in $G^*$, i.e., for all $(a,b) \in \cup_{i=0}^kS_i$. 
This is because either: (i)~both $a$ and $b$ are matched in $M$, in which case $\wt_{M^*}(a,b) = -2$  and $\beta_a + \beta_b \ge -2$ holds, or 
(ii)~both $a$ and $b$ are unmatched in $M$, in which case $(a,b) \in M^*$ and so 
$\wt_{M^*}(a,b) = 0 = \beta_a + \beta_b$ (recall that $\beta_a = -1$ and $\beta_b = 1$ in this case).

We will now show that these constraints hold for all the old edges as well, i.e., for any $(a,b) \in E$. Note that
$\alpha_a + \alpha_b \ge \wt_M(a,b)$ and also $\wt_{M^*}(a,b) = \wt_M(a,b)$ by construction. If $\beta_a = \alpha_a$
then either $\beta_b = \alpha_b$ or $\beta_b = \alpha_b + 1$, so $\beta_a + \beta_b \ge \alpha_a +\alpha_b \ge \wt_M(a,b) = \wt_{M^*}(a,b)$ for such an edge $(a,b)$. 
Else $\beta_a = \alpha_a - 1$.

We have two sub-cases here. In the first sub-case, $\beta_b = \alpha_b + 1$ and thus $\beta_a + \beta_b = \alpha_a + \alpha_b$.
Since $\wt_{M^*}(a,b) = \wt_M(a,b)$ for all $(a,b) \in E$, the statement follows. 
The other sub-case is  $\beta_b = \alpha_b$. So $\alpha_b \in \{\pm 1\}$, while $\alpha_a = 0$,
thus $\alpha_a + \alpha_b \in \{\pm 1\}$. 
We have $\wt_{M^*}(a,b) = \wt_M(a,b) \le \alpha_a + \alpha_b$. 
Note that for any edge $(a,b)$, the value $\wt_M(a,b)\in \{0,\pm 2\}$. Thus $\wt_M(a,b)$ is an even number, so 
the constraint $\wt_M(a,b) \le \alpha_a + \alpha_b$ is {\em slack} 
and we can tighten it to $\wt_M(a,b) \le \alpha_a + \alpha_b -1$. In other words, we have:
$\wt_{M^*}(a,b) = \wt_M(a,b) \le \alpha_a + \alpha_b -1 = \beta_a + \beta_b$.
Thus the edge covering constraints hold for all $(a,b) \in E$.
Hence $\vec{\beta}$ is a valid witness of $M^*$'s popularity in $G^*$.

Hence, $M^*$ is popular. Moreover, by construction, we have $\beta_a=0$ if and only if $a$ is unmatched in $M^*$ and $\beta_a = -1$ for every $a \in U_A \setminus U^*$. Hence, $M^*$ has a  witness $\vec{\beta}$ that satisfies Definition~\ref{constr-dom}, so $M^*$ is extra-dominant.   
\qed\endproof

\medskip

\proof{ (of Lemma~\ref{lemma3}).}
Let $N$ be any matching in $G$. Let $T' = T \cap E$.
Matchings $N$ and $T'$ can be viewed as matchings in $G^*$ as well since $E^* \supseteq E$. 
Since $T$ is popular in $G^*$, $\phi_{G^*}(T,N) \ge \phi_{G^*}(N,T)$, where $\phi_{G^*}$ is the function $\phi$ in the graph $G^*$.

Let $W$ be the set of unstable vertices in $G$ that get matched along {\em new} edges (i.e., those in $E^*\setminus E$) in $T$ 
but are left unmatched in $N$.

\begin{itemize}
	\item We have $\phi_{G^*}(T',N) = \phi_{G^*}(T,N) - |W|$.
	The vertices in $W$ used to vote for $T$ versus $N$, however they are now indifferent between $T'$ and $N$ because both $T'$ and $N$ leave them unmatched.
	\item We also have $\phi_{G^*}(N,T') = \phi_{G^*}(N,T)$.
	Indeed, if a vertex did not prefer $N$ to $T$, then it is either unmatched in $N$ or it was matched in $T$ by an edge in $E$ (since edges from $E^*\setminus E$ are at the end of every vertex's preference list), hence it is matched via the same edge in $T'$.
\end{itemize}

\begin{figure*}[ht]
	\centerline{\resizebox{0.6\textwidth}{!}{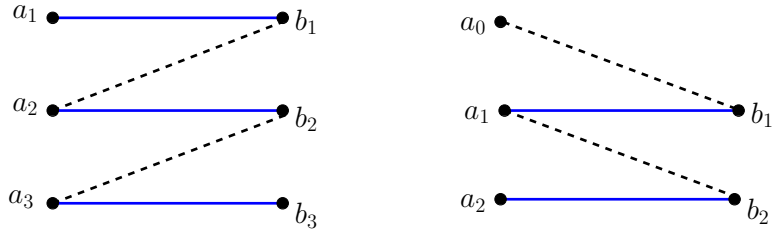}}
	\caption{\small The blue solid edges are in $T$ while the black dashed edges are in $N$. In the figure on the left, both the endpoints of $p$ are in $W$ while in the figure on the right, exactly one endpoint of $p$ (i.e., $a_2$) is in $W$.}
	\label{fig1:example}
\end{figure*}

Since both $T'$ and $N$ are matchings in $G$, we have $\phi_{G^*}(T', N) = \phi(T',N)$ and $\phi_{G^*}(N,T') = \phi(N,T')$.
Consider the subgraph of $G^*$ whose edges are given by $T \oplus N$. Its connected components are alternating paths / cycles. We are left to show that, for each such connected component $\rho$, 
we have $\phi(N\cap\rho,T'\cap\rho) \le 2\cdot\phi(T'\cap\rho,N\cap\rho)$.
This concludes the proof, since summing both sides over all connected components $\rho$ of $T \oplus N$, we get $\phi(N, T') \le 2\cdot\phi(T', N)$,  as $(T\setminus T')\cap N=\emptyset$.

First, let $\rho$ be an alternating cycle. We know from the popularity of $T$ in $G^*$ that, if we restrict to vertices of 
$\rho$, we have $\phi_{G^*}(T, N) \ge \phi_{G^*}(N, T)$. Every vertex in $\rho$ is matched in $N$, hence $\rho$ cannot contain any element 
of $W$. Thus $T' = T$ when restricted to $\rho$ and so $\phi(T',N) \ge \phi(N,T')$ when restricted to the vertices of $\rho$.

Now suppose $\rho$ is an alternating path, and denote this path by $p$.
As in the case of alternating cycles, we have $\phi_{G^*}(T,N) \ge \phi_{G^*}(N,T)$ restricted to vertices of $p$. 
If $p$ does not contain any element of $W$, then $T$ and $T'$ are identical on vertices of $p$ and so 
$\phi(T',N) \ge \phi(N,T')$ in $G$. 
Moreover, note that vertices from $W$ cannot have degree $2$ in $p$. So suppose one or both the endpoints of $p$ belong to $W$.

\medskip

\noindent{\em Case~1.} Both endpoints of $p$ are in $W$. Since, by definition, vertices of $W$ are not matched by $N$, exactly one endpoint of $p$ is in $A$. If $p$ consists of a single edge $(a,b)$, then $a,b \in W$
and these vertices are indifferent between $N$ and $T'$ since both these matchings leave them unmatched. 

So let $p = \langle a_1,b_1,\ldots,a_t,b_t\rangle$, where $(a_i,b_i) \in T$ for $1 \le i \le t$, where $t \ge 2$, and 
$a_1,b_t \in W$ (see Fig.~\ref{fig1:example}, left). 
Among the $2t$ vertices of $p$, we have: $\phi_{G^*}(T, N) = t_2 \ge t \ge t_1 = \phi_{G^*}(N, T)$.

The edges $(a_1,b_1)$ and $(a_t,b_t)$ are {\em new edges} and so neither of these edges is in $T'$. 
In the election between $T'$ and $N$ when restricted to the vertices of $p$, we have $\phi(T', N) = t_2 -2 \ge t-2$ 
(since $T'$ loses out on the votes of $a_1$ and $b_t$) and $\phi(N, T') = t_1 \le t$ (since $N$ preserves all its $t_1$ votes). Thus 
when restricted to the $2t$ vertices of $p$, we have $\phi(N, T')/{\phi(T', N)} \le t/(t-2)$. Note that $a_1$ and $b_t$ are indifferent
between $T'$ and $N$ and hence abstain from voting.

If $t \ge 4$ then we have the desired bound, i.e., $\phi(N, T')/{\phi(T', N)} \le t/(t-2) \le 2$.
We need to argue out the cases of $t = 2$ and $t= 3$ separately.

\smallskip

\begin{itemize}
	\item Suppose $t = 2$. Then $p = \langle  a_1,b_1,a_2,b_2\rangle$ and both $a_1$ and $b_2$ are in $W$. So their partners in $T$, i.e., 
	$b_1$ and $a_2$, are unstable vertices. However unstable vertices form an independent set in $G$, so $(a_2,b_1) \notin E$, contradicting 
	that $(a_2,b_1) \in N \subseteq E$. 
	\item Suppose $t = 3$. Then $p = \langle  a_1,b_1,a_2,b_2,a_3,b_3\rangle$, where $a_1,b_1,a_3,b_3$ are unstable vertices
	(see Fig.~\ref{fig1:example}, left). 
	We claim $\phi(N\cap p, T'\cap p) \le 2 \le \phi(T'\cap p, N\cap p)$. The matching $T$ is an extra-dominant matching 
	(see Definition~\ref{constr-dom}) and so it has a witness  $\vec{\alpha}$ in $G^*$ such that $\alpha_{a_3} = -1$. We use this to show that 
	at most 2 vertices among the 4 vertices $b_1,a_2,b_2,a_3$ prefer $N$ to $T$.
	
	\smallskip
	
	\begin{itemize}
		
		\item[(i)] Suppose $(a_2,b_1)$ is a blocking edge to $T$. Then 
		$\alpha_{b_1} = \alpha_{a_2} = 1$ and this also implies that $\alpha_{b_2} = -1$, because 
		$\alpha_{a_2} + \alpha_{b_2} = \wt_T(a_2,b_2) = 0$.
		So $\wt_T(a_3,b_2) \le \alpha_{a_3} + \alpha_{b_2} = -2$, i.e., $(a_3,b_2)$ is a $(-,-)$ edge wrt $T$. However the edge $(a_3,b_3)$ is 
		a new edge that was introduced in $G^*$, so $b_3$ is $a_3$'s {\em last choice} neighbor in $G^*$. Thus 
		$(a_3,b_2)$ cannot be a $(-,-)$ edge wrt $T$ and so $(a_2,b_1)$ cannot be a blocking edge to $T$. That is, this case never happens.
		
		\item[(ii)] Hence $(a_2,b_1)$ is {\em not} a blocking edge to $T$ and so at least one of $a_2,b_1$ prefers $T$ to $N$. We also know that
		$\wt_T(a_3,b_2) \le \alpha_{a_3} + \alpha_{b_2} \le 0$ since $\alpha_{a_3} = -1$. Thus at least one of $a_3,b_2$ prefers $T$ to $N$.
		So at least 2 vertices among the 4 vertices $b_1,a_2,b_2,a_3$ prefer $T$ to $N$. So $\phi(N\cap p, T'\cap p) \le 2 \le \phi(T'\cap p, N\cap p)$.
	\end{itemize}
\end{itemize}

\medskip

\noindent{\em Case~2.} Exactly one endpoint of $p$ is in $W$ and $p$ has an even number of vertices. So
$p = \langle a_1,b_1,\ldots,a_t,$ $b_t\rangle$, where $(a_i,b_i) \in T$ and exactly one of $a_1,b_t$ is in $W$, say $b_t \in W$. 
Among the $2t$ vertices of $p$ in $G^*$, we have: $\phi_{G^*}(T, N) = t_2 \ge t \ge t_1 = \phi_{G^*}(N, T)$. 

The edge $(a_t,b_t)$ is a new edge and so $(a_t,b_t) \notin T'$. 
In the election between $T'$ and $N$ when restricted to these $2t$ vertices, we have $\phi(N, T') = t_1 \le t$ 
(since $N$ preserves all its $t_1$ votes from $p$) and $\phi(T', N) = t_2 -1 \ge t-1$ votes (since $T'$ gets all the votes in favor of $T$ 
other than $b_t$'s vote). Thus when restricted to these $2t$ vertices, we have 
$\phi(N, T')/{\phi(T', N)} \le t/(t-1) \le 2$ since $t \ge 2$.

\medskip

\noindent{\em Case~3.} Exactly one endpoint of $p$ is in $W$ and $p$ has an odd number of vertices, say $2t+1$, 
i.e., let $p = \langle a_0,b_1,a_1,\ldots,a_t\rangle$, where $(a_i,b_i) \in T$ and $a_t \in W$ (see Fig.~\ref{fig1:example}, right). 
Among the $2t+1$ vertices in $p$, we have:  $\phi_{G^*}(T, N) = t_2 \ge t+1$ while $\phi_{G^*}(N, T) = t_1 \le t$. Again, the edge $(a_t,b_t)$ is a new edge and so $(a_t,b_t) \notin T'$. 

In the election between $T'$ and $N$ when restricted to these $2t+1$ vertices,
we have $\phi(T', N) = t_2 -1 \ge t$ (since $T'$ gets all the votes in favor of $T$ other than $a_t$'s vote) 
while $\phi(N, T') = t_1 \le t$. Thus when restricted to these $2t+1$ vertices, 
$\phi(N, T') \le t \le \phi(T', N)$.  \qed\endproof

\subsection{Proofs of Theorem~\ref{thm:extra-dominant-extn} and Theorem~\ref{thm:dominant-extn}}
\label{sec:polytope-proof} We know from Theorem~\ref{thm:xc} (proved in Section~\ref{sec:lb-xc}) that the popular matching polytope has no compact extended formulation. As mentioned in Section~\ref{sec:poly-overview}, this polytope has a compact extended formulation when $G$ admits a  {\em perfect} stable matching, i.e., 
when $G$ has no unstable vertex. 

\paragraph{\bf The Reduced Instance $H$.}
Let $H$  be the marriage instance obtained by deleting the vertices of $U^*$ and their incident edges from $G^* = (A \cup B, E^*)$. Recall that $U^*$ is the set of unstable vertices of $G^*$. So the 
vertex set of $H$ is $(A \cup B)\setminus U^*$ and its edge set $E_H = E^* \setminus \cup_{u \in U^*}\delta^*(u)$ (note that, for $u \in U^*$, $\delta^*(u)=\delta(u)$). Preferences in $H$ are 
induced by preferences in $G^*$, i.e., vertices in $U^*$ are deleted from the preference lists of their neighbors. Since $U^*$ is the set of unstable vertices in $G^*$, every stable matching in $H$ is perfect. So
every popular matching in $H$ is perfect and hence, dominant (by Theorem~\ref{thm:dominant-comb}).

The polytope ${\cal E}_H$ is defined by constraints~\eqref{sec5-constr3-bis}--\eqref{sec5-constr5-bis} given in Section~\ref{sec:poly-overview}, where
the vertex set in these constraints is now $(A \cup B)\setminus U^*$ and the edge set is $E_H$.
We moreover claim that $x_{(u,u)} = 0$ for all $u \in (A \cup B)\setminus U^*$. This holds because all vertices in $(A \cup B)\setminus U^*$ are stable, and 
a popular fractional matching $\vec{x}$ has to fully match each stable vertex to 
neighbors other than itself, otherwise $\wt_x(S) > 0$ for any stable matching $S$~\cite[footnote~2]{HK17}.

The following theorem will be useful. For any matching $M$,
let $I_M$ be its incidence vector.

\begin{theorem}[\cite{HK17}]
	\label{first-thm}
	In any marriage instance $H$ that admits a perfect 
	stable matching, each extreme point $(\vec{x},\vec{\alpha})$ of ${\cal E}_H$ is such that $\vec{x}=I_M$, 
	where $M$ is a popular matching in $H$,    
	$\vec{\alpha} \in \{\pm 1\}^{n_0}$ is a (dominant) witness of $M$, and $n_0$ is the number of vertices in $H$.
\end{theorem}

It is easy to see (from Theorem~\ref{first-thm}) that for any vertex $u$ in $H$, the constraint $\alpha_u \le 1$ is a valid inequality for ${\cal E}_H$.
Therefore ${\cal E}_H \cap \{\alpha_v = 1: \ \forall v \in \Nbr(U^*)\}$ is a face of ${\cal E}_H$, where 
$\Nbr(U^*) \subseteq B$ is the set of neighbors in $G^*$ (equivalently, in $G$) of vertices in $U^*$.
Call this face ${\cal Z}_H$.
Theorem~\ref{first-thm} and the definition of ${\cal Z}_H$ imply the following: (let $n_0 = |(A \cup B)\setminus U^*|$)

\smallskip

\begin{itemize}
	\item every extreme point of ${\cal Z}_H$ is of the form $(I_N, \vec{\gamma})$ where $I_N$ is the incidence vector of a popular matching $N$ in $H$ and $\vec{\gamma}\in\{\pm 1\}^{n_0}$ is a (dominant) witness of $N$ such that $\gamma_v = 1$ for all $v \in \Nbr(U^*)$. 
\end{itemize}

\smallskip

\proof{ (of Theorem~\ref{thm:dominant-extn}).}
Let $(\vec{x},\vec{\alpha}) \in {\cal C}_{G^*}$. It follows from constraint~\eqref{new-constr4-bis} that $x_e~=~0$ for all edges $e$ incident to vertices in $U^*$ and from constraint~\eqref{new-constr4} that $\alpha_a~=~0$ for all $a \in U^*$ (recall that $U^*\subseteq A$). Let $(\vec{x}',\vec{\alpha}')$ be obtained from $(\vec{x},\vec{\alpha})$ by projecting out coordinates $x_e$ for $e \in \delta^*(a), a \in U^*$ and $\alpha_a$ for $a \in U^*$.

It is straightforward to see that
$(\vec{x}',\vec{\alpha}')$ satisfies constraints~\eqref{sec5-constr3-bis}--\eqref{sec5-constr5-bis} that define ${\cal E}_H$.
Moreover, we have $\alpha_b \ge 1$ for all $b \in \Nbr(U^*)$ due to constraint~(\ref{new-constr1}) for the edge $(a,b)$ where $a \in U^*$. This is because $c_x(a,b) = 1$ for such an edge $(a,b)$ and $\alpha_a = 0$. 
Thus $(\vec{x}',\vec{\alpha}') \in {\cal Z}_H$.

Since ${\cal Z}_H$ is integral, $(\vec{x}',\vec{\alpha}')$ is a convex combination of $(I_{N_1},\vec{\gamma}^1),\ldots,(I_{N_k},\vec{\gamma}^k)$ where $N_1,\ldots,N_k$ are popular matchings in $H$ with respective dominant witnesses $\vec{\gamma}^1,\ldots,\vec{\gamma}^k$ and we have $\gamma^i_v = 1$ for all $v \in \Nbr(U^*)$ and $1 \le i \le k$.
For each $i$, matching $N_i$ can be viewed as a matching $\tilde{N}_i$ in $G^*$, so $I_{\tilde{N}_i}$ has value 0 in coordinates corresponding
to edges in $\delta^*(u)$ for $u \in U^*$. Similarly, we extend the witness vector $\vec{\gamma}^i \in \{\pm 1\}^{n_0}$ to a vector $\vec{\tilde{\gamma}}^i \in \{0,\pm 1\}^n$ by assigning $\tilde{\gamma}^i_u = 0$ for all $u \in U^*$. Thus $(\vec{x},\vec{\alpha})$ is a convex combination of $(I_{\tilde{N}_1},\vec{\tilde{\gamma}}^1),\ldots,(I_{\tilde{N}_k},\vec{\tilde{\gamma}}^k)$. Hence ${\cal C}_{G^*}$ is integral.

Note that $\vec{\tilde{\gamma}}^i$ is a dominant witness of $\tilde{N}_i$, thus $\tilde{N}_i$ is a dominant matching in $G^*$ (by Theorem~\ref{thm:dominant-witness}). 
Hence, all extreme points of ${\cal C}_{G^*}$ are of the form $(I_N,\vec{\gamma})$, where $N$ is a dominant matching in $G^*$ and $\vec{\gamma}$ is a dominant witness. 
The following claim completes the proof of Theorem~\ref{thm:dominant-extn}. \endproof

\begin{new-claim}
	\label{last-clm}
	For any dominant matching $M$ in $G^*$ and any dominant witness $\vec{\beta}$ of $M$, we have $(I_M,\vec{\beta}) \in {\cal C}_{G^*}$.
\end{new-claim}
\proof{} 
We need to show that $(I_M,\vec{\beta})$ satisfies constraints~(\ref{new-constr3-bis})-(\ref{constr5-first}).
It follows from the definition of witness that $\sum_{u \in A \cup B}\beta_u = 0$. We also know that $\beta_u \in \{\pm 1\}$ for
$u \in (A \cup B)\setminus U^*$ (by definition of dominant witness) and $\beta_u = 0$ for $u \in U^*$ (see Lemma~\ref{lemma1}). 
Since $M$ is a dominant matching, all popular vertices, i.e., those in
$(A \cup B)\setminus U^*$, are matched in $M$; also $M \subseteq E^*_D$ (the set of dominant edges in $G^*$). Hence all constraints in \eqref{new-constr3}-\eqref{constr5-first}
along with \eqref{new-constr3-bis} are satisfied by $(I_M,\vec{\beta})$.

Regarding the constraints in \eqref{new-constr1}, it follows from the definition of witness that $\beta_a + \beta_b \ge \wt_M(a,b)$ for all
$(a,b) \in E^*$. We have $c_M(a,b) = \wt_M(a,b)$ for all edges $(a,b)$ except when $a \in U^*$. For such an edge $(a,b) \in U^*\times\Nbr(U^*)$,
we have $\wt_M(a,b) = 0$ but $c_M(a,b) = 1$. Since $\beta_a = 0$ and $\beta_b\in\{\pm 1\}$, we deduce $\beta_b= 1$ and  $\beta_a + \beta_b = 1 = c_M(a,b)$.
Thus $(I_M,\vec{\beta})$ satisfies all the constraints defining ${\cal C}_{G^*}$.   
\qed\endproof

\medskip

\proof{ (of Theorem~\ref{thm:extra-dominant-extn}).}
The constraints $\alpha_a \ge -1$ for all $a \in U_A \setminus U^*$ are valid inequalities for ${\cal C}_{G^*}$, thus 
${\cal C}_{G^*} \cap \{\alpha_a = -1: a \in U_A \setminus U^*\}$ defines a face of ${\cal C}_{G^*}$. This face is the polytope ${\cal F}_{G^*}$.
Since ${\cal C}_{G^*}$ is integral, so is ${\cal F}_{G^*}$. Every extreme point of ${\cal F}_{G^*}$ has the form $(I_N,\vec{\gamma})$ 
where $N$ is a dominant matching and $\vec{\gamma}$ is a dominant witness of $N$ along with the extra condition that 
$\gamma_a = -1$ for $a \in U_A \setminus U^*$. Thus $N$ is an extra-dominant matching (see Definition~\ref{constr-dom}) and
$\vec{\gamma}$ is an {\em extra-dominant witness} of $N$.

Since $(I_M, \vec{\beta}) \in {\cal C}_{G^*}$ for every dominant matching $M$ and every dominant witness $\vec{\beta}$ of $M$, we have
$(I_N,\vec{\gamma}) \in {\cal F}_{G^*}$ for every extra-dominant matching $N$ and every extra-dominant witness $\vec{\gamma}$ of $N$.
Thus ${\cal F}_{G^*}$ is an  extension of the extra-dominant matching polytope of $G^*$.  \qed\endproof

\section{Quasi-Popular Matching via the Popular Fractional Matching Polytope}\label{sec:fractional-popular} In this section we will show another proof of Theorem~\ref{thm:algo}. In fact, our algorithm in this section computes a quasi-popular
matching of cost at most that of a min-cost popular fractional matching. We will first devise an understanding of {\em witness} for
quasi-popular matchings.

\subsection{Witness for Quasi-Popular Matchings}\label{sec:witness-quasi} Let $G = (A\cup B, E)$ be a marriage instance and let $M$ be any matching in $G$. 
Recall the edge labels in $\{(\pm, \pm)\}$ from Section~\ref{prelims}. 
Based on these labels, we will now define an edge weight function $g_M$ in $G$ as follows:
\begin{equation}\label{eq:gm}
	g_M(e) = \begin{cases}  +2 & \text{if\ $e$\ is\ labeled\ $(+,+)$}\\
		-4 & \text{if\ $e$\ is\ labeled\ $(-,-)$}\\
		-1 & \text{if\ $e$\ is\ labeled\ $(+,-)$ or $(-,+)$}\\
		0 & \text{if\ $e \in M$.}
	\end{cases}                            
\end{equation}

Note that in the above edge weight function, a vote of ``$-$'' gets scaled by a factor of 2 to become a $-2$ while a  vote of ``$+$'' 
remains a +1. This is because in an election between $M$ and a rival matching, a ``$-$'' is a vote for $M$
and against the rival matching while a ``$+$'' is a vote for the rival matching.
Also, $g_M(e) = 0$ for all $e \in M$ since both vertices vote ``0'' for each other: both endpoints of $e$ are indifferent between $M$ and
the rival matching since each of them gets assigned the same partner in both the matchings.

Let $\tilde{G}$ be the graph $G$ augmented with self-loops, i.e., each vertex is assumed to be its own last choice neighbor. 
Thus any matching $M$ in $G$ becomes a perfect matching $\tilde{M}$ in $\tilde{G}$ by including self-loops at all 
vertices left unmatched in $M$. The edge weight function $g_M$ can be extended to self-loops as well:
let $g_M(u,u) = 0$ if $u$ is matched to itself in $\tilde{M}$, else $g_M(u,u) = -2$.

Thus $g_M$ is an edge weight function in $\tilde{G}$ and Claim~\ref{clm1} follows easily from the definition of $g_M$.

\begin{new-claim}
	\label{clm1}
	For any matching $N$ in $G$, we have $g_M(\tilde{N}) = \phi(N,M) - 2\phi(M,N)$.
\end{new-claim}

Hence, a matching $M$ in $G$ satisfies $u(M) \le 2$ if and only if every perfect matching in $\tilde{G}$ has weight at most $0$ with respect to the $g_M$.
Consider the max-weight perfect matching LP in $\tilde{G} = (A \cup B, \tilde{E})$: this is \eqref{LP3} given below in variables $x_e$ for $e \in \tilde{E}$;
recall that $\tilde{\delta}(u) = \delta(u) \cup \{(u,u)\}$.
The linear program \eqref{LP4} is the dual of \eqref{LP3}. The dual variables are $\alpha_u$ for $u \in A \cup B$.

\vspace{-.3cm}

\begin{linearprogram}
	{
		\label{LP3}
		\maximize \sum_{e \in \tilde{E}} g_M(e)\cdot x_e  
	}
	\qquad\sum_{e \in {\tilde{\delta}}(u)}x_e \ & = \ \ 1  \ \ \, \forall\, u \in A \cup B \notag\\
	x_e  \ & \ge \ \ 0   \ \ \ \forall\, e \in \tilde{E}. \notag
\end{linearprogram}

\begin{linearprogram}
	{
		\label{LP4}
		\minimize \sum_{u \in A \cup B}\alpha_u
	}
	\alpha_{a} + \alpha_{b} \ & \ge \ \ g_{M}(a,b) \ \ \ \ \forall\, (a,b)\in E \notag\\
	\alpha_u \ & \ge \ \ g_M(u,u) \ \ \ \, \forall\, u \in A \cup B. \notag
\end{linearprogram}

Given a matching $M$, if we can show a a feasible solution $\vec{\alpha} \in \mathbb{R}^n$ to \eqref{LP4}
such that $\sum_{u\in A \cup B}\alpha_u~=~0$, then the optimal value of \eqref{LP3} is at most 0, so we have
$\phi(N,M) - 2\phi(M,N) \le 0$ for all $N$, i.e., $u(M) \le 2$.  Let us call such a vector $\vec{\alpha}$ a {\em witness} for
quasi-popularity. Interestingly, every quasi-popular matching admits a simple witness as shown below.

\begin{lemma}
	\label{lemma:quasi-witness}
	If $M$ is a quasi-popular matching in $G$ then $M$ has a witness $\vec{\alpha} \in \{0, \pm 1, \pm 2\}^n$.
\end{lemma}
\proof{} 
Since $M$ is quasi-popular, the optimal value of \eqref{LP3} is at most 0 and
so $\tilde{M}$ is an optimal solution to \eqref{LP3} (because $g_M(\tilde{M}) = 0$). 
The graph $G$ is bipartite, hence the constraint matrix of \eqref{LP4} is totally unimodular (note that adding self-loops does not change this fact). Thus \eqref{LP4} has an integral optimal solution $\vec{\alpha}$. 

We have $\alpha_u \ge g_M(u,u) \ge -2$ for all $u$.  Complementary slackness implies  
$\alpha_u + \alpha_v = g_M(u,v) = 0$ for each edge $(u,v) \in M$. 
Thus $\alpha_u = -\alpha_v \le 2$ for every vertex $u$ matched in $M$. 
Regarding any vertex $u$ that is left unmatched in $M$, we have $(u,u) \in \tilde{M}$, so $\alpha_u = g_M(u,u) = 0$ (by complementary slackness). 
Hence $\vec{\alpha} \in \{0, \pm 1, \pm 2\}^n$.   
\qed\endproof

\subsection{A Min-Cost Popular Fractional Matching}\label{sec:pop-frac-matching} The popular fractional matching polytope ${\cal L}_G$ was briefly discussed in Section~\ref{sec:poly-overview}. 
The polytope ${\cal L}_G$ is the set of all perfect fractional matchings $\vec{x}$ in $\tilde{G}$ that satisfy $\wt_x(N) 
\le 0$ for all perfect matchings $N$ in $\tilde{G}$. 

The function $\wt_x(N)$ defined below is linear in $\vec{x}$: (recall the function $\vote_u(v,\vec{x})$ defined in Section~\ref{sec:poly-overview})
\begin{equation*}
	\wt_x(N) \ = \ \sum_{e \in N}\wt_x(e) \ = \ \sum_{u \in A \cup B}\vote_u(N(u),\vec{x}) \ = \ \sum_{u\in A \cup B}\left(\sum_{v': v' \prec_u N(u)}x_{(u,v')} - \sum_{v': v' \succ_u N(u)}x_{(u,v')}\right).
\end{equation*}

A compact extended formulation ${\cal E}_G$ of the polytope ${\cal L}_G$ was given in Section~\ref{sec:poly-overview}. 
Thus we can find a min-cost popular fractional matching $\vec{q}$ in $G$ in polynomial time by optimizing over ${\cal E}_G$. 
We can, in fact, assume that $\vec{q}$ is a vertex of ${\cal L}_G$, hence half-integral (recall that ${\cal L}_G$ is a half-integral polytope~~\cite[Theorem~1]{HK17}). 
We will now show how we can use $\vec{q}$ to obtain a quasi-popular matching of cost at most $\cost(\vec{q})$.

We will show that $\vec{q}=(I_{N_1} + I_{N_2})/2$ 
where $N_1, N_2$ are quasi-popular matchings in $G$ and $I_N$ is the edge incidence vector of matching $N$.
Since $\cost(N_1) + \cost(N_2) = 2\cdot\cost(\vec{q})$, one of $N_1, N_2$ has cost at most $\cost(\vec{q})$.
Thus the matching in $\{N_1,N_2\}$ with lower cost is our desired matching.

\paragraph{\bf A New Graph $G'$.}
The following useful property on the min-cost popular fractional matching $\vec{q}$ follows from the proof of half-integrality of ${\cal L}_G$ given in \cite{HK17}: for every $u \in A \cup B$ we have
$\sum_{e \in \delta(u)}q_e \in \{0,1\}$. 
In other words, every vertex $u$ is either ``completely matched'' or 
``completely unmatched'' in $\vec{q}$ (so $q_{(u,u)} = 1$ in the latter case), i.e., no vertex is {\em half-matched} in $\vec{q}$.

Let $U_A$ be the set of vertices in $A$ that are left unmatched in $\vec{q}$ and let $U_B$ be the set of vertices in $B$ that are left unmatched in $\vec{q}$.
We now augment the graph $G$ with new vertices and new edges as follows.

Corresponding to each $u \in U_A \cup U_B$, there will be a new vertex $u'$. Thus the vertex set of the augmented graph (call it $G'$) is $A' \cup B'$ where
$A' = A \cup U'_B$
and $B' = B \cup U'_A$. Here  $U'_A = \{u': u \in U_A\}$, i.e., $U'_A$ is a copy of $U_A$, and  $U'_B = \{u': u \in U_B\}$ is a copy of $U_B$.

\begin{figure}[h]
	\centerline{\resizebox{0.23\textwidth}{!}{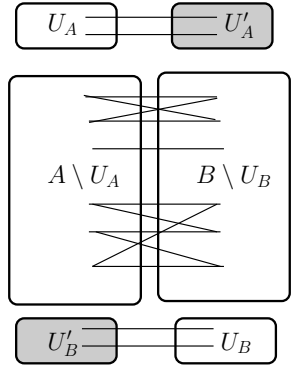}}
	\caption{Vertices in gray sets, i.e., in $U'_A \cup U'_B$, are the new vertices in $G'$ and the new edges are $(u,u')$ for $u \in U_A \cup U_B$. The edges shown here are in the half-integral matching $\vec{q'}$.}
	\label{fig1}
\end{figure}

The edge set $E'$ of $G'$ is $E \cup \{(u,u'): u \in U_A\cup U_B\}$. The vertex $u'$ is the least preferred neighbor of $u$ for every $u \in U_A \cup U_B$ and
$u$ is $u'$'s only neighbor. 
Fig.~\ref{fig1} will be helpful. 

\paragraph{\bf The Fractional Matching $\vec{q'}$.}
We extend $\vec{q}$ to a half-integral matching $\vec{q'}$ in $G'$ as follows.
\begin{itemize}
	\item for each $e \in E$: assign $q'_e = q_e$;
	\item for every $u \in U_A \cup U_B$: assign $q'_{(u,u')} = 1$. 
\end{itemize}
Thus $\vec{q'} = \vec{q} \cup \{(u,u'): u \in U_A \cup U_B\}$.
Observe that the fractional matching $\vec{q'}$ is {\em perfect} in $G'$. This is because all vertices in $A \cup B$ that are outside $U_A \cup U_B$ were already
(completely) matched in $\vec{q}$. The vertices left unmatched in $\vec{q}$ are precisely those in $U_A \cup U_B$, and each $u \in U_A \cup U_B$ is now matched to its {\em twin} $u' \in U'_A \cup U'_B$ in $\vec{q'}$.
The following lemma will be useful to us. Let $n' = |A'\cup B'|$.

\begin{lemma}\label{lem:q'-popular}
	The half-integral matching $\vec{q'}$ is popular in $G'$, and it has a witness $\vec{\alpha'} \in \{0,\pm 1\}^{n'}$.
\end{lemma}
\proof{} 
It follows from the {\em special} half-integrality of $\vec{q}$ in $G$ (no vertex is half-matched) that $\wt_q(e) \in \mathbb{Z}$ for all $e \in E$;
also $\wt_q(u,u) \in \{0,-1\}$ for all vertices $u \in A \cup B$. Consider the LP with objective function $\min \sum_{u\in A \cup B}\alpha_u$ and constraints
\eqref{sec5-constr1}--\eqref{sec5-constr3}, with $\vec{q}$ replacing $\vec{x}$ in these constraints. The total unimodularity of this system 
implies that $\vec{q}$ has an integral witness $\vec{\alpha}$; more precisely, $\vec{\alpha} \in \{0, \pm 1\}^n$.

Also $\alpha_u = \wt_q(u,u) = q_{(u,u)}-1 = 0$ for every $u \in U_A \cup U_B$ (by complementary slackness on \eqref{sec5-constr3}): this is because 
$\vec{\alpha}$ is an optimal solution to the above LP and $\vec{q}$ is an optimal solution to the dual LP that maximizes $\wt_q(\vec{x})$
subject to constraints~\eqref{sec5-constr5}-\eqref{sec5-constr5-bis}.  We refer to \cite{HK17} for details. 

In order to prove the popularity of $\vec{q'}$, we will define a witness $\vec{\alpha'}$ for $\vec{q'}$ as follows:
let $\alpha'_u = \alpha_u$ for all $u \in A \cup B$ and $\alpha'_u = 0$ for $u \in U'_A \cup U'_B$. As $\vec{\alpha} \in \{0,\pm 1\}^{n}$, we have $\vec{\alpha'} \in \{0,\pm 1\}^{n'}$. We need to show that $\vec{q'}$ and
$\vec{\alpha'}$ satisfy constraints~\eqref{sec5-constr3-bis}-\eqref{sec5-constr5-bis}. We already know that $\vec{q'}$ satisfies \eqref{sec5-constr5}-\eqref{sec5-constr5-bis}. We will now show that \eqref{sec5-constr3-bis}-\eqref{sec5-constr3} are satisfied.

\begin{enumerate}
	\item $\sum_{u \in A' \cup B'}\alpha'_u = 0$: this follows from the definition of $\vec{\alpha}'$ and the fact that  
	$\sum_{u \in A \cup B}\alpha_u = 0$.
	
	\smallskip
	
	\item $\alpha'_a + \alpha'_b \ge \wt_{q'}(a,b)$ for all $(a,b) \in E'$: this follows from the fact that
	$\alpha'_a + \alpha'_b = \alpha_a + \alpha_b \ge \wt_{q}(a,b) = \wt_{q'}(a,b)$ for all $(a,b) \in E$.
	The new edges in $G'$ are $(u,u')$ for $u \in U_A \cup U_B$ and we have $\alpha'_u + \alpha'_{u'} = 0 =  \wt_{q'}(u,u')$.
	
	\smallskip
	
	\item $\alpha'_v \ge -1 = \wt_{q'}(v,v)$ for all $v \in A' \cup B'$. 
	
\end{enumerate}

Thus the half-integral matching $\vec{q'}$ is popular in $G'$ with witness $\vec{\alpha}'$.  
\qed\endproof

\medskip

It was shown in \cite{HK17} that any popular fractional matching that is perfect can be written as a convex combination of popular
matchings; moreover, these popular matchings can be efficiently computed. Theorem~\ref{lemma:new} follows from this proof in \cite{HK17} which, in turn, is based on the proof of integrality of the stable matching polytope in \cite{TS98}.
We include in the appendix a proof of this theorem for our case, i.e., for the half-integral matching $\vec{q'}$, showing that $\vec{q'}$ can be written as the convex combination of two popular matchings; moreover, these two matchings can be efficiently computed.

Note that this theorem does not hold for the popular half-integral matching $\vec{x}$ (which is not a convex combination of popular matchings) 
in the instance shown in Fig.~\ref{fig:non-int} since $\vec{x}$ is not perfect. 

\begin{theorem}
	\label{lemma:new}
	The popular (perfect) half-integral matching $\vec{q'}$ in $G'$ can be written as $\vec{q'} = (I_{M_1} + I_{M_2})/2$ for popular matchings $M_1$ and $M_2$ in $G'$. 
\end{theorem}

The popularity of matchings $M_1$ and $M_2$ will be shown by defining witnesses of popularity $\vec{\beta}$ and $\vec{\gamma}$ as follows.
Recall that $\vec{\alpha'}$ is the witness of $\vec{q'}$ constructed in the proof of Lemma~\ref{lem:q'-popular}.
For any $u \in A' \cup B'$: 

\begin{itemize}
	\item if $\alpha'_u \in \{\pm 1\}$ then let $\beta_u = \gamma_u = \alpha'_u$; else (so $\alpha'_u = 0$)
	
	\smallskip
	
	\begin{itemize}
		\item if $u \in A'$ then $\beta_u = -1$ and $\gamma_u = 1$; else (so $u \in B'$) $\beta_u = 1$ and $\gamma_u = -1$.
	\end{itemize}
\end{itemize}

\smallskip

Thus $\beta_u, \gamma_u \in \{\pm 1\}$ for every $u \in A' \cup B'$. 
Corollary~\ref{cor1} follows from the fact that $\alpha'_a = \alpha'_b = 0$ for $a \in U_A \subseteq A'$ and $b \in U_B \subseteq B'$.

\begin{corollary}
	\label{cor1}
	We have $\beta_a = -1$  and $\gamma_{a} = 1$ $\forall a \in U_A$ and
	$\beta_b = 1$ and $\gamma_b = -1$ $\forall b \in U_B$.
\end{corollary}

Let $N_1$
(resp., $N_2$) be the matching obtained by deleting edges $(u,u')$ for $u \in U_A \cup U_B$ from $M_1$ (resp., $M_2$). 
Thus $N_1$ and $N_2$ are matchings in $G$. Note that $\vec{q} = (I_{N_1} + I_{N_2})/2$.

\begin{figure}[h]
	\centerline{\resizebox{0.66\textwidth}{!}{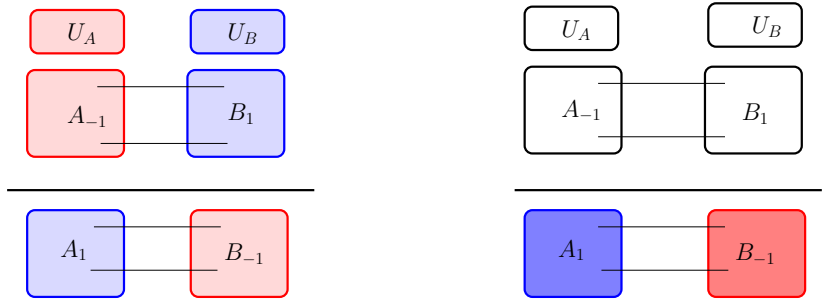}}
	\caption{The figure on left represents $\beta$-values and the one on right represents $\beta'$-values. On left, elements in blue (resp., red) sets have $\beta$-value 1 (resp., $-1$). We have $\beta'_a = \beta_a + 1 \ \forall \, a \in A$ and $\beta'_b = \beta_b - 1 \ \forall \, b \in B$.}
	\label{fig2}
\end{figure}

\begin{lemma}
	\label{lemma1bis}
	The matchings $N_1$ and $N_2$ are quasi-popular in $G$.
\end{lemma}  
\proof{} 
In order to prove the quasi-popularity of $N_1$, it will be useful to partition $A \setminus U_A$ into $A_{-1} \cup A_1$ and $B \setminus U_B$ into $B_1 \cup B_{-1}$ as follows. For $i = \pm 1$:
\begin{itemize}
	\item let $A_{i}$ be the set of those $a \in A \setminus U_A$ such that $\beta_a = i$.
	\item let $B_{i}$ be the set of those $b \in B \setminus U_B$ such that $\beta_b = i$.
\end{itemize}  
Since $\beta_a + \beta_b = \wt_{M_1}(a,b) = 0$ for every edge $(a,b) \in M_1$, we have $N_1 \subseteq (A_1 \times B_{-1}) \cup (A_{-1} \times B_1)$. 
Fig.~\ref{fig2} will be useful.
We now define a certificate $\vec{\beta}'$ of $N_1$'s quasi-popularity. 
\begin{itemize}
	\item Let $\beta'_a = 2$ for all $a \in A_1$ and let  $\beta'_b = -2$ for all $b \in B_{-1}$. Let $\beta'_u = 0$ for all other vertices $u$.
\end{itemize}

We need to show that $\vec{\beta}'$ obeys the three constraints given below, see~\eqref{LP3}--\eqref{LP4}.

\begin{minipage}{.5\textwidth}
	\begin{eqnarray}
		\beta'_a + \beta'_b & \ \ge \ & g_{N_1}(a,b) \ \ \ \ \ \forall (a,b) \in E\label{quasi-constr1}\\
		\beta'_u & \ \ge \ & g_{N_1}(u,u) \ \ \ \ \ \forall u \in A \cup B\label{quasi-constr2}\end{eqnarray}
	
\end{minipage}\begin{minipage}{.5\textwidth}
\begin{eqnarray}
	\sum_{u \in A \cup B} \beta'_u & \ = \ & 0,\label{quasi-constr3}\\
	\nonumber 
\end{eqnarray}
\end{minipage}

\medskip

where $g_{N_1}(\cdot)$ is defined as in~\eqref{eq:gm}, with $M=N_1$. For any $u \in A \cup B$, $g_{N_1}(u,u) = 0$ if $u$ is unmatched in $N_1$ and $g_{N_1}(u,u) = -2$ if $u$ is matched in $N_1$.

\medskip

\noindent{\em Constraint~\eqref{quasi-constr1}.} For any edge $(a,b)$, we have 
$\beta'_a + \beta'_b = \beta_a + \beta_b$: this is because $\beta'_a = \beta_a + 1$ for $a \in A$ and $\beta'_b = \beta_b - 1$ for $b \in B$.
We have $\beta'_a + \beta'_b = \beta_a + \beta_b \ge  \wt_{M_1}(a,b) \ge g_{N_1}(a,b)$ for all edges 
$(a,b) \in E,$\footnote{If $\wt_{M_1}(a,b) = 0$ then $g_{N_1}(a,b) \in \{0,-1\}$; if $\wt_{M_1}(a,b) = -2$ then $g_{N_1}(a,b) = -4$; if $\wt_{M_1}(a,b)~=~2$ then $g_{N_1}(a,b) = 2$.} where $\beta_a + \beta_b \ge \wt_{M_1}(a,b)$ holds for each edge $(a,b) \in E$ due to the popularity of $M_1$ in $G'$
with witness $\vec{\beta}$.

\medskip

\noindent{\em Constraint~\eqref{quasi-constr2}.}
It follows from Corollary~\ref{cor1} that $\beta_a = -1$ for $a \in U_A$ and  $\beta_b = 1$ for $b \in U_B$, so 
$\beta'_u = 0 =  g_{N_1}(u,u)$ for an unmatched vertex $u$. We also have 
$\beta'_u \ge -2 = g_{N_1}(u,u)$ for a matched vertex $u$. Thus $\beta'_u \ge g_{N_1}(u,u)$ for all vertices $u$.

\medskip

\noindent{\em Constraint~\eqref{quasi-constr3}.}
For every edge $(a,b) \in N_1$, observe that $\beta'_a + \beta'_b = \beta_a + \beta_b = \wt_{M_1}(a,b) = 0$. We also observed above that $\beta'_u = 0$ for every unmatched vertex $u$. 
So $\sum_{u \in A \cup B} \beta'_u = 0$.
Thus constraints~\eqref{quasi-constr1}-\eqref{quasi-constr3} are satisfied. We conclude that $N_1$ is quasi-popular in $G$.

The quasi-popularity of $N_2$ in $G$ can be shown similarly.
\qed\endproof

\medskip

Thus both $N_1$ and $N_2$ are quasi-popular matchings in $G$. Since $\vec{q} = (I_{N_1} + I_{N_2})/2$, one of $N_1,N_2$ has cost at most $\cost(\vec{q})$. Since $\vec{q}$ can be computed in polynomial time and $N_1, N_2$ can be efficiently obtained from $\vec{q}$ (see the proof of Theorem~\ref{lemma:new}), we can conclude the following theorem.

\begin{theorem}
	\label{thm:quasi-new}
	Given a marriage instance $G = (A \cup B, E)$ with a function $\cost: E \rightarrow \mathbb{R}$, there is a polynomial time algorithm 
	to compute a quasi-popular matching $N$ such that $\cost(N) \le \cost(\vec{q})$, where $\vec{q}$ is a min-cost popular fractional matching in $G$.
\end{theorem}

\section{Hardness of Finding a Min-Cost Quasi-Popular Matching}\label{sec:hardness} We prove Theorem~\ref{thm:hardness} in this section. We show a reduction from 3SAT. Given a 3SAT formula $\psi$ on $n$ variables, we
construct a marriage instance $G_{\psi}$ and a cost function on the edges. At a high level, this construction resembles the structure of the instance used in~\cite{CFKP18} to show the hardness of deciding if a marriage instance admits a stable matching that is also dominant. However, our proof is much more involved, and requires the introduction of new tools. 

In fact, the hardness reductions from~\cite{CFKP18} heavily rely on the combinatorial characterizations of stable/popular matchings in terms of certain forbidden induced structures. As mentioned in Section~\ref{intro}, such a characterization is not known for quasi-popular matchings. 
Recall from Section~\ref{sec:witness-quasi} that a matching is quasi-popular if and only if it admits a witness in $\{0, \pm 1, \pm 2\}^n$.
We will use such witnesses in our reduction.

\subsection{Our Reduction from 3SAT} As done in \cite{CFKP18}, it will be useful to perform a simple transformation of the 3SAT formula $\psi$ so that the transformed formula
has a unique occurrence of each negative literal. This property is easily accomplished: let $X_1,\ldots,X_n$ be the $n$ variables in $\psi$. 
For each $i$, replace all occurrences of $\neg X_i$ in $\psi$ with $X_{n+i}$, i.e., a single new variable. 
The clauses $X_i \vee X_{n+i}$ and $\neg X_i \vee \neg X_{n+i}$ are added to capture $\neg X_i \iff X_{n+i}$.

So if the original formula $\psi$ was $C_1 \wedge \cdots \wedge C_m$, the transformed formula 
$\psi = C_1 \wedge \cdots \wedge C_m \wedge C_{m+1} \wedge\cdots\wedge C_{m+n} \wedge D_{m+n+1} \wedge \cdots \wedge D_{m+2n}$,
where $C_1,\ldots,C_m$ are the original $m$ clauses with negated literals substituted by new variables and for $1\le i\le n$,
$C_{m+i}$ is the clause $X_i \vee X_{n+i}$ and $D_{m+n+i}$ is the clause $\neg X_i \vee \neg X_{n+i}$.

$G_{\psi}$ is essentially the parallel composition of \emph{clause gadgets}, all sharing exactly two vertices $s$ and $t$. 
A \emph{positive} clause gadget corresponds to one of the $C_i$'s and a \emph{negative} clause gadget corresponds to one of the $D_i$'s.  Each clause gadget is a series composition of {\em literal gadgets} and
the gadgets corresponding to positive and negative literals are different (see Fig.~\ref{newfig4:example}).

\begin{figure*}[h]
	\centerline{\resizebox{0.9\textwidth}{!}{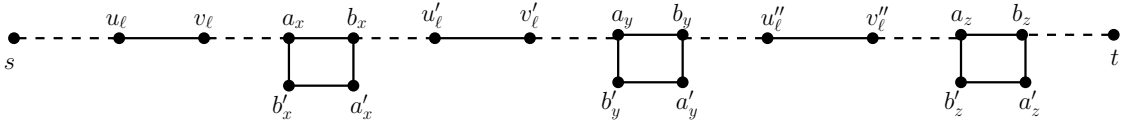}}
	\caption{\small A clause gadget corresponding to a positive clause $C_{\ell} = x \vee y \vee z$. The endpoints of this path, i.e., $s$ and $t$, are common to all clauses. The solid edges have cost 0 while the dashed edges have cost 1.}
	\label{newfig2:example}
\end{figure*}

We now describe the preference lists of vertices in a positive clause $C_{\ell} = x \vee y \vee z$ (see Fig.~\ref{newfig2:example}).
We first describe the preference lists of $u_{\ell},v_{\ell}, u'_{\ell},v'_{\ell}, u''_{\ell},v''_{\ell}$.

\begin{minipage}[c]{0.45\textwidth}
	
	\centering
	\begin{align*}
		&  u_{\ell} : \ v_{\ell} \succ s    \quad \quad && u'_{\ell} : \ v'_{\ell} \succ b_x \quad \quad && u''_{\ell} : \ v''_{\ell} \succ b_y  \\
		&  v_{\ell} : \ a_x \succ u_{\ell}  \quad \quad && v'_{\ell} : \ a_y \succ u'_{\ell} \quad \quad && v''_{\ell} : \ a_z \succ u''_{\ell}\\
	\end{align*}
\end{minipage}

Thus $u_{\ell}$ has 2 neighbors: $v_{\ell}$ (top choice) and $s$ (second choice). Other preference lists are analogous. 
Recall that we assumed $x$ to be the first literal, $y$ to be the second literal, and $z$ to be the third literal in $C_{\ell}$.
We next describe the preference lists of the 4 vertices $a_x,b_x,a'_x,b'_x$ that occur in $x$'s gadget in $C_{\ell}$.

\begin{minipage}[c]{0.45\textwidth}
	
	\centering
	\begin{align*}
		&  a_x : \ b'_x \succ v_{\ell} \succ b_x             \quad \quad && a'_x : \ b_x \succ b'_x \succ \underline{d'_x}\\
		&  b_x : \  u'_{\ell} \succ a_x \succ \underline{c_x} \succ a'_x  \quad \quad && b'_x : \ a'_x \succ a_x \\
	\end{align*}
\end{minipage}

The underlined vertices $c_x$ and $d'_x$ will occur in the gadget of $\neg x$ (similarly for vertices $c_y, d'_y, c_z, d'_z$ below).
Below are the preference lists of the 4 vertices $a_y,b_y,a'_y,b'_y$ that occur in $y$'s gadget in $C_{\ell}$.

\begin{minipage}[c]{0.45\textwidth}
	
	\centering
	\begin{align*}
		&  a_y : \ b'_y \succ v'_{\ell} \succ b_y                       \quad \quad && a'_y : \ b_y \succ b'_y \succ  \underline{d'_y}\\
		&  b_y : \ u''_{\ell} \succ a_y \succ \underline{c_y} \succ a'_y          \quad  \quad && b'_y : \ a'_y \succ a_y \\
	\end{align*}
\end{minipage}

The preference lists of the 4 vertices $a_z,b_z,a'_z,b'_z$ that occur in $z$'s gadget in
$C_{\ell}$ are given below.

\begin{minipage}[c]{0.45\textwidth}
	
	\centering
	\begin{align*}
		&  a_z : \ b'_z \succ v''_{\ell} \succ b_z               \quad \quad && a'_z : \ b_z \succ b'_z \succ \underline{d'_z}\\
		&  b_z : \ a_z \succ \underline{c_z} \succ a'_z \succ t       \quad \quad && b'_z : \ a'_z \succ a_z \\
	\end{align*}
\end{minipage}

The preference lists of $s$ and $t$ are arbitrary (they do not matter for our arguments). The preference lists of vertices that occur in a clause gadget with 2 positive literals will be totally analogous to the
preference lists of vertices in a clause gadget with 3 positive literals. In more detail,
the edge $(u''_{\ell},v''_{\ell})$ will be missing and so will the third literal gadget. In particular, the vertex $t$
will be adjacent to the $b$-vertex in the gadget of the second literal and $t$ will be the last choice of this $b$-vertex.

\begin{figure*}[h]
	\centerline{\resizebox{0.65\textwidth}{!}{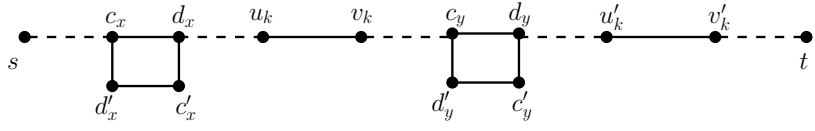}}
	\caption{\small A clause gadget corresponding to a negative clause $D_k = \neg x \vee \neg y$: every negative clause has only 2 literals.}
	\label{newfig3:example}
\end{figure*}

We will now describe the preference lists of vertices in a negative clause $k$---the overall picture here (see Fig.~\ref{newfig3:example}) 
will be the {\em mirror image} of a clause gadget with  2 positive literals.
For a gadget $D_k = \neg x \vee \neg y$, we have:

\begin{minipage}[c]{0.45\textwidth}
	\centering
	\begin{align*}
		&  u_k : \ d_x\succ v_k       \qquad\qquad && u'_k : \ d_y \succ  v'_k\\
		&  v_k : \ u_k \succ c_y      \qquad\qquad && v'_k : \ u'_k \succ  t\\
	\end{align*}
\end{minipage}
\noindent 

We describe below the preference lists of the 4 vertices $c_x,d_x,c'_x,d'_x$ that occur in the first literal gadget in the
clause $D_k$, which is $\neg x$'s (unique) gadget.

\begin{minipage}[c]{0.45\textwidth}
	\centering
	\begin{align*}
		&  c_x : \ d_x\succ \underline{b_{x,i}}\succ \underline{b_{x,j}}\succ\cdots \succ \underline{b_{x,\ell}}\succ d'_x\succ s    \quad \quad &&   c'_x : \ d'_x \succ d_x\\
		&  d_x : \ c'_x \succ u_k \succ c_x   \quad \quad &&  d'_x : \ \underline{a'_{x,i}} \succ \underline{a'_{x,j}} \succ \cdots \succ \underline{a'_{x,\ell}} \succ c_x \succ c'_x\\
	\end{align*}
\end{minipage}

The underlined vertices $b_{x,i},b_{x,j},\ldots,b_{x,\ell}$ in $c_x$'s preference list are the $b$-vertices in 
the $x$-gadgets in various clauses. The order among these vertices in $c_x$'s preference list is not important. Similarly, 
$a'_{x,i},a'_{x,j},\ldots,a'_{x,\ell}$ in $d'_x$'s preference list are the $a'$-vertices in the gadgets of $x$ that occur in various clauses
and the order among these is not important.

We describe next the preference lists of the 4 vertices $c_y,d_y,c'_y,d'_y$ that occur in the second literal gadget in $D_k$,
which is $\neg y$'s (unique) gadget.

\begin{minipage}[c]{0.45\textwidth}			
	\centering
	\begin{align*}
		&  c_y : \ v_k \succ d_y\succ \underline{b_{y,i'}}\succ \underline{b_{y,j'}}\succ\cdots \succ \underline{b_{y,\ell'}} \succ d'_y \quad \quad &&  c'_y : \ d'_y \succ d_y\\
		&  d_y : \ c'_y \succ u'_k \succ c_y                                        \quad \quad &&      d'_y : \  \underline{a'_{y,i'}} \succ \underline{a'_{y,j'}} \succ \cdots \succ \underline{a'_{y,\ell'}} \succ c_y \succ c'_y\\
	\end{align*}
\end{minipage}

\begin{figure*}[h]
	\centerline{\resizebox{0.65\textwidth}{!}{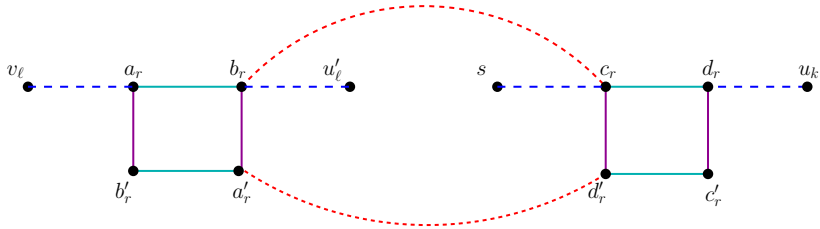}}
	\caption{\small A gadget corresponding to a positive literal (left) and to a negative literal (right) of a variable $r$. We use $c_r,d_r,c'_r,d'_r$ to denote the 4 vertices in the negative gadget of $r$ (there is a unique clause with $\neg r$).
		For the sake of convenience, we use $a_r,b_r,a'_r,b'_r$ to denote the 4 vertices in the positive gadget of $r$ in the $\ell$-th clause. Dashed edges have cost $1$ and solid edges have cost $0$.} 
	\label{newfig4:example}
\end{figure*}

\noindent{\bf  Edge Costs.} For each edge $e$ in $G_{\psi}$, we will set $\cost(e) \in \{0,1\}$. 
\begin{itemize}
	\item Set $\cost(e) = 0$ where $e$ is any of the $u$-$v$, $u'$-$v'$, and $u''$-$v''$ edges.
	\item Set $\cost(e) = 0$ where $e$ is an intra-gadget edge of any literal.
	\item For all other edges $e$, set $\cost(e) = 1$. 
\end{itemize}

\smallskip

In particular, for all edges $e$ incident to $s$ and $t$, we have $\cost(e) = 1$. 
Similarly, for any edge $e$ between $r$'s gadget and $\neg r$'s gadget for any variable $r$, we have $\cost(e) = 1$. 
In our figures all dashed edges have cost~1 and the solid edges have cost 0.

\begin{theorem}
	\label{thm:correctness}
	If $\psi$ is satisfiable then $G_{\psi}$ admits a quasi-popular matching $M$ with $\cost(M) = 0$. 
\end{theorem}

\proof{} 
There is a natural way of constructing the matching $M$ in our instance $G_{\psi}$. Include all $u$-$v$, $u'$-$v'$, and $u''$-$v''$ edges in $M$. 
Then, we will use the satisfying assignment for $\psi$ to choose edges from each literal gadget. For any variable $r \in \{X_1,\ldots,X_{2n}\}$:
\begin{itemize}
	\item if $r = \mathsf{true}$ then take the pair of edges $(c_r,d_r),(c'_r,d'_r)$ from $\neg r$'s gadget and the pair of edges 
	$(a_{r,i},b'_{r,i})$, $(a'_{r,i},b_{r,i})$ from $r$'s gadget in clause~$i$ (for every clause that $r$ belongs to). 
	\item if $r = \mathsf{false}$ then take the pair of edges $(c_r,d'_r),(c'_r,d_r)$ from $\neg r$'s gadget and the pair of edges 
	$(a_{r,i},b_{r,i})$, $(a'_{r,i},b'_{r,i})$ from $r$'s gadget in clause~$i$ (for every clause that $r$ belongs to). 
\end{itemize}

It is easy to see that $\cost(M) = 0$. We will now show a witness $\vec{\alpha} \in \{0, \pm 1, \pm 2\}^n$ to prove $u(M) \le 2$. 
Since $s$ and $t$ are unmatched in $M$, we will set $\alpha_s = \alpha_t = 0$. This will ensure that $\alpha_u \ge g_M(u,u)$ for all vertices $u$ since $\alpha_v = 0 = g_M(v,v)$ for $v \in \{s,t\}$ while $\alpha_u \ge -2 = g_M(u,u)$ for all vertices $u$ matched in $M$.

We now set $\alpha$-values for vertices matched in $M$.
Consider any negative clause, say $D_k = \neg x \vee \neg y$. Note that a satisfying assignment sets exactly one of $x,y$ to be
$\mathsf{true}$ (this is by our transformation of the formula $\psi$: if $x = X_i$ then $y = X_{n+i} = \neg X_i$). Thus we have two cases here:
either (i)~$x = \mathsf{true}$ and $y = \mathsf{false}$, or (ii)~$x = \mathsf{false}$ and $y = \mathsf{true}$.

Let $c_x,d_x,c'_x,d'_x$ be the 4 vertices in $\neg x$'s gadget and let $c_y,d_y,c'_y,d'_y$ be the 4 vertices in $\neg y$'s gadget.
The $\alpha$-values of these 8 vertices along with those of $u_k, v_k, u'_k, v'_k$ in cases~(i) and (ii) are given below.

\begin{minipage}[b]{0.3\linewidth}
	\centering
	\begin{eqnarray*}
		\setlength{\arraycolsep}{0.5ex}\setlength{\extrarowheight}{0.25ex}
		\begin{array}{@{\hspace{1ex}}c@{\hspace{1ex}}||@{\hspace{1ex}}c@{\hspace{1ex}}|@{\hspace{1ex}}c@{\hspace{1ex}}|@{\hspace{1ex}}c@{\hspace{1ex}}|@{\hspace{1ex}}c@{\hspace{1ex}}|@{\hspace{1ex}}c@{\hspace{1ex}}|@{\hspace{1ex}}c@{\hspace{1ex}}|@{\hspace{1ex}}c@{\hspace{1ex}}|@{\hspace{1ex}}c@{\hspace{1ex}}|@{\hspace{1ex}}c@{\hspace{1ex}}|@{\hspace{1ex}}c@{\hspace{1ex}}|@{\hspace{1ex}}c@{\hspace{1ex}}|@{\hspace{1ex}}c@{\hspace{1ex}}}
			\  & \alpha_{c_x} \ & \alpha_{d_x} \ &  \alpha_{c'_x}\ & \alpha_{d'_x} \ & \alpha_{u_k} \ & \alpha_{v_k} \ & \alpha_{c_y} \  & \alpha_{d_y} \ & \alpha_{c'_y} \  & \alpha_{d'_y} \  & \alpha_{u'_k} \   & \alpha_{v'_k} \ \\[.5ex] \hline
			case~(i) \ & -1 \ & 1 \ & -1 \ & 1 \  & 1 \ & -1 \ & 1  \ & -1 \ & 1 \ & -1 \  & 0 \ & 0 \ \\[.5ex] \hline
			case~(ii) \ & 1 \ & -1 \ & 1 \ & -1 \ & 0 \ & 0  \ & -1 \ & 1  \ & -1 \ & 1 \  & 1 \ & -1 \ \\[.5ex] 
		\end{array}
	\end{eqnarray*}
\end{minipage}
\medskip
\smallskip

\begin{figure*}[h]
	\begin{center}
		\centerline{\resizebox{0.62\textwidth}{!}{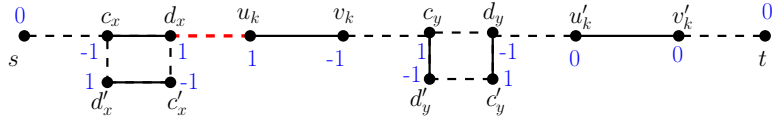}}
	\end{center}
	\caption{\small The $\alpha$-values we assigned in case~(i) for the gadget corresponding to $D_k = \neg x \vee \neg y$ are indicated next to vertices. The solid edges are in $M$ and $(d_x,u_k)$ is a blocking edge to $M$.}
	\label{newfig30:example}
\end{figure*}

It is easy to check that all intra-gadget edges (those with both endpoints in $D_k$) are {\em covered} by the above assignment of $\alpha$-values, i.e., $\alpha_p + \alpha_q \ge g_M(p,q)$ for every edge $(p,q)$. See Fig.~\ref{newfig30:example} for an illustration of $\alpha$-values in case~(i).

\begin{itemize}
	\item The edges $(u_k,v_k)$ and $(u'_k,v'_k)$ are covered. Similarly, the edges in the 4-cycles of gadgets of $\neg x$ and $\neg y$
	are covered: for any such edge $e$, we have $g_M(e) \in \{-1,0\}$ and sum of $\alpha$-values of endpoints of $e$ is 0.
	\item When  
	$x = \mathsf{true}$, the edge $(c_x,d_x) \in M$ and so $g_M(d_x,u_k) = 2$ and we have $\alpha_{d_x} = \alpha_{u_k} = 1$.
	Then  
	$y = \mathsf{false}$, the edges $(c_y,d'_y),(c'_y,d_y)$ are in $M$ and so $g_M(c_y,v_k) = g_M(d_y,u'_k) = -1$.
	It is easy to see that these edges are covered. 
	\item When  
	$x = \mathsf{false}$, the edge $(c'_x,d_x) \in M$ and so $g_M(d_x,u_k) = -1 = \alpha_{d_x} + \alpha_{u_k}$.
	Then  
	$y = \mathsf{true}$ and the edge $(c_y,d_y) \in M$ and so $g_M(c_y,v_k) = -1 = \alpha_{v_k} +  \alpha_{c_y}$ while 
	$g_M(d_y,u'_k) = 2 =  \alpha_{d_y} +  \alpha_{u'_k}$.
	\item The edges incident to $s$ and $t$ are covered in all cases. 
\end{itemize}

Consider any positive clause, say $C_{\ell} = x \vee y \vee z$ with 3 literals. 
We have three cases here:
(i)~$x = \mathsf{true}$, (ii) $x = \mathsf{false}$ and $y = \mathsf{true}$, (iii)~$x = y = \mathsf{false}$ and $z = \mathsf{true}$.
For a positive clause with 2 literals, only cases~(i) and (ii) occur.

Let $a_x,b_x,a'_x,b'_x$ be the 4 vertices in $x$'s gadget in $C_{\ell}$ and let $a_y,b_y,a'_y,b'_y$ be the 4 vertices in $y$'s gadget in $C_{\ell}$.
The $\alpha$-values of these 8 vertices along with those of $u_{\ell}, v_{\ell}, u'_{\ell}, v'_{\ell}$, and $u''_{\ell}, v''_{\ell}$ in cases~(i)-(iii) are given below.

\begin{minipage}[b]{0.3\linewidth}
	\centering
	\begin{eqnarray*}
		\setlength{\arraycolsep}{0.5ex}\setlength{\extrarowheight}{0.25ex}
		\begin{array}{@{\hspace{1ex}}c@{\hspace{1ex}}||@{\hspace{1ex}}c@{\hspace{1ex}}|@{\hspace{1ex}}c@{\hspace{1ex}}|@{\hspace{1ex}}c@{\hspace{1ex}}|@{\hspace{1ex}}c@{\hspace{1ex}}|@{\hspace{1ex}}c@{\hspace{1ex}}|@{\hspace{1ex}}c@{\hspace{1ex}}|@{\hspace{1ex}}c@{\hspace{1ex}}|@{\hspace{1ex}}c@{\hspace{1ex}}|@{\hspace{1ex}}c@{\hspace{1ex}}|@{\hspace{1ex}}c@{\hspace{1ex}}|@{\hspace{1ex}}c@{\hspace{1ex}}|@{\hspace{1ex}}c@{\hspace{1ex}}|@{\hspace{1ex}}c@{\hspace{1ex}}|@{\hspace{1ex}}c@{\hspace{1ex}}}
			\  & \alpha_{u_{\ell}} \ & \alpha_{v_{\ell}} \ &  \alpha_{a_x}\ & \alpha_{b_x} \ & \alpha_{a'_x} \ & \alpha_{b'_x} \ & \alpha_{u'_{\ell}} \ & \alpha_{v'_{\ell}} \ &  \alpha_{a_y}\ & \alpha_{b_y} \ & \alpha_{a'_y} \ & \alpha_{b'_y} \ & \alpha_{u''_{\ell}} \ & \alpha_{v''_{\ell}} \ \\[.5ex] \hline
			case~(i) \ & 0 \ & 0 \ & -1 \ & 1 \  & -1 \ & 1 \ & -2 \ & 2 \ & 0 \ & 0 \  & 0 \ & 0 \ & -1 \ & 1 \ \\[.5ex] \hline
			case~(ii) \ & -1 \ & 1 \ & 1 \ & -1 \ & 1 \ & -1 \ & 0 \ & 0 \ & 0 \ & 0 \ & 0 \ & 0 \  & -1 \ & 1 \ \\[.5ex] \hline
			case~(iii) \ & -1 \ & 1 \ & 1 \ & -1 \ & 1 \ & -1 \ & 0 \ & 0 \ & 2 \ & -2 \ & 2 \ & -2 \  & 1 \ & -1 \ \\[.5ex]
		\end{array}
	\end{eqnarray*}
\end{minipage}

\medskip

\smallskip

\begin{figure*}[h]
	\centerline{\resizebox{0.86\textwidth}{!}{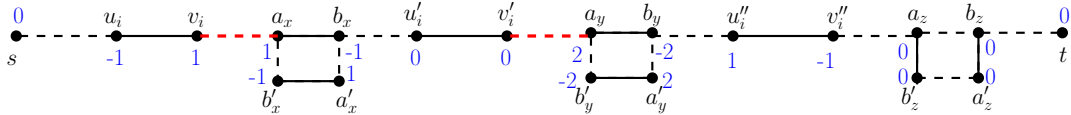}}
	\caption{\small The $\alpha$-values we assigned in case~(iii) for the gadget corresponding to $C_{\ell} = x \vee y \vee z$ are indicated next to vertices. The solid edges are in $M$ and the edges $(a_x,v_{\ell})$ and $(a_y,v'_{\ell})$ are blocking edges to $M$.}
	\label{newfig20:example}
\end{figure*}

Let $a_z,b_z,a'_z,b'_z$ be the 4 vertices in $z$'s gadget in $C_{\ell}$:
$\alpha$-values of these 4 vertices in cases~(i)-(iii) are described below.
\begin{itemize}
	\item In case~(i) and case~(ii), $z$ could be either $\mathsf{true}$ or $\mathsf{false}$: if 
	$z = \mathsf{true}$ then set $\alpha_{a_z} = \alpha_{b_z} = \alpha_{a'_z} = \alpha_{b'_z} = 0$, 
	else set $\alpha_{a_z} = \alpha_{a'_z} = 1$ and $\alpha_{b_z} = \alpha_{b'_z} = -1$.
	\item In case~(iii), $z = \mathsf{true}$: 
	set $\alpha_{a_z} = \alpha_{b_z} = \alpha_{a'_z} = \alpha_{b'_z} = 0$. 
\end{itemize}
See Fig.~\ref{newfig20:example} for the assignment of $\alpha$-values in case~(iii).

In the 3 cases above,
it is easy to check that all edges in the 4-cycles corresponding to literal gadgets are covered by the above assignment of 
$\alpha$-values. Similarly, the $u$-$v$, $u'$-$v'$, $u''$-$v''$ edges are covered and
the edges incident to $s$ and $t$ are also covered. 

Claim~\ref{clm3} below shows that every {\em link} edge (such as $(v_{\ell},a_x)$ and $(b_x,u'_{\ell})$) is also covered. 

\begin{new-claim}
	\label{clm3}
	Every link edge is covered in the 3 cases given above.
\end{new-claim}
\proof{} Consider case~(i): here the edges $(a_x,b'_x)$ and $(a'_x,b_x)$ are in $M$. So $g_M(a_x,v_{\ell}) = -1 =  \alpha_{v_{\ell}} +  \alpha_{a_x}$.
We have $g_M(b_x,u'_{\ell}) = -1 = \alpha_{b_x} + \alpha_{u'_{\ell}}$. Regarding the other two literal gadgets:

\smallskip

\begin{itemize}
	\item We have $g_M(a_y,v'_{\ell}) \le 2 = \alpha_{a_y} + \alpha_{v'_{\ell}}$. Similarly, $g_M(b_y,u''_{\ell}) = -1 = \alpha_{b_y} + \alpha_{u''_{\ell}}$.
	
	\item If $z = \mathsf{false}$, then $g_M(a_z,v''_{\ell}) = 2 = \alpha_{a_z} + \alpha_{v''_{\ell}}$; else  $g_M(a_z,v''_{\ell}) = -1 \le \alpha_{a_z} + \alpha_{v''_{\ell}}$.
\end{itemize}

\smallskip

Consider case~(ii): here the edge $(a_x,b_x) \in M$. So the edge $(a_x,v_{\ell})$ is a blocking edge to $M$ and we have 
$\alpha_{a_x} = \alpha_{v_{\ell}} = 1$. 

\begin{itemize}
	\item We have the edges $(a_y,b'_y)$ and $(a'_y,b_y)$ in $M$, thus $g_M(b_x,u'_{\ell}) = g_M(a_y,v'_{\ell}) = g_M(b_y,u''_{\ell}) = -1$.
	It is easy to check that all these edges are covered.
	\item If $z = \mathsf{false}$, then $g_M(a_z,v''_{\ell}) = 2 = \alpha_{a_z} + \alpha_{v''_{\ell}}$; else  $g_M(a_z,v''_{\ell}) = -1 \le \alpha_{a_z} + \alpha_{v''_{\ell}}$.
\end{itemize}

Consider case~(iii): here the edges $(a_x,b_x)$ and $(a_y,b_y)$ are in $M$. So both $(a_x,v_{\ell})$ and $(a_y,v'_{\ell})$ are blocking edges to $M$.
We have $\alpha_{a_x} = \alpha_{v_{\ell}} = 1$ and  $\alpha_{a_y} = 2$ and $\alpha_{v'_{\ell}} = 0$. So the blocking edges are covered.

\begin{itemize}
	\item  We also have $g_M(b_x,u'_{\ell}) = g_M(b_y,u''_{\ell}) = -1$ and we set $\alpha_{b_x} = -1, \alpha_{u'_{\ell}} = 0$ along with 
	$\alpha_{b_y} = -2$, $\alpha_{u''_{\ell}} = 1$. Thus these edges are covered.
	\item Finally, $g_M(a_z,v''_{\ell}) = -1$ and we set $\alpha_{a_z} = 0$ and  $\alpha_{v''_{\ell}} = -1$. Thus all link edges are covered. \qed  
\end{itemize}
\endproof

\medskip

Claim~\ref{new-clm2} below shows that 
edges between $r$'s gadget and $\neg r$'s gadget, for any variable~$r$,
are also covered. 
\begin{new-claim}
	\label{new-clm2}
	For any variable $r$, edges between $r$'s gadget and $\neg r$'s gadget are covered.
\end{new-claim}
\proof{} 
Suppose $r = \mathsf{false}$ in this assignment. So the edges $(a_r,b_r), (a'_r,b'_r)$, and $(c_r,d'_r)$ are in $M$. 
Note that $\alpha_{c_r} = 1$ and  $\alpha_{d'_r} = -1$;
moreover $\alpha_{b_r} \ge -2$. We have $g_M(b_r,c_r) = -1$, thus $\alpha_{b_r} + \alpha_{c_r} \ge -1 = g_M(b_r,c_r)$.
It is easy to check that, by construction of $\alpha$-values, when $r = \mathsf{false}$, we have $\alpha_{a'_r} \ge 0$. We have $g_M(a'_r,d'_r) = -1$,
so $\alpha_{a'_r} + \alpha_{d'_r} \ge -1 = g_M(a'_r,d'_r)$.

Suppose $r = \mathsf{true}$ in this assignment. So the edges $(a'_r,b_r), (c_r,d_r)$, and  $(c'_r,d'_r)$ are in $M$. 
Note that $\alpha_{c_r} = -1$ and  $\alpha_{d'_r} = 1$;
moreover $\alpha_{a'_r} \ge -2$ (in fact, $\alpha_{a'_r} \ge -1$). Thus $\alpha_{a'_r} + \alpha_{d'_r} \ge -1 = g_M(a'_r,d'_r)$.
It is easy to check that  when $r = \mathsf{true}$, we have $\alpha_{b_r} \ge 0$, thus 
$\alpha_{b_r} + \alpha_{c_r} \ge -1 = g_M(b_r,c_r)$.   
\qed\endproof

Thus $\vec{\alpha}$ is a feasible solution to \eqref{LP4}. Moreover,
$\sum_w \alpha_w = 0$ since for each edge $(p,q) \in M$, we have $\alpha_p + \alpha_q = 0$, also $\alpha_s = \alpha_t = 0$.
This finishes the proof of Theorem~\ref{thm:correctness}.   
\qed\endproof

\medskip

\noindent{\bf The Converse.}
We also need to show that if $G_{\psi}$ admits a quasi-popular matching of cost $0$, then $\psi$ admits a satisfying assignment.
This is the easy side of the reduction. 
\begin{lemma}
	\label{lem:consistency}
	Let $M$ be any quasi-popular matching in $G_{\psi}$ with $\cost(M) =  0$. Then $M$ contains all $u$-$v$, $u'$-$v'$, and $u''$-$v''$ edges. For any $r \in \{X_1,\ldots,X_{2n}\}$, we have:
	\begin{enumerate}
		\item From a gadget of $r$ (on vertices $a_r,b_r,a'_r,b'_r$), either (i)~$(a_r,b'_r),(a'_r,b_r)$ are in $M$ or (ii)~$(a_r,b_r)$, $(a'_r,b'_r)$ are in $M$. 
		
		-- If (i) happens, we say that $r$ is in \emph{true state}. If (ii) happens, we say that $r$ is in \emph{false state}.
		
		\item From a gadget of $\neg r$ (on vertices  $c_r,d_r,c'_r,d'_r$), either (i)~$(c_r,d'_r),(c'_r,d_r)$ are in $M$ or (ii)~$(c_r,d_r)$, $(c'_r,d'_r)$ are in $M$. 
		
		-- If (i) happens, we say that $\neg r$ is in \emph{true state}. If (ii) happens, we say that $\neg r$ is in \emph{false state}.
		\item $r$ and $\neg r$ cannot be simultaneously in true state.
	\end{enumerate}
\end{lemma}
\proof{} 
Since $\cost(M) = 0$, $M$ is forbidden to use any edge other than the $u$-$v$ edges, the $u'$-$v'$ edges, the $u''$-$v''$ 
edges, and the 4 edges in the gadget of any literal. Moreover, since $M$ is quasi-popular, $M$ cannot leave two adjacent vertices unmatched.
Thus $M$ contains all $u$-$v$ edges, $u'$-$v'$ edges, and $u''$-$v''$ edges; similarly, points 1 and 2 follow.

The preferences of the vertices are set such that if both $(a'_r,b_r)$ and $(c_r,d'_r)$ are in $M$ then the alternating cycle 
$\rho = a'_r - b_r - c_r - d'_r - a'_r$ 
has a blocking edge $(b_r,c_r)$ in it.

Consider $M \oplus \rho$ versus $M$. The 3 vertices $b_r,c_r$, and $d'_r$ prefer $M\oplus\rho$ to $M$ while the vertex $a'_r$
prefers $M$ to $M\oplus\rho$. This means $M$'s unpopularity factor is at least 3, i.e., $u(M) \ge 3$. However $M$ is quasi-popular, i.e.,
$u(M) \le 2$, thus both $(a'_r,b_r)$ and $(c_r,d'_r)$ 
cannot simultaneously be in $M$ for any $r \in \{X_1,\ldots,X_{2n}\}$.  
\qed\endproof

Lemma~\ref{lem:consistency} indicates a natural way of defining an assignment for $\psi$ using a matching $M$ with cost 0 and unpopularity factor at most 2.

\begin{lemma}
	\label{lem:correctness}
	If $G_{\psi}$ has a quasi-popular matching $M$ with $\cost(M) = 0$ then $\psi$ is satisfiable. 
\end{lemma}
\proof{} 
For any variable $r \in \{X_1,\ldots,X_{2n}\}$ consider the edges in $\neg r$'s gadget that are in $M$. If $(c_r,d'_r)$ and $(c'_r,d_r)$ 
are in $M$ then set $r = \mathsf{false}$ else set $r = \mathsf{true}$. 

Lemma~\ref{lem:consistency} tells us that when $r = \mathsf{false}$ then for every clause $i$ that $r$ is present in, the edges 
$(a_{r,i},b_{r,i})$ and $(a'_{r,i},b'_{r,i})$ from $r$'s gadget in the $i$-th clause are in $M$ (where $a_{r,i},b_{r,i},a'_{r,i},b'_{r,i}$ are the 4 
vertices from $r$'s gadget in the $i$-th clause). We now need to show that every clause 
has at least one literal  set to $\mathsf{true}$. Suppose not. We have 3 cases here.

\begin{enumerate}
	
	\item Let $C_i = x \vee y \vee z$. Suppose $x, y, z$ are in false state. Consider the alternating path $\rho$ wrt $M$: 
	\[s - (u_i,v_i) - (a_{x,i},b_{x,i}) - (u'_i,v'_i) - (a_{y,i},b_{y,i}) - (u''_i,v''_i) - (a_{z,i},b_{z,i}) - t.\] 
	In the election between $M \oplus \rho$ and $M$, the 10 vertices $s, v_i, a_{x,i}, b_{x,i}, v'_i, a_{y,i}, b_{y,i}, v''_i,a_{z,i}$, and $t$ 
	vote for $M \oplus \rho$ while the 4 vertices $u_i,u'_i,u''_i$, and $b_{z,i}$ vote for $M$. Thus $\phi(M\oplus\rho, M) = 10$ and 
	$\phi(M, M\oplus\rho) = 4$. Hence $u(M) \ge 10/4$ contradicting that $u(M) \le 2$.
	
	\item Let $C_j = x \vee y$, i.e., this is a positive clause with 2 literals. Suppose both $x$ and $y$ are in false state.
	Consider the following alternating path $\rho$ wrt $M$:
	\[s - (u_j,v_j) - (a_{x,j},b_{x,j}) - (u'_j,v'_j) - (a_{y,j},b_{y,j}) - t.\] 
	In the election between $M \oplus \rho$ and $M$, the 7 vertices $s, v_j, a_{x,j}, b_{x,j}, v'_j, a_{y,j}$, and $t$ vote for $M \oplus \rho$ 
	while the 3 vertices $u_j,u'_j$, and $b_{y,j}$ vote for $M$. Thus $\phi(M\oplus\rho, M) = 7$ and $\phi(M, M\oplus\rho) = 3$. Hence 
	$u(M) \ge 7/3$ contradicting that $u(M) \le 2$.
	
	\item Let $D_k = \neg x \vee \neg y$. Suppose both $\neg x$ and $\neg y$ are in false state.
	Consider the following alternating path $\rho$ wrt $M$: 
	\[s - (c_x,d_x) - (u_k,v_k) -(c_y,d_y) - (u'_k,v'_k) - t.\] 
	In the election between $M \oplus \rho$ and $M$, the 7 vertices $s, d_x, u_k, c_y,d_y,u'_k$, and $t$ vote for $M \oplus \rho$ 
	while the 3 vertices $c_x,v_k$, and $v'_k$ vote for $M$. Thus $\phi(M\oplus\rho, M) = 7$ and $\phi(M, M\oplus\rho) = 3$. Hence
	$u(M) \ge 7/3$ contradicting that $u(M) \le 2$.  \qed
\end{enumerate} \endproof

This finishes the proof of correctness of our reduction. Hence Theorem~\ref{thm:hardness} follows. 

\section{Lower Bounds on the Extension Complexity}\label{sec:lb-xc} We prove Theorem~\ref{thm:xc} in this section.
Recall that, given polytopes ${\cal P}_1\subseteq \mathbb{R}^k$, ${\cal P}_2 \subseteq \mathbb{R}^{\ell}$ and an affine map $f:\mathbb{R}^{\ell} \rightarrow \mathbb{R}^k$ such that $f({\cal P}_2) = {\cal P}_1$, we say that ${\cal P}_2$ is an \emph{extension} of ${\cal P}_1$. The number of facets of ${\cal P}_2$ (or equivalently, the number of inequalities in a minimal linear description of ${\cal P}_2$ that are not valid at equality) is its \emph{size}, and the minimum size of an extension of ${\cal P}_1$ is the \emph{extension complexity} of ${\cal P}_1$, denoted by $xc({\cal P}_1)$. 

Our starting point in the proof of Theorem~\ref{thm:xc} is the following lower bound on the extension complexity of the independent set polytope. 
Let ${\cal I}_H$ be the independent set polytope of a graph $H$.

\begin{theorem}[\cite{Goos-et-al}]
	\label{thr:goos}
	There exists an (explicitly constructed) family of graphs $H = (V_H,E_H)$ with bounded degree such that $xc({\cal I}_H)=2^{\Omega(n_H/\log n_H)}$, where $|V_H| = n_H$. 
\end{theorem}

\subsection{A Lower Bound on the Extension Complexity of ${\cal P}$}\label{sec:ec-P} Here we lower bound the extension complexity of the popular matching polytope.
Given a graph $H = (V_H,E_H)$ which is an instance of the independent set problem, we will build an instance
$G = (A\cup B,E)$ with strict preference lists as follows. Let $V_H = \{1,\ldots,n_H\}$. 

\smallskip

\begin{itemize}
	\item For every edge $e \in E_H$, there will be a gadget $Y_e$ in $G$ on 6 vertices $s_e,t_e,s'_e,t'_e,s''_e,t''_e$.
	\item For every vertex $i \in V_H$, there will be a gadget $Z_i$ in $G$ on 4 vertices $a_i,b_i,a'_i,b'_i$.
	\item There are 2 more vertices in $G$: these are $a_0$ and $b_0$.
\end{itemize}

\smallskip

\begin{figure}[h]
	\centerline{\resizebox{0.6\textwidth}{!}{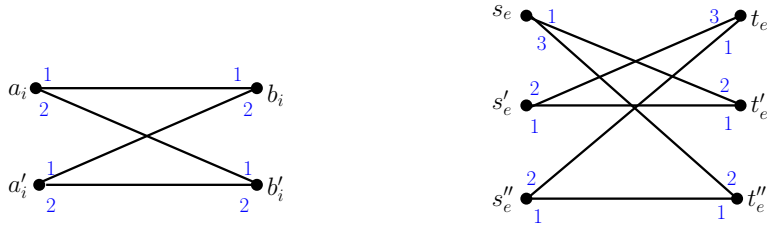}}
	\caption{To the left is the gadget $Z_i$ on vertices $a_i,b_i,a'_i,b'_i$ and to the right is the gadget $Y_e$ on vertices $s_e,t_e,s'_e,t'_e,s''_e,t''_e$. The numbers on edges denote vertex preferences: 1 is top choice and so on.} 
	\label{fig:hardness1}
\end{figure}

We let $A = \{a_0\} \cup \{a_i,a'_i: i \in V_H\} \cup \{s_e,s'_e,s''_e: e \in E_H\}$ and
$B = \{b_0\} \cup \{b_i,b'_i: i \in V_H\}\cup$ $\{t_e,t'_e,t''_e: e \in E_H\}$. 
We now describe the edge set of $G$.
We first describe the preference lists of the 6~vertices in $Y_e$, where $e = (i,j)$ and $i < j$ (see Fig.~\ref{fig:hardness1}, right).

Recall that $e = (i,j)$ and $i < j$. The preference lists of $s_e$ and $t_e$ are given below:
\[s_e\colon \, t'_e \succ \underline{b_j} \succ t''_e \ \ \ \ \ \ \ \text{and}\ \ \ \ \ \ \  t_e\colon  \, s''_e \succ \underline{a_i} \succ s'_e.\]

The underlined vertices $a_i$ and $b_j$ belong to the gadgets $Z_i$ and $Z_j$, respectively.
The preference lists of $s'_e, s''_e$ and $t'_e, t''_e$ are given below:
\[s'_e\colon \, t'_e \succ t_e \ \ \ \ \ \ \text{and} \ \ \ \ \ \ s''_e \colon \, t''_e \succ t_e \ \ \ \ \ \ \text{and}\ \ \ \ \ \ t'_e\colon  \, s'_e \succ s_e\ \ \ \ \ \ \text{and}\ \ \ \ \ \ t''_e\colon  \, s''_e\succ s_e. \]

We now describe the preference lists of the 4 vertices $a_i,b_i,a'_i,b'_i$ in $Z_i$ (see Fig.~\ref{fig:hardness1}, left).

\begin{eqnarray*}
	a'_i\colon \, b_i \succ b'_i \hspace*{1in}  a_i\colon \, b_i \succ b'_i \succ b_0 \succ \cdots \\
	b'_i\colon \, a_i \succ a'_i \hspace*{1in}  b_i\colon \, a_i \succ a'_i \succ a_0 \succ \cdots
\end{eqnarray*}  

The vertex $a_i$ has $b_0$ as its third choice followed by all the vertices 
$t_{e_1},t_{e_2},\ldots$  where $i$ is the {\em lower-indexed}
endpoint of $e_1,e_2,\ldots$. Similarly, $b_i$ has  $a_0$ as its third choice followed by all the vertices $s_{e'_1},s_{e'_2},\ldots$ where $i$ is the {\em higher-indexed} endpoint of $e'_1,e'_2,\ldots$.

The order among the vertices $t_{e_1},t_{e_2},\ldots$ (similarly, 
$s_{e'_1},s_{e'_2},\ldots$) in the preference list of $a_i$ (resp., $b_i$) does not matter;
hence these are represented as ``$\cdots$'' above. 

The vertex $a_0$ (resp.~$b_0$) has $b_1,\ldots, b_{n_H}$ (resp.~$a_1,\ldots, a_{n_H}$) as its neighbors and its preference list is some arbitrary 
permutation of them . 
Thus the number of vertices in $G$ is $n = 6m_H + 4n_H + 2$, where $m_H = |E_H|$.

\paragraph{\bf Unpopular Vertices and Edges.}
It is easy to check that a max-size popular matching in $G$ leaves $a_0$ and $b_0$ unmatched and matches the remaining $n-2$ vertices. For instance, one could run the 
max-size popular matching algorithm from~\cite{Kav14}: it first computes the stable matching $\cup_i\{(a_i,b_i),(a'_i,b'_i)\} \cup_e\{(s_e',t_e'),(s_e'',t_e'')\}$, and finally outputs the matching $\cup_i\{(a_i,b'_i),(a'_i,b_i)\} \cup_e\{(s_e,t'_e),(s'_e,t_e),(s''_e,t''_e)\}$.
Thus $a_0$ and $b_0$ are {\em unpopular} vertices, i.e., $a_0$ and $b_0$ are left unmatched in all popular matchings, while $\{s_e,t_e\}_{e \in E_H}$ are {\em unstable} vertices, i.e., they are left unmatched in every stable matching. Moreover, for each $i$: all the four vertices of $Z_i$ are contained in the same connected component of the popular subgraph $G_0$.

Consider any edge $e = (i,j)$.
Since $a_i$ prefers $b_0$ to $t_e$ and similarly, $b_j$ prefers $a_0$ to $s_e$, the unpopularity of vertices $a_0,b_0$ implies that no popular matching contains 
$(a_i,t_e)$ or $(s_e,b_j)$. Thus $(a_i,t_e)$ and $(s_e,b_j)$ are unpopular edges.
So for any $i \in [n_H]$: every popular matching $M$ in $G$ includes either the pair of edges $(a_i,b_i),(a'_i,b'_i)$ or the pair of edges $(a_i,b'_i),(a'_i,b_i)$ and similarly, for every $e \in E_H$, every maximum size popular matching contains either the pair of edges $(s_e,t'_e), (s'_e,t_e)$ or the pair of edges $(s_e,t''_e), (s''_e,t_e)$.

Let $M$ be any max-size popular matching in $G$. It matches all vertices in $A \cup B$ other than $a_0,b_0$.
Let $\vec{\alpha} \in \{0, \pm 1\}^n$ be a witness of $M$. Lemma~\ref{lemma1} (see Section~\ref{sec:algo}) implies the following two useful properties:

\smallskip

\begin{itemize}
	\item[(i)] For every $e \in E_H$: $\alpha_{s_e} = \alpha_{t_e} = -1$ (because $s_e$ and $t_e$ are unstable vertices).
	\item[(ii)] For every $i \in [n_H]$: either $\alpha_u = 0$ for all $u \in Z_i$ or $\alpha_u \in \{\pm 1\}$ for all $u \in Z_i$.
\end{itemize}

\smallskip

Let $S_M$ be the set of indices $i \subseteq [n_H]$ such that $\alpha_u = 0$ for all $u \in Z_i$.

\begin{lemma}
	\label{lem:redn1}
	The set $S_M$ is an independent set in the graph $H$.
\end{lemma}
\proof{} 
We will show that the set $U_M = [n_H] \setminus S_M$ is a vertex cover in $H$.
Note that $U_M$ is the set of indices $i$ such that $\alpha_u \in \{\pm 1\}$ for all $u \in Z_i$.
Consider any edge $e = (i,j) \in E_H$, let $i < j$. Since $M$ is a max-size popular matching in $G$, we know from what was discussed above that either the pair $(s_e,t'_e), (s'_e,t_e)$ or  the pair $(s_e,t''_e), (s''_e,t_e)$ is in $M$. 

\smallskip

\begin{itemize}
	\item If  $(s'_e,t_e) \in M$, then $\wt_M(a_i,t_e) = 0$ for $t_e$ prefers $a_i$ to $s'_e$ and $a_i$ prefers both $b_i$
	and $b'_i$ (its possible partners in $M$) to $t_e$. 
	Since $\alpha_{t_e} = -1$ (see property~(i) above), $\alpha_{a_i}=1$ because $\alpha_{a_i} + \alpha_{t_e} \ge 0$.
	
	\smallskip      
	
	\item If  $(s_e,t''_e) \in M$, then $\wt_M(s_e,b_j) = 0$ since $s_e$ prefers $b_j$ to $t''_e$ while $b_j$ prefers both $a_j$
	and $a'_j$ (its possible partners in $M$) to $s_e$. Since $\alpha_{s_e} = -1$ (see property~(i) above), $\alpha_{b_j}$ has to be 1 so that 
	$\alpha_{s_e} + \alpha_{b_j} \ge 0$.
\end{itemize}
Thus, for every edge $(i,j) \in E_H$, at least one of $i,j$ is in $U_M$, that is, $U_M$ is a vertex cover of $H$.  
\qed\endproof

\begin{lemma}
	\label{lem:redn2}
	Let $S \subseteq [n_H]$ be an independent set in $H$.
	Then there is a max-size popular matching $M_S$ in $G$ such that $M_S \supseteq \cup_{i\in S}\{(a_i,b_i)\}$.
\end{lemma}
\proof{} 
We will construct a matching $M_S$ that matches all vertices in $G$ other than $a_0,b_0$ along with a witness  $\vec{\alpha}$ of $M_S$'s popularity as follows. For every $i \in [n_H]$:

\smallskip

\begin{itemize}
	\item if $i \in S$ then include edges $(a_i,b_i)$ and $(a'_i,b'_i)$ in $M_S$ and set $\alpha_u = 0$ for all $u \in Z_i$;
	\item else include edges $(a_i,b'_i)$ and $(a'_i,b_i)$ in $M_S$ and set $\alpha_{a_i} = \alpha_{b_i} = 1$ and set $\alpha_{a'_i} = \alpha_{b'_i} = -1$.
\end{itemize}

\smallskip

For every $e = (i,j) \in E_H$, where $i < j$, we do as follows:
\begin{itemize}
	\item if $i \notin S$ then include the edges $(s_e,t'_e),(s'_e,t_e)$, and $(s''_e,t''_e)$ in $M_S$; 
	set $\alpha_{s_e} = \alpha_{t_e} = \alpha_{t''_e} = -1$
	and $\alpha_{s'_e} = \alpha_{t'_e} = \alpha_{s''_e} = 1$.
	\item else (so $j \notin S$ since $i \in S$) include the edges $(s_e,t''_e),(s'_e,t'_e)$, and $(s''_e,t_e)$ in $M_S$; 
	set $\alpha_{s_e} = \alpha_{t_e} = \alpha_{s'_e} = -1$
	and $\alpha_{t'_e} = \alpha_{s''_e} = \alpha_{t''_e} = 1$.
\end{itemize}

Set $\alpha_{a_0} = \alpha_{b_0} = 0$.
It is easy to check that $\alpha_u + \alpha_v = 0$ for every $(u,v) \in M_S$. Hence
$\sum_{u \in A\cup B}\alpha_u = 0$.  
We prove in Claim~\ref{clm:xc} below that $\alpha_u + \alpha_v \ge \wt_{M_S}(u,v)$
for every edge $(u,v)$ in $G$. Also, $\alpha_u = 0 = \wt_{M_S}(u,u)$ for $u \in \{a_0,b_0\}$ and
$\alpha_u \ge -1 = \wt_{M_S}(u,u)$ for all other vertices $u$.  Thus  $\vec{\alpha}$ is a witness of $M_S$'s popularity.  
\qed\endproof

\begin{new-claim}
	\label{clm:xc}
	$\alpha_u + \alpha_v \ge \wt_{M_S}(u,v)$ for every edge $(u,v)$ in $G$. 
\end{new-claim}
\proof{} 
It is easy to check that we have $\alpha_u + \alpha_v = \wt_{M_S}(u,v)$ when $u$ and $v$ belong to the same gadget, i.e., when $u,v \in Z_i$ for some $i \in [n_H]$ or 
$u,v \in Y_e$ for some $e \in E_H$. What we need to check now is that for any $e = (i,j)$ and $i < j$, the edges $(a_i,t_e)$ and $(s_e,b_j)$ are {\em covered}.

Suppose $i \in S$. Then $(s''_e,t_e) \in M_S$. So $\alpha_{a_i} + \alpha_{t_e} = 0-1 = -1 > -2 = \wt_{M_S}(a_i,t_e)$, since $t_e$ is matched to its top choice $s''_e$ 
and so $\wt_{M_S}(a_i,t_e) = -2$. Since $i \in S$, it follows that $j \notin S$ and so we have $\alpha_{a_j} = \alpha_{b_j} = 1$. Thus 
$\alpha_{s_e} + \alpha_{b_j} = -1+1 = 0 = \wt_{M_S}(s_e,b_j)$ since $s_e$ is matched to $t''_e$ and $s_e$ prefers $b_j$ to $t''_e$. 

If $i \notin S$ then $\alpha_{a_i} = 1$. So we have $\alpha_{a_i} + \alpha_{t_e} = 1-1 = 0 = \wt_{M_S}(a_i,t_e)$ since $(s'_e,t_e) \in M_S$ and 
$t_e$ prefers $a_i$ to $s'_e$. Also, $s_e$ is matched to its top choice $t'_e$ and so $\wt_{M_S}(s_e,b_j) = -2 < \alpha_{s_e} + \alpha_{b_j}$ since $\alpha_{b_j} \ge 0$.
Thus the edges $(a_i,t_e)$ and $(s_e,b_j)$ are covered in all cases.  
\qed\endproof

\medskip

\noindent{\bf Extension Complexity.} 
The max-size popular matching polytope of $G$ is a face $F'$ of the popular matching polytope ${\cal P}$.
Our goal is to define a linear surjective map $f$ from $F'$ to ${\cal I}_H$. 
Hence, given an extension ${\cal T}'$ for ${\cal P}$, we obtain that 
\[{\cal I}_H = \{\vec{x} : \ \vec{x}=f(\vec{z}), \ (\vec{z}, \vec{y}) \in {\cal T}',\ |\vec{z}|=\frac{n}{2}-1\}.\]
Therefore, an extension of small size for ${\cal P}$ would imply the existence of an extension of small size for ${\cal I}_H$. 
The lower bound on the extension complexity of ${\cal I}_H$ will therefore imply a lower bound on the extension complexity of ${\cal P}$.

\smallskip

We will need the following properties: 
\begin{enumerate}
	\item\label{uno-bis} Let $M$ be a max-size popular matching. Consider the following 0-1 assignment $\vec{x}$ to the vertices of $H$. For $i \in [n_H]$:
	\begin{equation*} 
		x_i = \begin{cases} 1   & \text{if\ $(a_i,b_i) \in M$};\\
			0    & \text{otherwise.}			
		\end{cases}
	\end{equation*}
	
	It follows from Lemma~\ref{lem:redn1} that $\vec{x}$ is the incidence vector of an independent set in $H$.
	
	\item\label{due-bis} Let $\vec{x}$ be the incidence vector of an independent set in $H$. It follows from the proof of Lemma~\ref{lem:redn2} that there is a max-size popular matching $M$ with the following properties: 
	\[\text{for}\ i \in [n_H]:  (a_i,b_i) \in M  \iff x_i = 1.\]  
\end{enumerate}

Define the mapping $f$ from $F'$ to ${\cal I}_H$ as: for $\vec{z} \in F'$, let
$f(\vec{z}) = \vec{x}$ where $x_i = z_{(a_i,b_i)}$ for $i \in [n_H]$. Part~\ref{uno-bis} above implies that, if $\vec{z}$ is a vertex of $F'$, then $f(\vec{z})$ is the incidence vector of an independent set in $H$. Together with part~\ref{due-bis}, we deduce $f(F') = {\cal I}_H$. 
Thus we can conclude that 
$xc({\cal P}) \ge 2^{\Omega(n_H/\log n_H)} = 2^{\Omega(m/\log m)}$ where $m$ is the number of edges in $G$ (note that $m = \Theta(n_H + m_H) = \Theta(n_H)$ since graphs $H$ have bounded degree). This finishes the proof of the first part of Theorem~\ref{thm:xc}.

\subsection{A Lower Bound on the Extension Complexity of ${\cal Q}$} \label{sec:lb-xc-quasi} We now prove the second part of Theorem~\ref{thm:xc}.
Given a SAT formula $\psi$ on $n$ variables, let $C(\psi)$ be the convex hull of all vectors $\vec{x} \in\{0,1\}^n$ such that $\psi(\vec{x})=1$ where, as usual, for $i \in [n]$, $x_i=1$ corresponds to the $i$-th variable being set to true. It is now easy to construct a family of 2SAT instances $\psi$ with $n$  variables and $O(n)$ clauses such that $xc(C(\psi))=2^{\Omega(n/\log n)}$: indeed, every independent set instance $H = (V,E)$ can be formulated as a 2SAT instance $\psi$ with $n = |V|$ variables $x_1,\dots,x_{n}$ and clauses $(\neg x_i \vee \neg x_j)$ iff $(i,j) \in E$. Since the hard instances of the independent set polytope from Theorem~\ref{thr:goos} come from graphs of bounded degree, the number of clauses in the 2SAT instance we create is linear in the number of variables.

Now apply the sequence of reductions from Section~\ref{sec:hardness}, transforming the 2SAT instance $\psi$ with $n$ variables and $\Theta(n)$ clauses into an equivalent 2SAT instance with $\Theta(n)$ variables and clauses, 
and then defining the min-cost quasi-popular matching instance $G_{\psi}$ on $\Theta(n)$ vertices and 
$\Theta(n)$ edges defined in Section~\ref{sec:hardness}. The reduction from Section~\ref{sec:hardness} assumed that $\psi$ was a 3SAT instance 
but it clearly applies to a 2SAT instance as well.

Recall the edge cost function $\cost$ defined in Section~\ref{sec:hardness}.
Consider the following face $F$ of the quasi-popular matching polytope ${\cal Q}$ of $G_{\psi}$, where $F=\{ \vec{z} \in {\cal Q}: \cost(\vec{z})=0\}$. 
Our goal is to define a linear surjective map $h$ from the face $F$ defined above to $C(\psi)$. 
Hence, given an extension ${\cal T}$ for ${\cal Q}$, we obtain that 
$$C(\psi)=\{ \vec{x} : \ \vec{x} = h(\vec{z}), \ (\vec{z}, \vec{y}) \in {\cal T}, \ \cost(\vec{z})=0\}.$$
Therefore, an extension of small size for ${\cal Q}$ would imply the existence of an extension of small size for $C(\psi)$. The lower bound on the extension complexity of $C(\psi)$ will therefore imply the claimed bound.
\smallskip

We will need the following properties: 
\begin{enumerate}
	\item\label{uno} Let $M$ be a matching that belongs to $F$. Thus $\cost(M) = 0$. Let $S$ be the following true/false assignment to the variables in $\psi$. For $i \in [n]$:
	\begin{equation*} 
		S(x_i) = \begin{cases} \mathsf{false}   & \text{if\ $(c_i,d'_i) \in M$};\\
			\mathsf{true}    & \text{otherwise.}			
		\end{cases}
	\end{equation*}

	It follows from Lemma~\ref{lem:correctness} that $S$ satisfies $\psi$. 
	
	\item\label{due} Let $S$ be a satisfying assignment for $\psi$. As shown in the proof of Theorem~\ref{thm:correctness}, there is a matching $M \in F$ with the following properties: 
	\[\text{for}\ i \in [n]:  (c_i,d'_i) \in M  \iff S(x_i) = \mathsf{false} .\]  
\end{enumerate}

Define the mapping $h$ from $F$ to $C(\psi)$ as: for $\vec{z} \in F$, let
$h(\vec{z}) = \vec{x}$ where $x_i=1-z_{(c_i,d'_i)}$ for $i \in [n]$. Part~\ref{uno} above implies that, if $\vec{z}$ is a vertex of $F$, then $h(\vec{z})$ is a satisfying assignment. Together with part~\ref{due}, we deduce $h(F)=C(\psi)$. 
Thus we can conclude that 
$xc({\cal Q}) \ge 2^{\Omega(n/\log n)}$. Since the number of edges of $G_\psi$ is $\Theta(n)$, the bound follows.

\section{The Dominant Matching Polytope}
\label{sec-new:polytopes} We now prove Theorem~\ref{thm:dom}. Section~\ref{sec:dom-mat-poly-ext} shows a compact extended formulation for the dominant matching polytope. In Section~\ref{sec:dominant-polytope}, we formulate it in the original space $\mathbb{R}^m$ and Section~\ref{sec:lb-facets} shows a lower bound on the number of its facets.

\subsection{A New Extended Formulation for the Dominant Matching Polytope} \label{sec:dom-mat-poly-ext} We now show how to modify Theorem~\ref{thm:dominant-extn} so as to obtain an extended formulation for ${\cal D}_G$ that holds for {\em any} marriage instance $G = (A \cup B, E)$. Let $U$ be the set of unpopular vertices of $G$. 

Consider the following variant ${\cal E}'_G$ of the extended formulation ${\cal E}_G$ for the popular fractional matching polytope of $G$~\cite{KMN09} recalled in Section~\ref{sec:poly-overview}:

\begin{minipage}{.5\textwidth}
	\begin{eqnarray}
		\sum_{u \in A \cup B}\alpha_u \ & = & \ 0 \label{last-constr3-bis-generic}\\
		\quad \alpha_a + \alpha_b \ & \ge & \ c_x(a,b) \ \ \ \ \ \ \ \ \ \forall (a,b) \in E\label{last-constr1-generic}\\
		\alpha_u  \ & \ge & \ - \sum_{e \in \delta(u)} x_{e}\ \ \ \ \ \forall u \in A \cup B\label{last-constr3-generic}
	\end{eqnarray}\end{minipage}\begin{minipage}{.5\textwidth}
	\begin{eqnarray}
		\sum_{e\in\tilde{\delta}(u)}x_e \ & = & \ 1 \ \ \ \ \ \ \ \ \ \ \ \ \ \ \ \ \ \ \forall u\in A\cup B \label{last-constr5-generic}\\  
		x_e \ & \ge & \ 0 \, \ \ \ \ \ \ \ \ \ \ \ \ \ \ \ \ \ \ \forall e \in \tilde{E}, \label{last-constr5-bis-generic}\\
		\nonumber \\ 
		\nonumber \end{eqnarray}
\end{minipage}

\smallskip

\noindent where $c_x(a,b) = \wt_x(a,b)$ for $a,b \notin U$, see~\eqref{eq:wt-x} for the definition of $\wt_x(a,b)$;  for $a \in U$ or $b\in U$, $c_x(a,b) = 1$. For a definition of $\tilde G = (A \cup B, \tilde{E})$, see Section~\ref{prelims}.

The only difference between ${\cal E}'_G$ and ${\cal E}_G$ is in the right side of constraints~\eqref{last-constr1-generic} 
for $a \in U$ or $b \in U$.

Let $E_D \subseteq E$ be the set of dominant edges of $G$, i.e., $e \in E_D$ if there is a dominant matching in $G$ containing edge $e$. 
Adding to~\eqref{last-constr3-bis-generic}--\eqref{last-constr5-bis-generic} the following constraints: (here $\tilde{E}_{D} = E_D \cup \{(u,u): u \in U\}$)
\begin{eqnarray}
	\ \ x_{e} & = & \ \ 0 \ \ \ \forall e \in \tilde{E} \setminus \tilde{E}_{D} \label{eq:add-4} 
\end{eqnarray}
gives a face, call it ${\cal R}_G$, of ${\cal E}'_G$. 

\begin{new-claim}
	\label{last-clm-2}
	For any dominant matching $M$ in $G$ and any dominant witness $\vec{\beta}$ of $M$, $(I_M,\vec{\beta}) \in {\cal R}_G$.
\end{new-claim}

As $M \subseteq \tilde{E}_{D}$, constraint~\eqref{eq:add-4} holds.
The proof of Claim~\ref{last-clm-2} is identical to that of Claim~\ref{last-clm}, with the addendum that \eqref{last-constr3-generic} is tight for all unstable vertices $u$ (see Lemma~\ref{lemma1}), thus $\beta_u = 0$ for $u \in U$ and this implies $\beta_v=1$ for $v \in \Nbr(U)$.
Here $\Nbr(U)$ are the neighbors in $G$ of vertices in $U$.

Now let $(\vec{x},\vec{\alpha}) \in {\cal R}_G$. We show it is a convex combination of vectors $(I_M,\vec{\beta}_M)$, where $M$ is a popular matching and $\beta_M$ is a witness of $M$, and that $\beta_v\in \{\pm 1\}$ for all $v \in (A\cup B)\setminus U$. The proof is completely analogous to the proof of Theorem~\ref{first-thm} (\cite{HK17}), hence omitted. We conclude the following.

\begin{theorem}\label{thm:ed-dominant}
	The polytope ${\cal R}_G$ defined by constraints~\eqref{last-constr3-bis-generic}--\eqref{eq:add-4} 
	is an extended formulation of the dominant matching polytope of $G$. 
\end{theorem}

\subsection{The Dominant Matching Polytope in $\mathbb{R}^m$}\label{sec:dominant-polytope} In this section we describe the convex hull of incidence vectors of dominant matchings in $G = (A \cup B, E)$.
Recall that all dominant matchings in $G$ match the same subset of vertices (see Section~\ref{prelims}).
We let again $U \subseteq A \cup B$ be the set of unpopular vertices in $G$. In particular, every vertex in $U$ is 
left unmatched in any dominant matching in $G$. 
Note that $U$ has to be an independent set in $G$.
We let again $E_D \subseteq E$ be the set of dominant edges in $G$.
Suppose $\vec{x} \in \mathbb{R}^m$ is a convex combination of dominant matchings. Then we have:

\begin{eqnarray}
	\sum_{e\in\delta(u)}x_e \ &  = & \ 1 \ \ \forall u\in (A\cup B)\setminus U, \nonumber \\ x_e  \ & = & \ 0 \ \ \forall e\in E \setminus E_D,  \label{dom-eqn1} \\  x_e \ & \ge & \ 0 \ \ \forall e \in E_D. \nonumber \end{eqnarray}

For any matching $M$ in $G$, let $k_M \ge 0$ be the number of vertices in $U$ that are matched in $M$. Consider the following family of constraints:
\begin{equation}
	\label{dom-eqn2}
	\wt_x(M) \ \ \le \ \ -k_M \ \ \ \forall \text{ matchings}\ M\ \text{in}\ G,
\end{equation} where 
$\wt_x(M) = \sum_{u \in A \cup B}\vote_u(M(u),\vec{x})$. 
Note that we are in the graph $G$ (and not in $\tilde{G}$) -- so neither $\vec{x}$ nor $M$ needs to be perfect in $G$. We define therefore $\vote_u(M(u),\vec{x})$, i.e., $u$'s vote
for its assignment in $M$ versus its assignment in $\vec{x}$, as follows.

\begin{itemize}
	\item For $u \in U$:  $\vote_u(M(u),\vec{x}) = 1$ if $u$ is matched in $M$, else it is 0;
	\item For $u \in (A\cup B)\setminus U$: if $u$ is matched in $M$ then $\vote_u(M(u),\vec{x}) = \sum_{v': v' \prec_u v}x_{(u,v')} - \sum_{v': v' \succ_u v}x_{(u,v')}$, where $v$ is $u$'s partner in $M$; else (i.e., $u$ is unmatched in $M$) it is $-1$.
\end{itemize}

\smallskip

For a fixed $M$, $\wt_x(M)$ is a linear function of $\vec{x}$, hence
the set ${\cal X}_G \subset \mathbb{R}^m$ of points that satisfy (\ref{dom-eqn1}) and (\ref{dom-eqn2}) is a polytope. We now prove that ${\cal X}_G$ is the convex hull of dominant matchings in $G$.

The starting point is the fact that there is an efficient separation oracle for the constraints that define ${\cal X}_G$. 
We show next that, given a point $\vec{x}$ that satisfies the constraints in (\ref{dom-eqn1}), we can efficiently determine
if $\vec{x}$ satisfies all the constraints in (\ref{dom-eqn2}) by solving the max-weight perfect matching problem in the graph
$\tilde{G}$ (this is $G$ augmented with self-loops) with the edge weight function $c_x$. Recall that $c_x$ is exactly the same as $\wt_x$ 
defined in (\ref{eq:wt-x}) for edges $(a,b)$ where neither $a$ nor $b$ is in $U$ and $c_x(a,b) = 1$ if $a$ or $b$ is in $U$. 
We extend the function $c_x$ to  self-loops as $c_x(u,u) = 0$ for $u \in U$ and $c_x(u,u) = -1$ for 
$u \notin U$. Note that $c_x$ is an affine function of $\vec{x}$.

\smallskip        

\begin{new-claim}
	\label{claim1}
	Let $\vec{x}$ satisfy all constraints in (\ref{dom-eqn1}). Then $\vec{x} \in {\cal X}_G$ if and only if $c_x(M) \le 0$ for every perfect matching $M$ in $\tilde{G}$.
\end{new-claim}

\smallskip

If we had assigned $c_x(a,b) = 0$ for edges $(a,b)$ where $a$ or $b$ is in $U$, then for $e \in E$, $c_x(e) = \wt_x(e)$, which is the sum of votes of the endpoints of $e$ for each other versus their respective assignments in $\vec{x}$. 
Then it would have been $c_x(M) = \wt_x(M)$ for any perfect matching $M$ in $\tilde{G}$. With the above assignment of weights, i.e.,
with $c_x(a,b) = 1$ for edges $(a,b)$ where $a$ or $b$ is in $U$, we now have $c_x(M) = \wt_x(M) + k_M$ for any perfect 
matching $M$ in $\tilde{G}$. Thus Claim~\ref{claim1} follows.

Consider the max-weight perfect matching LP in $\tilde{G} = (A\cup B, \tilde{E})$ with respect to $c_x$: this is \eqref{LP5} below in variables $y_e$ for $e \in \tilde{E}$. The linear program \eqref{LP6} in variables $\alpha_u$ for $u \in A \cup B$ is its dual. 

\begin{linearprogram}
	{
		\label{LP5}
		\maximize \sum_{e \in \tilde{E}} c_x(e)\cdot y_e  
	}
	\qquad\sum_{e \in {\tilde{\delta}}(u)}y_e \ & = \ \ 1  \ \ \, \forall\, u \in A \cup B \notag\\
	y_e  \ & \ge \ \ 0   \ \ \ \forall\, e \in \tilde{E}. \notag
\end{linearprogram}

\begin{linearprogram}
	{
		\label{LP6}
		\minimize \sum_{u \in A \cup B}\alpha_u
	}
	\alpha_{a} + \alpha_{b} \ & \ge \ \ c_{x}(a,b) \ \ \ \ \forall\, (a,b)\in E \notag\\
	\alpha_u \ & \ge \ \ c_x(u,u) \ \ \ \, \forall\, u \in A \cup B. \notag
\end{linearprogram}

So $\vec{x} \in \mathbb{R}^m$ that satisfies the constraints in (\ref{dom-eqn1}) also satisfies the constraints in (\ref{dom-eqn2}) if 
and only if the
optimal solution of \eqref{LP5} is 0, equivalently, if and only if the optimal solution of \eqref{LP6} is 0; thus $\vec{x}\in{\cal X}_G$ if and only if there exists a dual feasible $\vec{\alpha} \in \mathbb{R}^n$ such that $\sum_{u\in A\cup B}\alpha_u = 0$. 
Hence ${\cal X}_G$ has the following compact extended formulation:

\begin{minipage}{.5\textwidth}
	\begin{eqnarray}
		\sum_{u\in A \cup B}\alpha_u \ & = & \ 0 \label{old-new-constr2}\\
		\qquad  \alpha_a + \alpha_b \ & \ge & \ c_x(a,b) \ \ \ \ \forall (a,b) \in E \label{old-new-constr1}\\ 
		\alpha_u \ & \ge & \ c_x(u,u) \ \ \  \ \forall u \in A\cup B\label{old-new-constr1-bis}\\
		\nonumber
	\end{eqnarray}
\end{minipage}\begin{minipage}{.45\textwidth}
\begin{eqnarray}
	\sum_{e\in\delta(u)}x_e \ & = & \ 1 \ \ \ \forall u\in (A\cup B)\setminus U\label{old-new-constr2-bis}\\ 
	x_e \ & = & \ 0 \ \ \  \forall e \in E \setminus E_D \label{old-new-constr3}\\  x_e \ &  \ge &  \ 0   \ \ \ \forall e \in E_D. \label{old-new-constr3-bis}\\
	\nonumber
\end{eqnarray}\end{minipage}

Let ${\cal Y}_G$ be the polytope described by~\eqref{old-new-constr2}-\eqref{old-new-constr3-bis}.
We will show that the projection of ${\cal Y}_G$ onto the space of the $\vec{x}$ variables is the dominant matching polytope ${\cal D}_G$.  Thus ${\cal X}_G$ is the same as ${\cal D}_G$. 

We proved in
Section~\ref{sec:dom-mat-poly-ext} that constraints~\eqref{last-constr3-bis-generic}-\eqref{eq:add-4} define an extension ${\cal R}_G$ of ${\cal D}_G$ (see Theorem~\ref{thm:ed-dominant}). 
Claim~\ref{lem:polytope} settles the upper bound given in Theorem~\ref{thm:dom}.

\begin{new-claim}
	\label{lem:polytope}
	${\cal Y}_G$ is obtained from ${\cal R}_G$ by projecting out coordinates $x_{(u,u)}$ for $u \in A \cup B$.
\end{new-claim}
\proof{} 
Consider any point $(\vec{x},\vec{\alpha}) \in {\cal R}_G$. 
\begin{enumerate}
	\item Constraint~\eqref{last-constr3-bis-generic} and constraint~\eqref{old-new-constr2} are the same.
	\item Constraints~\eqref{last-constr5-generic} and \eqref{eq:add-4} imply that $\sum_{e\in\delta(u)}x_e = 1$ for all $u \in(A\cup B)\setminus U$ and $x_{(u,u)} = 1$ for all $u \in U$. Thus the right side of constraint~\eqref{last-constr3-generic} is exactly the same as $c_x(u,u)$ for all vertices $u$. Thus constraints~\eqref{old-new-constr1}-\eqref{old-new-constr2-bis} hold.
	\item Constraints~\eqref{old-new-constr3} and \eqref{old-new-constr3-bis} are implied by constraints~\eqref{last-constr5-bis-generic} and~\eqref{eq:add-4}. 
\end{enumerate}
Thus if ${\cal R}'_G$ is the projection of ${\cal R}_G$ obtained by projecting out coordinates $x_{(u,u)}$ -- i.e., corresponding to self-loops -- then ${\cal R}'_G\subseteq {\cal Y}_G$.

\smallskip

Conversely, consider any point $(\vec{x},\vec{\alpha}) \in {\cal Y}_G$. Constraints~\eqref{old-new-constr1-bis} and \eqref{old-new-constr2-bis} imply that
$\alpha_u \ge c_x(u,u) = -1 = -\sum_{e\in\delta(u)}x_e$ for $u \in (A \cup B)\setminus U$. We also have $\alpha_u \ge c_x(u,u) = 0 = -\sum_{e\in\delta(u)}x_e$ for $u \in U$ (the second equality holds because every edge $e$ incident to $u \in U$ is in $E \setminus E_D$ and constraint~\eqref{old-new-constr3} tells us that $x_e = 0$). Thus constraints~\eqref{last-constr3-bis-generic}-\eqref{last-constr3-generic} hold.

Let us augment $\vec{x}$ with coordinates $x_{(u,u)}$ for all vertices $u$ defined as $x_{(u,u)} = 1$ for all $u \in U$ and $x_{(u,u)} = 0$ for all $u \in (A \cup B) \setminus U$.
\begin{itemize}
	\item Since $\sum_{e \in \delta(u)} x_{e} = 1$ for $u \in (A \cup B)\setminus U$ and $\sum_{e \in \delta(u)} x_{e} = 0$ for $u \in U$,  constraint~\eqref{last-constr5-generic} is satisfied. 
	\item Constraint~\eqref{last-constr5-bis-generic} clearly holds.
	\item Moreover, constraint~\eqref{old-new-constr3} along with the above values of $x_{(u,u)}$ for all $u$ implies constraint~\eqref{eq:add-4}.
\end{itemize}
Thus the augmented $(\vec{x}, \vec{\alpha}) \in {\cal R}_G$, in other words, the
original $(\vec{x}, \vec{\alpha}) \in {\cal R}'_G$ (the projection of ${\cal R}_G$ defined above). That is,  $ {\cal Y}_G \subseteq {\cal R}'_G$. 
Hence, ${\cal Y}_G = {\cal R}'_G$.   
\qed\endproof

\subsection{A Lower Bound on the Number of Facets of the Dominant Matching Polytope}
\label{sec:lb-facets} We will now prove the second part of Theorem~\ref{thm:dom}, i.e., the dominant matching polytope ${\cal D}_G$ has $\Omega(c^m)$ facets, for some constant $c>1$. For any
$k \ge 2$, we will construct an instance $G$ on $n = 8k+2$ vertices and $m = \Theta(k)$ edges and a family ${\cal N}$ of $2^k$ max-size popular matchings that are
{\em not} dominant. 
Because of the results from Section~\ref{sec:dominant-polytope}, for every matching $N \in {\cal N}$, there has to be some constraint in
\eqref{dom-eqn1} or~\eqref{dom-eqn2} that is violated by $N$. For any distinct $N, N' \in {\cal N}$, we will show that among
the constraints in \eqref{dom-eqn1}-\eqref{dom-eqn2}, the ones violated by $N$ and those violated by $N'$ have no overlap. This will lead to the lower bound.

\begin{figure}[h]
	\centerline{\resizebox{0.9\textwidth}{!}{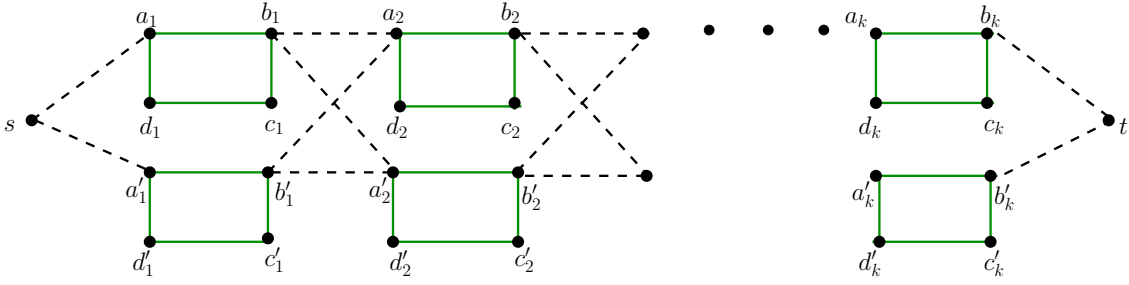}}
	\caption{The instance $G$ used in our lower bound argument. All matchings in ${\cal N}$ use only the solid edges.}
	\label{fig:lower-bound}
\end{figure}

Fig.~\ref{fig:lower-bound} shows the instance $G$ that will be used here. 
The preferences of $a'_i,b'_i,c'_i,d'_i$ for $1 \le i \le k$ will be totally
analogous to the preferences of the corresponding $a_i,b_i,c_i,d_i$, with $x'$ replacing $x$ and vice-versa
for every $x$ in $\cup_i\{a_i,b_i,c_i,d_i\}$.

We describe below the preference lists of $a_i,b_i,c_i,d_i$ for all $i$ other than $a_1$ and $b_k$.
\begin{eqnarray*}
	a_i\colon \, b_i \succ d_i \succ b_{i-1} \succ b'_{i-1} \hspace*{1in}  c_i\colon \, b_i \succ d_i  \\
	b_i\colon \, a_{i+1} \succ a'_{i+1} \succ a_i \succ c_i \hspace*{1in}  d_i\colon \, a_i \succ c_i.
\end{eqnarray*}  

Note that $b_i$ prefers both neighbors $a_{i+1}, a'_{i+1}$ that are outside its own gadget $\langle a_i,b_i,c_i,d_i\rangle$ 
to neighbors $a_i, c_i$ in its own gadget while
$a_i$ prefers both neighbors $b_i, d_i$ in its own gadget to neighbors $b_{i-1},b'_{i-1}$ that are outside its own gadget.
The preference lists of $a_1$ and $b_k$ are given below.
\[ a_1: \, b_1 \succ d_1 \succ s \hspace*{1in} b_k\colon \, a_k \succ c_k \succ t.\]

The preferences of $s$ and $t$ do not matter; in fact, they are {\em unpopular} vertices in $G$ since $M^* = \cup_{i=1}^k\{(a_i,d_i), (c_i,b_i), (a'_i,d'_i), (c'_i,b'_i)\}$ is a dominant matching in $G$.
To show this, we first claim that the vector $\vec{\alpha}$ with $\alpha_s=\alpha_t=0$, $\alpha_{u}=1$ for $u\in\{a_i,b_i,a'_i,b'_i\}$ for every $i \in [k]$, and  $\alpha_{u}=-1$ for $u\in\{c_i,d_i,c'_i,d'_i\}$ for every $i \in [k]$ is a witness for $M^*$. 

Observe that $\alpha_u=0=\wt_{M^*}(u,u)$ for $u\in\{s,t\}$, while $\alpha_u\geq -1=\wt_{M^*}(u,u)$ for all other vertices~$u$. Now take an edge $(u,v)$. If $(u,v) \in {M^*}$, then $\alpha_u + \alpha_v = 0 = \wt_{M^*}(u,v)$. If $u \in \{s,t\}$, then $\alpha_u+\alpha_v=1\geq \wt_{M^*}(u,v)=0$. If $u \in \{b_i,b_i'\}$ and $v \in \{a_{i+1},a_{i+1}'\}$ for some $i$, then $\alpha_u + \alpha_v = 2\geq \wt_{M^*}(u,v)$. Observe that for each $i \in [k]$, we have:
\begin{itemize}
	\item $\alpha_{a_i}+\alpha_{b_i}=2=\wt_{M^*}(a_i,b_i)$; \ $\alpha_{c_i}+\alpha_{d_i}=-2=\wt_{M^*}(c_i,d_i)$.
\end{itemize} The same equalities hold when we replace every vertex $x$ above with vertex $x'$. It is easy to see that $\sum_{u} \alpha_u = 0$ and this concludes the proof that $\vec{\alpha}$ is a witness for $M^*$. Hence $M^*$ is popular. Moreover, since edges $(c_1,d_1)$ and $(c'_1,d'_1)$ are labeled $(-,-)$, no $M^*$-alternating path starting from $s$ in $G_{M^*}$ can reach $t$. Hence, Theorem~\ref{thm:dominant-comb} implies that $M^*$ is a dominant matching.

We are now ready to define the family ${\cal N}$ of matchings. Any matching $N \in {\cal N}$ is 
constructed by choosing for each $i = 1,\ldots,k$:
\begin{itemize}
	\item either the 4 edges in $\{(a_i,b_i),(c_i,d_i),(a'_i,d'_i),(c'_i,b'_i)\}$ 
	\item or the  4 edges in $\{(a_i,d_i),(c_i,b_i),(a'_i,b'_i),(c'_i,d'_i)\}$.
\end{itemize}  

If the former happens, we say that $\langle a_i,b_i,c_i,d_i\rangle$ is in \emph{stable} state and $\langle a'_i,b'_i,c'_i,d'_i\rangle$ is in \emph{dominant} state. Else we say that $\langle a_i,b_i,c_i,d_i\rangle$ is in \emph{dominant} state and $\langle a'_i,b'_i,c'_i,d'_i\rangle$ is in \emph{stable} state.

Observe that when $\langle a_i,b_i,c_i,d_i\rangle$ is in dominant state, the edge $(c_i,d_i)$ is labeled $(-,-)$ wrt $N$ and
similarly, when $\langle a'_i,b'_i,c'_i,d'_i\rangle$ is in dominant state, the edge $(c'_i,d'_i)$ is labeled $(-,-)$ wrt $N$.
Recall that $G_N$ is $G$ with edges labeled $(-,-)$ wrt $N$ deleted. 
We now show that the following three properties are true in $G$:
\begin{enumerate}
	\item\label{label-z} every $N \in {\cal N}$ is a max-size popular matching and uses only dominant edges;
	\item\label{label-a} for each $N\in {\cal N}$, $G_N$ has exactly one $N$-augmenting path $\rho_N$;
	\item\label{label-b} for each matching $M$ in $G$, $M\supseteq (\rho_N\setminus N)$ for at most one $N \in {\cal N}$.
\end{enumerate}

\smallskip

\proof{ (of \eqref{label-z}.}
We will prove the popularity of $N$ by showing a witness $\vec{\alpha}$. Define $\vec{\alpha}$ as follows: first, let
$\alpha_s = \alpha_t = 0$. For $1 \le i \le k$:
\begin{itemize}
	\item if $N \supset \{(a_i,b_i),(c_i,d_i),(a'_i,d'_i),(c'_i,b'_i)\}$, i.e., if $\langle a_i,b_i,c_i,d_i\rangle$ is in stable state and $\langle a'_i,b'_i,c'_i,d'_i\rangle$ is in dominant state, then let $\alpha_{a_i} = \alpha_{b_i} = \alpha_{c_i} = \alpha_{d_i} = 0$ while
	$\alpha_{a'_i} = \alpha_{b'_i} = 1$ and $\alpha_{c'_i} = \alpha_{d'_i} = -1$.
	\item else let $\alpha_{a_i} = \alpha_{b_i} = 1$ and $\alpha_{c_i} = \alpha_{d_i} = -1$ while
	$\alpha_{a'_i} = \alpha_{b'_i} = \alpha_{c'_i} = \alpha_{d'_i} = 0$.
\end{itemize}

It is easy to check that every edge is {\em covered}, i.e., $\alpha_u + \alpha_v \ge \wt_N(u,v)$ for all edges $(u,v)$ in $G$. 
In particular, we have $\wt_N(e) = 0$ for every dashed edge $e$ in Fig.~\ref{fig:lower-bound} and $\alpha_u \ge 0$ for $u \in \cup_i\{a_i,b_i,a'_i,b'_i\}$.
We also have $\sum_u\alpha_u = 0$ and $\alpha_u \ge \wt_N(u,u)$ for all vertices $u$. Thus $\vec{\alpha}$ is a witness of $N$'s popularity in $G$. 
Moreover, $N$ matches all vertices other than $s$ and $t$, exactly as the dominant matching $M^*$ does.  Hence $N$ is a max-size popular matching in $G$.

Observe that the matchings $M = \cup_{i=1}^{k-1}\{(a_i,b_i),(c_i,d_i),(a'_i,b'_i),(c'_i,d'_i)\} \cup \{(a_k,d_k),(c_k,b_k),(a'_k,d'_k)$, $(c'_k,b'_k)\}$ and $M' = \{(a_1,d_1),(c_1,b_1),(a'_1,d'_1),(c'_1,b'_1)\}\cup_{i=2}^{k}\{(a_i,b_i),(c_i,d_i),(a'_i,b'_i),(c'_i,d'_i)\}$ are dominant. The proof is analogous to the one previously given for $M^*$, hence it is omitted.

Observe that every edge in $N$ is in $M^* \cup M \cup M'$. Thus $N$ uses only dominant edges. \qed\endproof

\smallskip

\proof{ (of \eqref{label-a}).}
The main observation here is that if a gadget, say $\langle a_i,b_i,c_i,d_i\rangle$, is in dominant state in $N$ then
no $N$-alternating path in $G_N$ can pass through $\langle a_i,b_i,c_i,d_i\rangle$. This is because the edge $(c_i,d_i)$ is {\em not} present in $G_N$.
Thus there is a unique $N$-augmenting path $\rho_N$ in $G_N$: for each $i$, $\rho_N$
uses either $(a_i,b_i)$ or $(a'_i,b'_i)$ depending on whether $\alpha_{a_i} = 0$ or $\alpha_{a_i} = 1$. If
$\alpha_{a_i} = 0$, then $(a_i,b_i) \in \rho_N$, else $(a'_i,b'_i) \in \rho_N$.
Note that the edges in $\rho_N \setminus N$ belong to $G_N$ since each such edge $e$ (a dashed edge in Fig.~\ref{fig:lower-bound}) 
satisfies $\wt_N(e) = 0$. \qed\endproof

\smallskip

\proof{ (of \eqref{label-b}).}
Let $M$ be any matching in $G$ and let $N, N' \in {\cal N}$. There is some index $i$ such that one of $N, N'$ 
contains the edge $(a_i,b_i)$ and the other contains the edge $(a'_i,b'_i)$. Let $i$ be the largest such index. Assume without loss of generality that
$(a_i,b_i) \in N$ and $(a'_i,b'_i) \in N'$.

\begin{itemize}
	\item If $i = k$, then the augmenting path $\rho_N$ ends with
	the edge $(b_k,t)$ while the augmenting path $\rho_{N'}$ ends with the edge $(b'_k,t)$. Since only one of $(b_k,t), (b'_k,t)$ can be present in $M$, 
	it follows that $M$ can be a superset of either $(\rho_N\setminus N)$ or $(\rho_{N'}\setminus N')$ but not both.
	
	\item Suppose $i < k$. 
	By the definition of index $i$, either $(a_{i+1},b_{i+1}) \in N \cap N'$ or $(a'_{i+1},b'_{i+1}) \in N \cap N'$.
	Assume the former (without loss of generality). So $\rho_N \setminus N$ contains $(b_i,a_{i+1})$ while $\rho_{N'} \setminus N'$ contains $(b'_i,a_{i+1})$. 
	Since only one of $(b_i,a_{i+1}), (b'_i,a_{i+1})$ can be present in $M$, it again follows that $M$ can be a superset of only one of $(\rho_N\setminus N)$,
	$(\rho_{N'}\setminus N')$. 
\end{itemize}
Thus $M\supseteq (\rho_N\setminus N)$ for at most one $N \in {\cal N}$. \qed\endproof

\paragraph{\bf Our Lower Bound Argument.}
For $N \in {\cal N}$, let ${\cal M}_N$ be the set of matchings $M$ such that $M \supseteq (\rho_N\setminus N)$. 
We show the following claim for every $N \in {\cal N}$ and $T \notin {\cal M}_N$. 

\begin{new-claim}
	\label{thm2:lb-claim}
	$N$ satisfies $\wt_N(T) \leq - k_T$ for all matchings $T \notin {\cal M}_N$. 
\end{new-claim}
\proof{} 
Let $T \notin {\cal M}_N$ and let $e \in (\rho_N \setminus N) \setminus T$. Let $G^e$ be the instance obtained from $G$ by removing edge $e$. Note that $T,N$ are matchings of $G^e$, and $N$ is popular in $G^e$ (since $N$ is popular in $G$).
There is no $N$-augmenting path in $G^e_N$, since $\rho_N$ was the only $N$-augmenting path in $G_N$ (from property~(\ref{label-a})). Hence,
from Theorem~\ref{thm:dominant-comb} we conclude that $N$ is dominant in $G^e$. That is, $N \in {\cal D}_{G^e}$. Since $e \notin T$, the value $k_T$ is the same in both $G$ and $G^e$. Hence $\wt_N(T) \leq - k_T$. 
\qed\endproof

\smallskip

Since $I_N \notin {\cal D}_G$, there must be some inequality from~\eqref{dom-eqn1} or~\eqref{dom-eqn2} that is not satisfied by $I_N$. 
We know from property~\eqref{label-z} that the incidence vector $I_N$ of matching $N$ satisfies all constraints in \eqref{dom-eqn1}.
Hence, because of Claim~\ref{thm2:lb-claim}, $I_N$ must be cut off by an inequality $\wt_N(M) \leq - k_M$ for some $M \in {\cal M}_N$.
Thus, in any minimal system contained in~\eqref{dom-eqn1}--\eqref{dom-eqn2}, at least one inequality $\wt_N(M) \leq - k_M$ for some $M \in {\cal M}_N$ is present.

Since ${\cal M}_N\cap {\cal M}_{N'}=\emptyset$ for $N'\neq N$ in ${\cal N}$ (from property~\eqref{label-b}), any such minimal system contains at
least $|{\cal N}|=2^k = 2^{\Theta(m)}$ inequalities.  
The lower bound in Theorem~\ref{thm:dom} follows from the fact that inequalities in a minimal system are in one-to-one correspondence with the facets of 
the polyhedron they describe, see e.g.~\cite[Theorem~3.30]{CCZ}. This finishes the proof of Theorem~\ref{thm:dom}.

\section{Conclusions and Open Problems}\label{sec:open-problems}  
Our algorithms show that one can circumvent the strong hardness result for computing a min-cost popular matching by relaxing the notion of popularity via the concept of unpopularity factor.  An open question is to show an efficient algorithm for computing matchings with a smaller unpopularity factor. 
Another open problem is to obtain a trade-off with respect to the two parameters: cost and unpopularity factor.

A future line of research is whether bi-criteria algorithms 
yield positive results in other matching problems under preferences. 
The landscape of matchings under preferences is rich with algorithmic results 
and there are also several intractability results. We refer to the books/monographs~\cite{GI89,knuth1976,manlove2013,roth1992} on this subject. 
It would be interesting to explore efficient bi-criteria algorithms for many well-known NP-hard problems here: such results may also be of
practical relevance.

\medskip

\noindent {\bf Acknowledgments.} 
We thank two anonymous reviewers of an earlier version of this paper for very helpful comments that improved the current presentation. We are indebted to one of them for the (simpler) proof of Theorem~\ref{thm:dominant-extn} included here.

\newpage
\pagenumbering{roman}
\section*{Appendix: Proof of Theorem~\ref{lemma:new} (derived from \cite{HK17})}

\proof{}
Consider any $a \in A'$. Since $\vec{q'}$ is half-integral, the vertex $a$ has at most 2 partners in $\vec{q'}$. Call these 2 partners $b$ and $b'$: it could be the case that $b = b'$. If $b = b'$ then we have $q'_{(a,b)} = 1$; else $q'_{(a,b)} = q'_{(a,b')} = 1/2$. We will form an array $X'_a$ with 2 cells (each of length $1/2$) as follows. In case $a$ has a single partner $b$ in $\vec{q'}$ then both cells of $X'_a$ have $b$ in them.
Else, recalling that $\vec{\alpha'} \in \{0,\pm 1\}^{n'}$ (where $n' = |A'\cup B'|$) and assuming without loss of generality that $b\prec_a b'$:
\begin{itemize}
	\item if $\alpha'_a \in \{ \pm 1\}$, then $X'_a=(b,b')$, i.e., $a$'s partners in $\vec{q'}$ are arranged in {\em increasing} order of $a$'s preference order. 
	\item if $\alpha'_a = 0$ then $X'_a=(b',b)$, i.e., $a$'s partners in $\vec{q'}$ are arranged in {\em decreasing} order of $a$'s preference order. 
\end{itemize}

Consider any $b \in B'$. Since $\vec{q'}$ is half-integral, the vertex $b$ has at most 2 partners in $\vec{q'}$. 
Call these 2 partners $a$ and $a'$: it could be the case that $a = a'$. If $a = a'$ then we have $q'_{(a,b)} = 1$; else $q'_{(a,b)} = q'_{(a',b)} = 1/2$. 
We will form an array $X'_b$ with 2 cells (each of length $1/2$) as follows. In case $b$ has a single partner $a$ in $\vec{q'}$ then both cells of $X'_b$ have $a$ in them.
Else, assuming without loss of generality $a\prec_b a'$: 
\begin{itemize}
	\item if $\alpha'_b \in \{ \pm 1\}$, then $X'_b=(a',a)$, i.e., $b$'s partners in $\vec{q'}$ are arranged in {\em decreasing} order of $b$'s preference order. 
	\item if $\alpha'_b = 0$, then $X'_b=(a,a')$, i.e., $b$'s partners in $\vec{q'}$ are arranged in {\em increasing} order of $b$'s preference order.
\end{itemize}

Define sets $M_1$ and $M_2$ as follows:
\begin{eqnarray*}
	M_1 & \ = \ & \{(u,v): u \in A'\cup B' \ \text{and}\ v \ \text{is\ in\ the\ first\ cell\ of}\ X'_u\}\\
	M_2 & \ = \ & \{(u,v): u \in A'\cup B' \ \text{and}\ v \ \text{is\ in\ the\ second\ cell\ of}\ X'_u\}.
\end{eqnarray*}
We will now adapt the proof from \cite{HK17}  to show in Lemma~\ref{clm:integral} and Lemma~\ref{clm:popular} that both $M_1$ and $M_2$ are popular
matchings in $G'$. Thus $\vec{q'} = (I_{M_1} + I_{M_2})/2$ for popular matchings $M_1$ and $M_2$ in $G'$.   
\qed\endproof

\begin{lemma}
	\label{clm:integral}
	$M_1$ and $M_2$ are matchings in $G'$.
\end{lemma}  
\proof{} 
For any vertex $u \in A \cup B$, we need to show that if $(a,b) \in M_1$ (resp., $M_2$) then there is no other edge incident to either $a$ or $b$ in $M_1$ (resp., $M_2$).
If $q'_{(a,b)} = 1$, then $X'_a$ has the vertex $b$ in both cells and similarly, $X'_b$ has the vertex $a$ in both cells. Thus $(a,b)$ is the only
edge incident to $a$ and to $b$ in both $M_1$ and $M_2$. 

Suppose $q'_{(a,b)} = 1/2$. We need to show that $b$ is in the first (resp., second) cell of $X'_a$ if and only if $a$ is in the first (resp., second) cell of $X'_b$.
It was shown in \cite{HK17} that the LP that leads to the formulation of the popular fractional matching polytope (constraints~\eqref{sec5-constr3-bis}-\eqref{sec5-constr5-bis}
for the graph $G'$) is {\em self-dual}, i.e., the dual LP is same as the primal LP. So $(\vec{q'},\vec{\alpha'})$ is both a primal optimal and a dual optimal solution. Complementary slackness implies that for every edge $(a,b)$ such that $q'_{(a,b)} > 0$, we have $\alpha'_a + \alpha'_b = \wt_{q'}(a,b)$; in other words, the edge
covering constraint for edge $(a,b)$ is {\em tight}. So we have:
\[ \alpha'_a + \alpha'_b \ = \ \sum_{b': b' \prec_a b}q'_{(a,b')} - \sum_{b': b' \succ_a b}q'_{(a,b')} \ + \ \sum_{a': a' \prec_b a}q'_{(a',b)} - \sum_{a': a' \succ_b a}q'_{(a',b)}.\]

Since $\vec{q'}$ is a half-integral matching and $q'_{(a,b)} = 1/2$, there is a unique neighbor $b' \ne b$ of $a$ such that $q'_{(a,b')} = 1/2$. Similarly, there is a unique
neighbor $a' \ne a$ of $b$ such that $q'_{(a',b)} = 1/2$.
Let $I_{b'\prec_a b}$ be the indicator variable that is 1 if $b'\prec_a b$ and it is 0 otherwise. Similarly for $I_{b'\succ_a b}, I_{a'\prec_b a}$, and $I_{a'\succ_b a}$.
Hence:
\begin{eqnarray*}
	\alpha'_a + \alpha'_b \ =  \ \frac{1}{2}\cdot\left(I_{b'\prec_a b} - I_{b'\succ_a b} + I_{a'\prec_b a} - I_{a'\succ_b a}\right)  
	\ & = & \ \left(2\cdot I_{b'\prec_a b} \ - 1 + \ 2\cdot I_{a'\prec_b a} \ - \ 1\right)/2\\
	\ & = & \ I_{b'\prec_a b} \ + \ I_{a'\prec_b a} \ - \ 1.
\end{eqnarray*}
Thus $(\alpha'_a + \alpha'_b) \in \{0, \pm 1\}$. 

\medskip

\noindent{\em Case 1.} $\alpha'_a + \alpha'_b = 0$. So either (i)~$I_{b'\prec_a b} = 1$ and $I_{a'\prec_b a} = 0$ 
or  (ii)~$I_{b'\prec_a b} = 0$ and $I_{a'\prec_b a} = 1$.
In other words, in the half-integral matching $\vec{q'}$: either (i)~$b$ is the better of $a$'s two partners in $\vec{q}$ and $a$ is the worse of $b$'s two partners in $\vec{q}$ or (ii)~$b$ is the worse of $a$'s two partners in $\vec{q}$ and $a$ is the better of $b$'s two partners in $\vec{q}$.

Since $\alpha'_a + \alpha'_b = 0$, either $\alpha'_a = \alpha'_b = 0$ or $\{\alpha_a, \alpha_b\} = \{1, -1\}$. 

\begin{itemize}
	\item[(1.1)] Suppose $\alpha'_a = \alpha'_b = 0$.
	In case~(i) above, $b$ is in the {\em first} cell of $X'_a$ and similarly,  $a$ is in the {\em first} cell of $X'_b$.
	In case~(ii), $b$ is in the {\em second} cell of $X'_a$ and similarly, $a$ is in the {\em second} cell of $X'_b$. 
	
	\item[(1.2)] Suppose $\{\alpha_a, \alpha_b\} = \{1, -1\}$. In case~(i), $b$ is in the {\em second} cell of $X'_a$ and similarly, 
	$a$ is in the {\em second} cell of $X'_b$. 
	In case~(ii), $b$ is in the {\em first} cell of $X'_a$ and similarly, $a$ is in the {\em first} cell of $X'_b$. 
\end{itemize}

\medskip

\noindent{\em Case 2.} $\alpha'_a + \alpha'_b = 1$. So $I_{b' \prec_a b} = I_{a' \prec_b a} = 1$.
Hence, in the half-integral matching $\vec{q'}$: $b$ is the better of $a$'s two partners in $\vec{q}$ and $a$ is the better of $b$'s two partners in $\vec{q}$. 
Since $\alpha'_a + \alpha'_b = 1$, $\{\alpha'_a, \alpha'_b\} = \{0,1\}$. 
\begin{itemize}
	\item[(2.1)] If $\alpha'_a = 0$ and $\alpha'_b = 1$, then $b$ is in the first cell of $X'_a$
	and similarly, $a$ is in the first cell of $X'_b$. 
	\item[(2.2)] If $\alpha'_a = 1$ and $\alpha'_b = 0$, then $b$ is in the second cell of $X'_a$ and similarly, $a$ is in the second cell of $X'_b$. 
\end{itemize}

\medskip

\noindent{\em Case 3.} $\alpha'_a + \alpha'_b = -1$. So $I_{b' \prec_a b} = I_{a' \prec_b a} = 0$. So in the half-integral matching $\vec{q'}$: $b$ is the worse of $a$'s two partners in $\vec{q}$ and $a$ is the worse of $b$'s two partners in $\vec{q}$. 
Since $\alpha'_a + \alpha'_b = -1$, $\{\alpha'_a, \alpha'_b\} = \{0,-1\}$. 
\begin{itemize}
	\item[(3.1)] If $\alpha'_a = 0$ and $\alpha'_b = -1$, then $b$ is in the second cell of $X'_a$
	and $a$ is in the second cell of $X'_b$. 
	\item[(3.2)] If $\alpha'_a = -1$ and $\alpha'_b = 0$, then $b$ is in the first cell of $X'_a$ and $a$ is in the first cell of $X'_b$. 
\end{itemize}
Thus in all cases, we have seen that the cell with $b$ in $X'_a$ and the cell with $a$ in $X'_b$ are exactly aligned with each other.  
\qed\endproof

\begin{lemma}
	\label{clm:popular}
	$M_1$ and $M_2$ are popular in $G'$.
\end{lemma}  
\proof{} 
We will use the vectors $\vec{\beta}$ and $\vec{\gamma}$ defined in Section~\ref{sec:pop-frac-matching} below the statement of Theorem~\ref{lemma:new}.
For all $v \in A \cup B$, we have: $\beta_v \ge -1 = \wt_{M_1}(v,v)$ and $\gamma_v \ge -1 = \wt_{M_2}(v,v)$. 
Claim~\ref{clm:zero} and Claim~\ref{clm:cover} show that $\vec{\beta}$ and $\vec{\gamma}$ satisfy the other properties of witnesses we need to show.
Hence $M_1$ and $M_2$ are popular in $G'$.  
\qed\endproof

\begin{new-claim}
	\label{clm:zero}   
	$\sum_{u \in A'\cup B'} \beta_u = 0$ and $\sum_{u \in A'\cup B'} \gamma_u = 0$.
\end{new-claim}
\proof{} 
For every edge $(a,b) \in M_1$, we will show that $\beta_a + \beta_b = 0$. Since $M_1$ is a perfect matching, this implies $\sum_{u \in A'\cup B'} \beta_u = 0$. 
Let $(a,b) \in M_1$. 

Suppose $q'_{(a,b)} = 1$. Then using complementary slackness as discussed in the proof of Lemma~\ref{clm:integral}, we deduce $\alpha'_a + \alpha'_b = \wt_{q'}(a,b) = 0$. So either $\alpha'_a = \alpha'_b = 0$ or $\{\alpha'_a,\alpha'_b\} = \{-1,1\}$.
In the former case, $\beta_a = -1$ and $\beta_b = 1$; in the latter case, $\beta_a = \alpha'_a$ and $\beta_b = \alpha'_b$, thus $\{\beta_a,\beta_b\} = \{-1,1\}$.
Hence in all cases, we have $\beta_a + \beta_b = 0$. 

Suppose $q'_{(a,b)} = 1/2$. Consider all cases in the proof of Lemma~\ref{clm:integral}. 
\begin{itemize}
	\item Consider case~(1.1). In sub-case~(i): $\beta_a = -1$ and $\beta_b = 1$; in sub-case~(ii): $\beta_a = 1$ and $\beta_b = -1$.
	\item In case~(1.2), we have $\beta_a = \alpha'_a$ and $\beta_b = \alpha'_b$. Since $\{\alpha'_a, \alpha'_b\} = \{1, -1\}$, we have
	$\{\beta_a, \beta_b\} = \{1, -1\}$.
	\item In case~(2.1) and in case~(3.2), we have $\beta_a = -1$ and $\beta_b = 1$.
	\item In case~(2.2) and in case~(3.1), we have $\beta_a = 1$ and $\beta_b = -1$.
\end{itemize}

Thus in all cases, we have $\beta_a + \beta_b = 0$.
It can similarly be shown that for every edge $(a,b) \in M_2$, we have $\gamma_a + \gamma_b = 0$.
Thus $\sum_{u \in A'\cup B'} \gamma_u = 0$.  
\qed\endproof

\begin{new-claim}
	\label{clm:cover}
	For every edge $(a,b)$ in $G'$, we have: $\beta_a + \beta_b \ge \wt_{M_1}(a,b)$ and $\gamma_a + \gamma_b \ge \wt_{M_2}(a,b)$.
\end{new-claim}
\proof{} 
Let $(a,b)$ be any edge in $G'$. We will show that $\beta_a + \beta_b \ge \wt_{M_1}(a,b)$. The proof that $\gamma_a + \gamma_b \ge \wt_{M_2}(a,b)$ will be analogous.
We know that $\alpha'_a + \alpha'_b \ge \wt_{q'}(a,b)$ (since $(\vec{q'},\vec{\alpha'}) \in {\cal L}_G$). Let us consider the following cases:
\begin{enumerate}
	\item $\alpha'_a = \alpha'_b = 1$: in this case $\beta_a = \beta_b = 1$. We have $\beta_a + \beta_b = 2 \ge \wt_{M_1}(a,b)$.
	
	\smallskip
	
	\item $\alpha'_a = \alpha'_b = -1$: since $\wt_{q'}(a,b) \le \alpha'_a + \alpha'_b$, this means $\wt_{q'}(a,b) \le -2$, i.e., $\wt_{q'}(a,b) = -2$. 
	
	\smallskip
	
	\begin{itemize}
		\item So $a$ prefers both its partners in $\vec{q'}$ to $b$ and similarly, $b$ prefers both its partners in $\vec{q'}$ to $a$. Hence both $a$ and $b$ prefer their
		respective partners in $M_1$ to each other. Thus $\wt_{M_1}(a,b) = -2$. Since $\beta_a, \beta_b \ge -1$, we have $\beta_a + \beta_b \ge \wt_{M_1}(a,b)$.
	\end{itemize}
	
	\begin{figure}[h]
		\centerline{\resizebox{0.58\textwidth}{!}{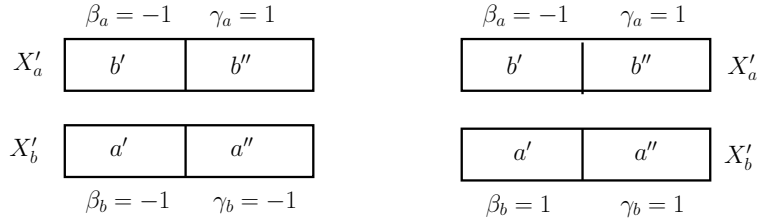}}
		\caption{The case where $\alpha'_a = 0$ and $\alpha'_b = -1$ is shown on the left and the case where $\alpha'_a = 0$ and $\alpha'_b = 1$ is shown on the right. In both    cases, we have $b' \succ_a b''$ and $a' \succ_b a''$.}
		\label{fig22}
	\end{figure}
	
	\vspace*{-0.3cm}
	
	\item $\alpha'_a = 0$ and $\alpha'_b = -1$: so $\wt_{q'}(a,b) \le \alpha'_a + \alpha'_b = -1$.  We have $\beta_a = \beta_b = -1$ here
	(see Fig.~\ref{fig22}, left).
	
	\smallskip
	
	\begin{itemize}
		\item Since $\wt_{q'}(a,b) \le -1$, either both $\vote_a(b,\vec{q'}), \vote_b(a,\vec{q'})$ are at most $-1/2$ or one of $\vote_a(b,\vec{q'})$, $\vote_b(a,\vec{q'})$ is 0 and the other is $-1$.
		In both cases, $a$ and $b$ prefer their more preferred partners in $\vec{q'}$ to each other.
		Moreover, $a$ and $b$ are matched in $M_1$ to their more preferred partners in $\vec{q'}$. Thus $\wt_{M_1}(a,b) = -2 = \beta_a + \beta_b$.
	\end{itemize}
	
	\smallskip
	
	\item $\alpha'_a = 0$ and $\alpha'_b = 1$: so $\wt_{q'}(a,b) \le \alpha'_a + \alpha'_b = 1$.  Here $\beta_a = -1$ and $\beta_b = 1$ here (see Fig.~\ref{fig22}, right).
	
	\smallskip
	
	\begin{itemize}
		\item If $q'_{(a,b)} > 0$ then $\wt_{q'}(a,b) = \alpha'_a + \alpha'_b = 1$, i.e., $q'_{(a,b)} = 1/2$ (since $\wt_{q'}(a,b)$ is odd). 
		So $\vote_a(b,\vec{q'}) = \vote_b(a,\vec{q'}) = 1/2$.
		In other words, $a$ and $b$ are each other's more preferred partners in $\vec{q'}$. Observe that when $\alpha'_a = 0$ and $\alpha'_b = 1$, $a$ and $b$ are matched in $M_1$ to their more preferred partners in $\vec{q'}$. Thus $(a,b) \in M_1$ and we have $\beta_a + \beta_b = 0 = \wt_{M_1}(a,b)$.
		
		\item Suppose $q'_{(a,b)} = 0$. Since $\wt_{q'}(a,b) \le 1$, we have $\vote_a(b,\vec{q'}) \le 0$ or $\vote_b(a,\vec{q'}) \le 0$. This means 
		one of $a, b$ prefers its more preferred partner in $\vec{q'}$ to the other. 
		Since $a$ and $b$ are matched in $M_1$ to their more preferred partners in $\vec{q'}$, we have $\wt_{M_1}(a,b) \le 0 = \beta_a + \beta_b$.
	\end{itemize}
	
	\smallskip
	
	The cases where $\alpha'_a \in \{\pm 1\}$ and $\alpha'_b = 0$ are analogous to case~3 and case~4.
	
	\medskip
	
	\item $\{\alpha'_a,\alpha'_b\} = \{-1,1\}$. Suppose $\alpha'_a = -1$ and $\alpha'_b = 1$.
	We have $\beta_a = -1$ and $\beta_b = 1$ here, also $\wt_{q'}(a,b) \le 0$. 
	
	\smallskip
	
	\begin{itemize}
		\item If $q'_{(a,b)} > 0$ then $\alpha'_a + \alpha'_b = 0 = \wt_{q'}(a,b)$.
		So $b$ is matched in $M_1$ either to $a$ or to a neighbor preferred to $a$. Hence, $\wt_{M_1}(a,b) \le 0 = \beta_a + \beta_b$.
		
		\smallskip
		
		\item Suppose $q'_{(a,b)} = 0$. Since $\wt_{q'}(a,b) \le 0$,
		either (i)~one of $a, b$ prefers both its partners in $\vec{q'}$ to the other or (ii)~both $a,b$ like their more preferred partner in $\vec{q'}$ to each other. In both cases, one of $a,b$ is matched in $M_1$ to a neighbor preferred to the other. Thus $\wt_{M_1}(a,b) \le 0$, hence $\beta_a + \beta_b = 0 \ge \wt_{M_1}(a,b)$.
	\end{itemize}
	
	\smallskip
	
	The case where $\alpha'_a = 1$ and $\alpha'_b = -1$ is analogous.
	
	\medskip
	
	\item $\alpha'_a = \alpha'_b = 0$: here we have $\beta_a = -1$ and $\beta_b = 1$, also $\wt_{q'}(a,b) \le 0$.  
	
	\smallskip
	
	\begin{itemize}
		\item If $q'_{(a,b)} > 0$, then $\alpha'_a + \alpha'_b = 0 = \wt_{q'}(a,b)$. 
		So $a$ is matched in $M_1$ either to $b$ or to a neighbor preferred to $b$. Hence, $\wt_{M_1}(a,b) \le 0 = \beta_a + \beta_b$.
		
		\smallskip
		
		\item Suppose $q'_{(a,b)} = 0$. Since $\wt_{q'}(a,b) \le 0$,
		either (i)~one of $a, b$ prefers both its partners in $\vec{q'}$ to the other or (ii)~both $a,b$ like their more preferred partner in $\vec{q'}$ to each other. In both cases, one of $a,b$ is matched in $M_1$ to a neighbor preferred to the other. Thus, $\wt_{M_1}(a,b) \le 0$ and so $\beta_a + \beta_b = 0 \ge \wt_{M_1}(a,b)$.
	\end{itemize}
	
	\smallskip
	
	This finishes the proof of the claim.  \qed
\end{enumerate}
\endproof

\end{document}